\documentclass[zpreprint,zbstdefault]{./zeus_paper}
\usepackage[english]{babel}
\usepackage[running]{lineno}

\chardef\usc=95
\chardef\til=126
\catcode`\@=11 % @ signs are now treated as letters
\DeclareRobustCommand\xdotspace{\futurelet\@let@token\@xdotspace}
\def\@xdotspace{%
  \ifx\@let@token.\else
  \ifx\@let@token\bgroup.\else
  \ifx\@let@token\egroup.\else
  \ifx\@let@token\/.\else
  \ifx\@let@token\ .\else
  \ifx\@let@token~.\else
  \ifx\@let@token!.\else
  \ifx\@let@token,.\else
  \ifx\@let@token:.\else
  \ifx\@let@token;.\else
  \ifx\@let@token?.\else
  \ifx\@let@token/.\else
  \ifx\@let@token'.\else
  \ifx\@let@token).\else
  \ifx\@let@token-.\else
  \ifx\@let@token\@xobeysp.\else
  \ifx\@let@token\space.\else
  \ifx\@let@token\@sptoken.\else
   .\space
   \fi\fi\fi\fi\fi\fi\fi\fi\fi\fi\fi\fi\fi\fi\fi\fi\fi\fi}
\catcode`\@=12 % @ signs are no longer letters

\newcommand{\stru}[2]{%
   \relax\ifmmode\hbox{\vrule height#1 depth#2 width0pt}%
   \else\vrule height#1 depth#2 width0pt\fi}
\newcommand{\Ronum}[1]{\uppercase\expandafter{\romannumeral#1}}
\newcommand{\ronum}[1]{\expandafter{\romannumeral#1}}
\DeclareRobustCommand{\LaTeXZ}{%
  \LaTeX\kern-.05em4\kern-.1em
  {\raisebox{-0.2ex}{$\scriptstyle\text{ZEUS}$}}\xspace}
\DeclareMathAlphabet{\mathbf}{OT1}{cmr}{bx}{sl}
\newcommand{\eVdist}{\kern-0.06667em}

\newcommand{\Mev}{{\text{Me}\eVdist\text{V\/}}}
\newcommand{\Gev}{{\text{Ge}\eVdist\text{V\/}}}

\newcommand{\gev}{{\,\text{Ge}\eVdist\text{V\/}}}

\newcommand{\pbi}{\,\text{pb}^{-1}}

\newcommand{\cm}{\,\text{cm}}

\newcommand{\slashfrac}[2]{%
  \raisebox{0.5ex}{\ensuremath #1}\kern-0.12em/\kern-0.08em
  \raisebox{-.8ex}{\ensuremath #2}}
\newcommand{\sqr}[3]{%
    {\vcenter{\hrule height.#3ex\hbox{\vrule width.#2ex height#1ex
     \kern#1ex\vrule width.#3ex}\hrule height.#2ex}}}

\catcode`\@=11 % @ signs are now treated as letters
\newcommand{\parenbar}{\mathpalette\p@renb@r}
\def\p@renb@r#1#2{\vbox{%
  \ifx#1\scriptscriptstyle \dimen@.7em\dimen@ii.2em\else
  \ifx#1\scriptstyle \dimen@.8em\dimen@ii.25em\else
  \dimen@1em\dimen@ii.4em\fi\fi \offinterlineskip
  \ialign{\hfill##\hfill\cr
    \vbox{\hrule width\dimen@ii}\cr
    \noalign{\vskip-.3ex}%
    \hbox to\dimen@{$\mathchar300\hfil\mathchar301$}\cr
    \noalign{\vskip-.3ex}%
    $#1#2$\cr}}}
\catcode`\@=12 % @ signs are no longer letters

\newcommand{\IP}{{\rm I$\kern-0.01667em$P}\xspace}

\mathchardef\qsm=63
\mathchardef\pls=43
\mathchardef\mns=512
\mathchardef\plm=518
\mathchardef\eql=61
\mathchardef\smallleft=300
\mathchardef\smallright=301
\mathchardef\les=316
\mathchardef\gre=318
\mathchardef\leq=532
\mathchardef\grq=533
\catcode`\@=11 % @ signs are now treated as letters
\newcounter{pict@width}
\newcounter{pict@height}
\newlength{\pict@scale}
\newcommand{\psfigadd}[4]{%
\setcounter{pict@width}{1*\ratio{#2+\pict@scale/2}{\pict@scale}}
\setcounter{pict@height}{1*\ratio{#3+\pict@scale/2}{\pict@scale}}
\setlength{\unitlength}{\pict@scale}
\hbox to #2{\hspace{-\fill}\begin{picture}(\thepict@width,\thepict@height)
\put(0,0){\psfig{figure=#1,width=#2,height=#3,clip=}}
\SetScale{0.283466457}
\SetWidth{1.763889}
{#4}
\end{picture}}
}
\newcounter{pict@widthfst}
\newcounter{pict@widthscd}
\newcounter{pict@widthtot}
\newcommand{\psfigaddtwo}[7]{%
\setcounter{pict@widthfst}{1*\ratio{#2+\pict@scale/2}{\pict@scale}}
\setcounter{pict@widthscd}{1*\ratio{#2+#4+\pict@scale/2}{\pict@scale}}
\setcounter{pict@widthtot}{1*\ratio{#2+#4+#6+\pict@scale/2}{\pict@scale}}
\setcounter{pict@height}{1*\ratio{#3+\pict@scale/2}{\pict@scale}}
\setlength{\unitlength}{\pict@scale}
\hbox{\hspace{-\fill}\begin{picture}(\thepict@widthtot,\thepict@height)
\put(0,0){\psfig{figure=#1,width=#2,height=#3,clip=}}
\put(\thepict@widthscd,0){\psfig{figure=#5,width=#6,height=#3,clip=}}
\SetScale{0.283466457}
\SetWidth{1.763889}
{#7}
\end{picture}}
}
\newcommand{\psfigror}[4]{%
\setcounter{pict@width}{1*\ratio{#2+\pict@scale/2}{\pict@scale}}
\setcounter{pict@height}{1*\ratio{#3+\pict@scale/2}{\pict@scale}}
\setlength{\unitlength}{\pict@scale}
\hbox{\begin{picture}(\thepict@width,\thepict@height)
\put(0,\thepict@height){\psfig{figure=#1,width=#3,height=#2,clip=,angle=270}}
\SetScale{0.283466457}
\SetWidth{1.763889}
{#4}
\end{picture}}
}
\newcommand{\psfigrol}[4]{%
\setcounter{pict@width}{1*\ratio{#2+\pict@scale/2}{\pict@scale}}
\setcounter{pict@height}{1*\ratio{#3+\pict@scale/2}{\pict@scale}}
\setlength{\unitlength}{\pict@scale}
\hbox{\begin{picture}(\thepict@width,\thepict@height)
\put(0,0){\psfig{figure=#1,width=#3,height=#2,clip=,angle=90}}
\SetScale{0.283466457}
\SetWidth{1.763889}
{#4}
\end{picture}}
}
\catcode`\@=12 % @ signs are no longer letters
\newlength\listtextwidth

\catcode`\@=11 % @ signs are now treated as letters
\newlength{\@tabfninsert}
\newlength{\@tabfnwidth}
\newcommand{\tabfootnote}[2]{%
  \setlength{\@tabfninsert}{0.8em}
  \setlength{\@tabfnwidth}{\textwidth}
  \addtolength{\@tabfnwidth}{-\@tabfninsert}
  \addtolength{\@tabfnwidth}{-0.4em}
  \noindent\makebox[\@tabfninsert][r]{\footnotesize$^{#1}$\hfil}\hfill%
  \parbox[t]{\@tabfnwidth}{\footnotesize #2\hfill}}
\catcode`\@=12 % @ signs are no longer letters

\def\citeCTD{{\cite{%
nim:a279:290,*npps:b32:181,*nim:a338:254%
}}\xspace}
\def\citeCAL{{\cite{%
nim:a309:77,*nim:a309:101,*nim:a321:356,*nim:a336:23%
}}\xspace}
\def\citeMVD{{\cite{%
nim:a581:656%
}}\xspace}
\def\citeSRTD{{\cite{%
nim:a401:63%
}}\xspace}
\def\citePRES{{\cite{%
nim:a382:419%
}}\xspace}
\begin{document}

%\linenumbers
%------------------------------------------------------------------------------
%       Title sheet
%------------------------------------------------------------------------------

\prepnum{{DESY-09-229}}

\title{
Scaled momentum spectra in deep inelastic scattering at HERA
}                                                       
                    
\author{ZEUS Collaboration}
%\draftversion{5.1}
\date{December 2009}
%\date{\today}

\abstract{
Charged particle production has been studied in neutral current deep inelastic 
$ep$ scattering with the ZEUS detector at HERA using 
an integrated luminosity of 0.44\,fb$^{-1}$.
Distributions of scaled momenta in the Breit frame are presented
for particles in the current fragmentation region.
The evolution of these spectra with the photon virtuality, $Q^2$,
is described in the kinematic region $10<Q^2<41000\gev^2$. 
Next-to-leading-order and modified leading-log-approximation
QCD calculations as well as predictions from
Monte Carlo models are compared to the data.
The results are also compared to $e^+e^-$ annihilation data.
The dependences of the pseudorapidity distribution of the particles
on $Q^2$ and on the energy in the $\gamma p$ system, $W$,
are presented and interpreted in the context of 
the hypothesis of limiting fragmentation.
}

\makezeustitle

%------------------------------------------------------------------------------
%       Authors
%------------------------------------------------------------------------------
%\include{paper-aut}

%===================================================================
%
%  MEMBER NAME  AUTH162a (ZEUS)     M  TEX
% 
%  JH.: transformed to a format, which is suited as input for
%       CONVERT, which automatically creates author-indices
%
%  Don't remove lines starting with a percent sign %,
%  CONVERT may need them urgently !
%  
%=====================================================================

                                                   %
\begin{center}
{                      \Large  The ZEUS Collaboration              }
\end{center}

%       members:

H.~Abramowicz$^{44, ad}$, I.~Abt$^{34}$, L.~Adamczyk$^{13}$, M.~Adamus$^{53}$, S.~Antonelli$^{4}$, P.~Antonioli$^{3}$, \\
A.~Antonov$^{32}$, M.~Arneodo$^{49}$, V.~Aushev$^{26, y}$, Y.~Aushev$^{26, y}$
% ...

%       institutes:

\makebox[3em]{$^{1}$}
\begin{minipage}[t]{14cm}
{\it Argonne National Laboratory, Argonne, Illinois 60439-4815, USA}~$^{A}$

\end{minipage}\\
\makebox[3em]{$^{2}$}
\begin{minipage}[t]{14cm}
{\it Andrews University, Berrien Springs, Michigan 49104-0380, USA}

\end{minipage}\\
\makebox[3em]{$^{3}$}
\begin{minipage}[t]{14cm}
{\it INFN Bologna, Bologna, Italy}~$^{B}$

\end{minipage}\\
\makebox[3em]{$^{4}$}
\begin{minipage}[t]{14cm}
{\it University and INFN Bologna, Bologna, Italy}~$^{B}$

\end{minipage}\\
\makebox[3em]{$^{5}$}
\begin{minipage}[t]{14cm}
{\it Physikalisches Institut der Universit\"at Bonn,
Bonn, Germany}~$^{C}$

\end{minipage}\\
\makebox[3em]{$^{6}$}
\begin{minipage}[t]{14cm}
{\it H.H.~Wills Physics Laboratory, University of Bristol,
Bristol, United Kingdom}~$^{D}$

\end{minipage}\\
\makebox[3em]{$^{7}$}
\begin{minipage}[t]{14cm}
{\it Panjab University, Department of Physics, Chandigarh, India}

\end{minipage}\\
\makebox[3em]{$^{8}$}
\begin{minipage}[t]{14cm}
{\it Calabria University,
Physics Department and INFN, Cosenza, Italy}~$^{B}$

\end{minipage}\\
\makebox[3em]{$^{9}$}
\begin{minipage}[t]{14cm}
{\it Chonnam National University, Kwangju, South Korea}

\end{minipage}\\
\makebox[3em]{$^{10}$}
\begin{minipage}[t]{14cm}
{\it Jabatan Fizik, Universiti Malaya, 50603 Kuala Lumpur, Malaysia}~$^{E}$

\end{minipage}\\
\makebox[3em]{$^{11}$}
\begin{minipage}[t]{14cm}
{\it Nevis Laboratories, Columbia University, Irvington on Hudson,
New York 10027, USA}~$^{F}$

\end{minipage}\\
\makebox[3em]{$^{12}$}
\begin{minipage}[t]{14cm}
{\it The Henryk Niewodniczanski Institute of Nuclear Physics, Polish Academy of Sciences, Cracow,
Poland}~$^{G}$

\end{minipage}\\
\makebox[3em]{$^{13}$}
\begin{minipage}[t]{14cm}
{\it Faculty of Physics and Applied Computer Science,
AGH-University of Science and \mbox{Technology}, Cracow, Poland}~$^{H}$

\end{minipage}\\
\makebox[3em]{$^{14}$}
\begin{minipage}[t]{14cm}
{\it Department of Physics, Jagellonian University, Cracow, Poland}

\end{minipage}\\
\makebox[3em]{$^{15}$}
\begin{minipage}[t]{14cm}
{\it Deutsches Elektronen-Synchrotron DESY, Hamburg, Germany}

\end{minipage}\\
\makebox[3em]{$^{16}$}
\begin{minipage}[t]{14cm}
{\it Deutsches Elektronen-Synchrotron DESY, Zeuthen, Germany}

\end{minipage}\\
\makebox[3em]{$^{17}$}
\begin{minipage}[t]{14cm}
{\it INFN Florence, Florence, Italy}~$^{B}$

\end{minipage}\\
\makebox[3em]{$^{18}$}
\begin{minipage}[t]{14cm}
{\it University and INFN Florence, Florence, Italy}~$^{B}$

\end{minipage}\\
\makebox[3em]{$^{19}$}
\begin{minipage}[t]{14cm}
{\it Fakult\"at f\"ur Physik der Universit\"at Freiburg i.Br.,
Freiburg i.Br., Germany}~$^{C}$

\end{minipage}\\
\makebox[3em]{$^{20}$}
\begin{minipage}[t]{14cm}
{\it Department of Physics and Astronomy, University of Glasgow,
Glasgow, United \mbox{Kingdom}}~$^{D}$

\end{minipage}\\
\makebox[3em]{$^{21}$}
\begin{minipage}[t]{14cm}
{\it Department of Engineering in Management and Finance, Univ. of
the Aegean, Chios, Greece}

\end{minipage}\\
\makebox[3em]{$^{22}$}
\begin{minipage}[t]{14cm}
{\it Hamburg University, Institute of Exp. Physics, Hamburg,
Germany}~$^{C}$

\end{minipage}\\
\makebox[3em]{$^{23}$}
\begin{minipage}[t]{14cm}
{\it Imperial College London, High Energy Nuclear Physics Group,
London, United \mbox{Kingdom}}~$^{D}$

\end{minipage}\\
\makebox[3em]{$^{24}$}
\begin{minipage}[t]{14cm}
{\it Institute of Particle and Nuclear Studies, KEK,
Tsukuba, Japan}~$^{I}$

\end{minipage}\\
\makebox[3em]{$^{25}$}
\begin{minipage}[t]{14cm}
{\it Institute of Physics and Technology of Ministry of Education and
Science of Kazakhstan, Almaty, \mbox{Kazakhstan}}

\end{minipage}\\
\makebox[3em]{$^{26}$}
\begin{minipage}[t]{14cm}
{\it Institute for Nuclear Research, National Academy of Sciences, and
Kiev National University, Kiev, Ukraine}

\end{minipage}\\
\makebox[3em]{$^{27}$}
\begin{minipage}[t]{14cm}
{\it Kyungpook National University, Center for High Energy Physics, Daegu,
South Korea}~$^{J}$

\end{minipage}\\
\makebox[3em]{$^{28}$}
\begin{minipage}[t]{14cm}
{\it Institut de Physique Nucl\'{e}aire, Universit\'{e} Catholique de
Louvain, Louvain-la-Neuve, \mbox{Belgium}}~$^{K}$

\end{minipage}\\
\makebox[3em]{$^{29}$}
\begin{minipage}[t]{14cm}
{\it Departamento de F\'{\i}sica Te\'orica, Universidad Aut\'onoma
de Madrid, Madrid, Spain}~$^{L}$

\end{minipage}\\
\makebox[3em]{$^{30}$}
\begin{minipage}[t]{14cm}
{\it Department of Physics, McGill University,
Montr\'eal, Qu\'ebec, Canada H3A 2T8}~$^{M}$

\end{minipage}\\
\makebox[3em]{$^{31}$}
\begin{minipage}[t]{14cm}
{\it Meiji Gakuin University, Faculty of General Education,
Yokohama, Japan}~$^{I}$

\end{minipage}\\
\makebox[3em]{$^{32}$}
\begin{minipage}[t]{14cm}
{\it Moscow Engineering Physics Institute, Moscow, Russia}~$^{N}$

\end{minipage}\\
\makebox[3em]{$^{33}$}
\begin{minipage}[t]{14cm}
{\it Moscow State University, Institute of Nuclear Physics,
Moscow, Russia}~$^{O}$

\end{minipage}\\
\makebox[3em]{$^{34}$}
\begin{minipage}[t]{14cm}
{\it Max-Planck-Institut f\"ur Physik, M\"unchen, Germany}

\end{minipage}\\
\makebox[3em]{$^{35}$}
\begin{minipage}[t]{14cm}
{\it NIKHEF and University of Amsterdam, Amsterdam, Netherlands}~$^{P}$

\end{minipage}\\
\makebox[3em]{$^{36}$}
\begin{minipage}[t]{14cm}
{\it Physics Department, Ohio State University,
Columbus, Ohio 43210, USA}~$^{A}$

\end{minipage}\\
\makebox[3em]{$^{37}$}
\begin{minipage}[t]{14cm}
{\it Department of Physics, University of Oxford,
Oxford United Kingdom}~$^{D}$

\end{minipage}\\
\makebox[3em]{$^{38}$}
\begin{minipage}[t]{14cm}
{\it INFN Padova, Padova, Italy}~$^{B}$

\end{minipage}\\
\makebox[3em]{$^{39}$}
\begin{minipage}[t]{14cm}
{\it Dipartimento di Fisica dell' Universit\`a and INFN,
Padova, Italy}~$^{B}$

\end{minipage}\\
\makebox[3em]{$^{40}$}
\begin{minipage}[t]{14cm}
{\it Department of Physics, Pennsylvania State University,
University Park, Pennsylvania 16802, USA}~$^{F}$

\end{minipage}\\
\makebox[3em]{$^{41}$}
\begin{minipage}[t]{14cm}
{\it Polytechnic University, Sagamihara, Japan}~$^{I}$

\end{minipage}\\
\makebox[3em]{$^{42}$}
\begin{minipage}[t]{14cm}
{\it Dipartimento di Fisica, Universit\`a 'La Sapienza' and INFN,
Rome, Italy}~$^{B}$

\end{minipage}\\
\makebox[3em]{$^{43}$}
\begin{minipage}[t]{14cm}
{\it Rutherford Appleton Laboratory, Chilton, Didcot, Oxon,
United Kingdom}~$^{D}$

\end{minipage}\\
\makebox[3em]{$^{44}$}
\begin{minipage}[t]{14cm}
{\it Raymond and Beverly Sackler Faculty of Exact Sciences,
School of Physics, Tel Aviv University, Tel Aviv, Israel}~$^{Q}$

\end{minipage}\\
\makebox[3em]{$^{45}$}
\begin{minipage}[t]{14cm}
{\it Department of Physics, Tokyo Institute of Technology,
Tokyo, Japan}~$^{I}$

\end{minipage}\\
\makebox[3em]{$^{46}$}
\begin{minipage}[t]{14cm}
{\it Department of Physics, University of Tokyo,
Tokyo, Japan}~$^{I}$

\end{minipage}\\
\makebox[3em]{$^{47}$}
\begin{minipage}[t]{14cm}
{\it Tokyo Metropolitan University, Department of Physics,
Tokyo, Japan}~$^{I}$

\end{minipage}\\
\makebox[3em]{$^{48}$}
\begin{minipage}[t]{14cm}
{\it Universit\`a di Torino and INFN, Torino, Italy}~$^{B}$

\end{minipage}\\
\makebox[3em]{$^{49}$}
\begin{minipage}[t]{14cm}
{\it Universit\`a del Piemonte Orientale, Novara, and INFN, Torino,
Italy}~$^{B}$

\end{minipage}\\
\makebox[3em]{$^{50}$}
\begin{minipage}[t]{14cm}
{\it Department of Physics, University of Toronto, Toronto, Ontario,
Canada M5S 1A7}~$^{M}$

\end{minipage}\\
\makebox[3em]{$^{51}$}
\begin{minipage}[t]{14cm}
{\it Physics and Astronomy Department, University College London,
London, United \mbox{Kingdom}}~$^{D}$

\end{minipage}\\
\makebox[3em]{$^{52}$}
\begin{minipage}[t]{14cm}
{\it Warsaw University, Institute of Experimental Physics,
Warsaw, Poland}

\end{minipage}\\
\makebox[3em]{$^{53}$}
\begin{minipage}[t]{14cm}
{\it Institute for Nuclear Studies, Warsaw, Poland}

\end{minipage}\\
\makebox[3em]{$^{54}$}
\begin{minipage}[t]{14cm}
{\it Department of Particle Physics, Weizmann Institute, Rehovot,
Israel}~$^{R}$

\end{minipage}\\
\makebox[3em]{$^{55}$}
\begin{minipage}[t]{14cm}
{\it Department of Physics, University of Wisconsin, Madison,
Wisconsin 53706, USA}~$^{A}$

\end{minipage}\\
\makebox[3em]{$^{56}$}
\begin{minipage}[t]{14cm}
{\it Department of Physics, York University, Ontario, Canada M3J
1P3}~$^{M}$

\end{minipage}\\

%       references concerning institutes:

\begin{tabular}[h]{rp{14cm}}

$^{ A}$  &  supported by the US Department of Energy\\
$^{ B}$  &  supported by the Italian National Institute for Nuclear Physics (INFN) \\
$^{ C}$  &  supported by the German Federal Ministry for Education and Research (BMBF), under
 contract Nos. 05 HZ6PDA, 05 HZ6GUA, 05 HZ6VFA and 05 HZ4KHA\\
$^{ D}$  &  supported by the Science and Technology Facilities Council, UK\\
$^{ E}$  &  supported by an FRGS grant from the Malaysian government\\
$^{ F}$  &  supported by the US National Science Foundation. Any opinion,
 findings and conclusions or recommendations expressed in this material
 are those of the authors and do not necessarily reflect the views of the
 National Science Foundation.\\
$^{ G}$  &  supported by the Polish State Committee for Scientific Research, project No.
 DESY/256/2006 - 154/DES/2006/03\\
$^{ H}$  &  supported by the Polish Ministry of Science and Higher Education
 as a scientific project (2009-2010)\\
$^{ I}$  &  supported by the Japanese Ministry of Education, Culture, Sports, Science and Technology
 (MEXT) and its grants for Scientific Research\\
$^{ J}$  &  supported by the Korean Ministry of Education and Korea Science and Engineering
 Foundation\\
$^{ K}$  &  supported by FNRS and its associated funds (IISN and FRIA) and by an Inter-University
 Attraction Poles Programme subsidised by the Belgian Federal Science Policy Office\\
$^{ L}$  &  supported by the Spanish Ministry of Education and Science through funds provided by
 CICYT\\
$^{ M}$  &  supported by the Natural Sciences and Engineering Research Council of Canada (NSERC) \\
$^{ N}$  &  partially supported by the German Federal Ministry for Education and Research (BMBF)\\
$^{ O}$  &  supported by RF Presidential grant N 1456.2008.2 for the leading
 scientific schools and by the Russian Ministry of Education and Science through its
 grant for Scientific Research on High Energy Physics\\
$^{ P}$  &  supported by the Netherlands Foundation for Research on Matter (FOM)\\
$^{ Q}$  &  supported by the Israel Science Foundation\\
$^{ R}$  &  supported in part by the MINERVA Gesellschaft f\"ur Forschung GmbH, the Israel Science
 Foundation (grant No. 293/02-11.2) and the US-Israel Binational Science Foundation \\

\end{tabular}

%  references concerning mebers;

\begin{tabular}[h]{lp{14cm}}

$^{ a}$ & also affiliated with University College London,
 United Kingdom\\
$^{ b}$ & now at University of Salerno, Italy\\
$^{ c}$ & now at Queen Mary University of London, United Kingdom\\
$^{ d}$ & also working at Max Planck Institute, Munich, Germany\\
$^{ e}$ & also Senior Alexander von Humboldt Research Fellow at Hamburg University,
 Institute of \mbox{Experimental} Physics, Hamburg, Germany\\
$^{ f}$ & supported by Chonnam National University, South Korea, in 2009\\
$^{ g}$ & now at Institute of Aviation, Warsaw, Poland\\
$^{ h}$ & supported by the research grant No. 1 P03B 04529 (2005-2008)\\
$^{ i}$ & This work was supported in part by the Marie Curie Actions Transfer of Knowledge
 project COCOS (contract MTKD-CT-2004-517186)\\
$^{ j}$ & now at DESY group FS-CFEL-1\\
$^{ k}$ & now at DESY group FEB, Hamburg, Germany\\
$^{ l}$ & also at Moscow State University, Russia\\
$^{ m}$ & now at University of Liverpool, United Kingdom\\
$^{ n}$ & on leave of absence at CERN, Geneva, Switzerland\\
$^{ o}$ & now at CERN, Geneva, Switzerland\\
$^{ p}$ & also at Institute of Theoretical and Experimental
 Physics, Moscow, Russia\\
$^{ q}$ & also at INP, Cracow, Poland\\
$^{ r}$ & also at FPACS, AGH-UST, Cracow, Poland\\
$^{ s}$ & partially supported by Warsaw University, Poland\\
$^{ t}$ & partially supported by Moscow State University, Russia\\
$^{ u}$ & also affiliated with DESY, Germany\\
$^{ v}$ & now at Japan Synchrotron Radiation Research Institute (JASRI), Hyogo, Japan\\
$^{ w}$ & also at University of Tokyo, Japan\\
$^{ x}$ & now at Kobe University, Japan\\
$^{ y}$ & supported by DESY, Germany\\
$^{ z}$ & partially supported by Russian Foundation for Basic
 Research grant \mbox{No. 05-02-39028-NSFC-a}\\
$^{\dagger}$ &  deceased \\
$^{aa}$ & STFC Advanced Fellow\\
$^{ab}$ & nee Korcsak-Gorzo\\

\end{tabular}

\begin{tabular}[h]{lp{14cm}}

$^{ac}$ & This material was based on work supported by the
 National Science Foundation, while working at the Foundation.\\
$^{ad}$ & also at Max Planck Institute, Munich, Germany, Alexander von Humboldt
 Research Award\\
$^{ae}$ & now at Nihon Institute of Medical Science, Japan\\
$^{af}$ & now at SunMelx Co. Ltd., Tokyo, Japan\\
$^{ag}$ & now at Osaka University, Osaka, Japan\\
$^{ah}$ & now at University of Bonn, Germany\\
$^{ai}$ & also at \L\'{o}d\'{z} University, Poland\\
$^{aj}$ & member of \L\'{o}d\'{z} University, Poland\\
$^{ak}$ & now at Lund University, Lund, Sweden\\
$^{al}$ & also at University of Podlasie, Siedlce, Poland\\

\end{tabular}

%------------------------------------------------------------------------------
%       Text
%------------------------------------------------------------------------------
\pagenumbering{arabic} 
\pagestyle{plain}

\newcommand{\gammah}    {\mbox{$\gamma_{_{H}}$}}
\newcommand{\gammahc}   {\mbox{$\gamma_{_{PT}}$}}
\newcommand{\sleq} {\raisebox{-.6ex}{${\textstyle\stackrel{<}{\sim}}$}}
\newcommand{\sgeq} {\raisebox{-.6ex}{${\textstyle\stackrel{>}{\sim}}$}}
\newcommand{\leff}{\Lambda_{\mathrm{eff}}} 
% ----------------------------------------------------------------------------
%       Introduction
% ----------------------------------------------------------------------------
\section{Introduction}
\label{sec-int}

Quark fragmentation has previously been studied experimentally 
in deep inelastic $ep$ scattering (DIS) at HERA   
using observables such as multiplicity moments, scaled
momentum distributions and fragmentation
functions~\cite{zfp:c67:93,*pl:b414:428,epj:c11:251,h1:frag}. 
The results were compared to those obtained in $e^+e^-$ and $p\overline p$
collisions. In general, universal behaviour has been established and
scaling violations of the fragmentation functions\cite{alt79,*webnas}
observed. It has also been observed that perturbative 
Quantum Chromodynamics (pQCD) calculations using 
the modified leading-log-approximation 
(MLLA)~\cite{Fong:1989qy,*Fong:1990nt,*Dokshitzer:1991ej,MLLA}
and assuming local parton-hadron duality (LPHD)~\cite{Azimov:1984np} 
do not provide a full description of the data.

In this paper, multiplicity distributions of charged hadrons 
in the current region in the Breit frame\footnote{
The Breit frame is defined as the frame in which
the four-vector of the exchanged photon becomes (0,0,0,-Q).}
are presented as functions of the virtuality of the 
exchanged boson, $Q^2$ per unit of the scaled momentum, 
$x_p = {2P_{\rm Breit}}/{Q}$, and the variable $\ln (1/{x_p})$ 
in bins of $Q^2$.
Here,  $P_{\rm  Breit}$ denotes the momentum of a hadron in the Breit frame.
The data sample collected with the ZEUS detector between 
1996--2007,  comprising $0.44\,\rm{fb^{-1}}$, 
enables the study to be extended to $Q^2$ as high as $41\,000\,\gev^2$.
Predictions from next-to-leading-order (NLO) QCD calculations 
that combine full NLO matrix elements with fragmentation 
functions (FF) obtained from fits to $e^+e^-$ annihilation
data~\cite{Kretzer:2000yf,Kniehl:2000cr,Albino:2005me,*Albino:2008fy,Florian:2007aj,*Florian:2007hc} 
are compared to the measurements.  Predictions from
MLLA+LPHD~\cite{MLLA,Khoze:1996dn,Khoze:1996py} and
leading-order plus parton-shower  Monte Carlo programs 
are also considered.

In addition,  the measurements are compared to previous $ep$ 
results~\cite{zfp:c67:93,*pl:b414:428,epj:c11:251,h1:frag}
and to $e^+e^-$ annihilation 
data~\protect\cite{mark2:1988,*Braunschweig:1990yd,*Li:1989sn,*delphi:1993}.
The hadronisation in the current region in the Breit frame 
in $ep$ scattering can be compared directly to the hadronisation 
in one hemisphere of $e^+e^-$ annihilation events.
There, particle momenta are scaled 
to half of the centre-of-mass energy, $E^*=\sqrt{s}/2$. Previous studies on  
DIS hadronisation~ \cite{zfp:c67:93,*pl:b414:428,epj:c11:251,h1:frag,:2008hy} 
have shown good agreement with $e^+e^-$ annihilation at medium $Q^2$.  
At lower $Q^2$, $Q^{2}<40\,{\Gev}^2$, the two processes were
observed to behave differently~\cite{epj:c11:251}. 
This can be explained by higher-order QCD processes such 
as boson-gluon fusion (BGF) and initial-state QCD Compton radiation
occurring as part of the hard
interaction in $ep$ scattering but not in $e^+e^-$  annihilation.

Finally, the density of charged particles is studied as a
function of the particle pseudorapidity, $\eta^{\rm Breit}$,
in bins of $Q^2$, 
and as a function of the total $\gamma^\star p$ centre-of-mass energy, $W$.  
The data are used to test the hypothesis of limiting
fragmentation~\cite{limiting}, in which there has recently been renewed
interest, most notably in relativistic heavy ion 
collisions~\cite{Back:2005hs,*Adams:2005cy,*PHOBOS,*Bearden:2001qq,*Bearden:2001xw}.

\section{Experimental setup}
%%%%%%%%%%%%%%%%%%%%%%%%%%%%%%%%%%%%%%%%%%%%%%%%%%%%%%%%%%%%%%%%%%%

A detailed description of the ZEUS detector can be found
elsewhere\cite{zeus:1993:bluebook}. A brief outline of the
components most relevant for this analysis is given below.  Charged
particles were tracked in the central tracking detector (CTD)\citeCTD,
which operated in a magnetic field of 1.43~T provided by a thin
superconducting solenoid. The CTD consisted of 72 cylindrical
drift-chamber layers, organized in nine superlayers covering the
polar-angle\footnote{ The ZEUS coordinate system is  a
right-handed Cartesian system, with the $Z$ axis pointing in the
proton beam direction, referred to as the ``forward direction'', and
the $X$ axis pointing towards the centre of HERA. The coordinate
origin is at the nominal interaction point.}  region $15^\circ <
\theta < 164^\circ$ and the radial range from $18.2$~cm to $79.4$~cm. 
Before the 2003--2007 running period, the ZEUS tracking system was
upgraded with a silicon microvertex detector~(MVD)\citeMVD.

The high-resolution uranium--scintillator calorimeter~(CAL)~\citeCAL
covered 99.7\% of the total solid angle and consisted of three parts:
the forward~(FCAL), the barrel~(BCAL) and the rear~(RCAL)
calorimeters. Each part was subdivided transversely into towers and
longitudinally into one electromagnetic section~(EMC) and either
one~(in RCAL) or two~(in BCAL and FCAL) hadronic sections~(HAC). The
smallest subdivision of the CAL was called a cell. Under test-beam
conditions, the CAL single-particle relative energy resolutions were
$\sigma_E/E=0.18/\sqrt{E}$ for electrons and
$\sigma_E/E=0.35/\sqrt{E}$ for hadrons, with $E$ in GeV.

The position of the scattered electron was determined by combining
information from the CAL, the small-angle rear tracking
detector~(SRTD)~\citeSRTD and the presampler~(PRES)~\citePRES, both
mounted on the face of the RCAL.

The luminosity was measured using 
the Bethe-Heitler process $ep\to e\gamma p$
by a luminosity detector which consisted of 
a lead--scintillator~\cite{desy-92-066,zfp:c63:391,acpp:b32:2025} calorimeter
and, after 2002, an independent magnetic spectrometer~\cite{nim:a565:572}.
The fractional systematic
uncertainty on the measured luminosity was 2\% 
and 2.6\% for the 1996--2000 and 2004--2007 running periods, respectively.

%       Event selection
% ----------------------------------------------------------------------------
\section{Event sample}
\label{sec-exp}

The data presented here were collected with the ZEUS detector at HERA
between 1996 and 2007 and correspond to an integrated luminosity of $0.44
{\rm \ fb^{-1}}$. During 1995--97 (1998-2007) HERA operated with
protons with an energy of $E_p = 820\,\gev$ ($920\,\gev$) and
electrons\footnote{The term ``electron" is used for both
electrons and positrons.} with an energy of $E_e =
27.5\,\gev$, resulting in a centre-of-mass energy of $\sqrt{s} =
300\,\gev$ ($318\,\gev$).

A three-level trigger system~\cite{zeus:1993:bluebook,Smith:1992im,gtt} 
was used to select events online.  It relied on the presence of an energy
deposition in the CAL compatible with that of a scattered electron.
At the third level, an electron, identified using the pattern of energy
deposits in the CAL~\cite{nim:a365:508} and having an energy larger than
$4\gev$, was required.

The kinematic variables, $Q^2$ and the Bjorken scaling variable, $x$,
as well as the boost vector to the Breit frame were reconstructed using the
double angle (DA) method~\cite{proc:hera:1991:23} based on the
angles of the scattered electron and of the hadronic system. 
The total energy of the $\gamma p$-system, $W$, was calculated using
$W^2=Q_{\rm DA}^2(1-x_{\rm DA})/x_{\rm DA}$.

The tracks used in the analysis had to be  associated 
with the primary interaction vertex and were required 
to be in the region of high CTD acceptance, 
$|\eta|<1.75$, where $\eta=-\ln(\tan\theta/2)$  is the pseudorapidity 
of the track in the laboratory frame with $\theta$ being the polar angle 
of the measured track with respect to the proton direction. 
The tracks had to pass through at least three CTD superlayers
and were required to have a transverse momentum, 
$P_{T}^{\rm track}>150\,\Mev$. 
The details of the event reconstruction are similar to those in
a previous ZEUS publication~\cite{epj:c11:251} 
 and are described in detail elsewhere~\cite{Beata}.  

To reconstruct $x_p$,  the momentum four-vector of each track was boosted
to the Breit frame assuming the particle to have the pion mass. 

The analysis of the scaled momenta was restricted to events 
with $Q^{2}>160\,{\Gev}^2$.
A well reconstructed  neutral current DIS sample was selected
by requiring the following: 

\begin{itemize}

\item[$\bullet$]
$E^{\prime}_{e}>10\,\Gev$, where $E^\prime_e$ is the
energy of the scattered electron;

\item[$\bullet$]
$y_e\,\leq\,0.95$, where $y_e$ is  the inelasticity, $y=Q^2/sx$,
estimated from the energy and angle of the scattered electron;
this reduces the photoproduction background;

\item[$\bullet$] 
$y_{\rm JB}\,\geq\,0.04$, where $y_{\rm JB}$ is estimated by
the Jacquet-Blondel method~\cite{proc:epfacility:1979:391};
this rejects events for which the DA method
gives a poor reconstruction;

\item[$\bullet$]
 $35\,\leq\,\delta\,\leq\,60\,\Gev$, where
$\delta=\sum(E_i-P_{Z_i})$ and $E_i$  
is the energy of the $i$-th calorimeter cell, 
$P_{Z_i}$ is its momentum along the $Z$ axis and the sum runs over all cells; 
this removes photoproduction and events with initial-state radiation;

\item[$\bullet$]
$|Z_{\rm vertex}|<50\,\cm$, where $Z_{\rm vertex}$ is the $Z$ component 
of the position of the primary interaction vertex; 
this reduces background from events not originating from  $ep$ collisions;

\item[$\bullet$]
the position $(X,Y)$ of the scattered electron candidates in the RCAL was 
required to satisfy $\sqrt{X^2+Y^2}>35$~cm.

\end{itemize}

The limiting-fragmentation analysis was extended to events with lower  
$Q^{2}$ values, $Q^{2}>10\,{\Gev}^2$, using data
collected during 1996--2000 corresponding to an integrated 
luminosity of 77\,$\pbi$. 
The same selection as described  above was used but, \, for events with
$10 < Q^2 < 160~{\Gev}^2$, the last requirement was modified and 
one additional requirement introduced:

\begin{itemize}
\item[$\bullet$]
depending on experimental conditions, the position $(X,Y)$ of 
the scattered electron candidate in the RCAL was required 
to satisfy $|X|>12$~cm and $|Y|>6$~cm  or $|X|>14$~cm and $|Y|>14$~cm
or $\sqrt{X^2+Y^2}>35$~cm.

\item[$\bullet$]

${\eta}_{\rm max}>3.2$, where ${\eta}_{\rm max}$ 
is the pseudorapidity in the laboratory frame of 
the most forward energy deposit with at least 400~MeV;
this removes diffractive events. Even before the cut, 
for $Q^{2}>100\,{\Gev}^2$, the contamination with diffractive events
is well below 5\%, and the resulting corrections to the multiplicity
are below 2\%.   In the lowest $Q^2$ bin, the contribution of diffractive events
would grow up to 10\% without the cut.

\end{itemize}
Only  tracks with $P_Z^{\rm  Breit}<0$ enter the analysis 
of the scaled momenta, whereas this restriction does not apply
to the limiting-fragmentation studies.  The total number 
of events is listed in Table~\ref{tab:every}.

% ----------------------------------------------------------------------------
%       Models
% ----------------------------------------------------------------------------
\section{Models and Monte Carlo Simulations}
\label{sec-theory}

The NLO perturbative QCD calculations considered, combine the full NLO matrix
elements with NLO fragmentation functions obtained from fits to $e^+e^-$
data~\cite{Kretzer:2000yf,Kniehl:2000cr,Albino:2005me,*Albino:2008fy,Florian:2007aj,*Florian:2007hc}.
The resulting predictions were obtained using
the {\sc Cyclops} program~\cite{albino-2007-75}.

The MLLA
calculations~\cite{Bassetto:1982ma,Mueller:1982cq,Webber:1983if,Khoze:1996dn,MLLA}
describe parton production in terms of a shower evolution. 
%energy scale chosen~\cite{Fong:1989qy,*Fong:1990nt,*Dokshitzer:1991ej} and
They depend on two parameters only, the effective QCD scale,  $\leff$,  
and the infrared cutoff scale, $Q_0$, at which the parton cascade is stopped.
The calculations intrinsically include colour coherence and gluon
interference effects. Leading collinear and infrared singularities are removed 
and energy--momentum conservation is obeyed.

To connect predictions 
at the parton level to the hadron-level data, 
LPHD~\cite{Azimov:1984np} was assumed, which leaves only one free parameter,
the hadronisation constant, $K_h$. 
The conversion from energy to momentum spectra for the final-state hadrons was
performed assuming an effective hadron mass,
$m_{\mathrm{eff}}=Q_0$~\cite{Khoze:1996py}.
The input parameters  for the calculations were obtained by fits to LEP
$e^+e^-$ data. Conservative uncertainties, equivalent to three
standard deviations of the experimental uncertainty,
were assumed for the parameters:
$Q_0= \leff = 270\pm 20$~MeV and $K_h=1.31\pm0.03$.
The $\leff $ value agrees  with the value
$\leff=275\pm 4 {\rm (stat.)} ^{+4}_{-8} {\rm (syst.)} $~MeV
deduced from a ZEUS analysis of scaled momenta
in dijet photoproduction~\cite{Chekanov:2009kb}.
The usage of input parameters deduced from LEP data is justified
by the assumed equivalence of hadronisation
in one hemisphere of $e^+e^-$ annihilation and in the current region
of ep interactions in the Breit frame.

The predictions from
several Monte Carlo (MC) models were compared to the measurements. 
Neutral current DIS events were generated using 
the leading-order QCD {\sc Ariadne}\,4.12 program\,\cite{cpc:71:15}.  
The QCD cascade was simulated using the colour-dipole 
model (CDM)~\cite{CDM} inside {\sc Ariadne}. Additional samples
were generated with the MEPS model of {\sc Lepto}~6.5~\cite{cpc:101:108}. 
Both MC~programs, {\sc Ariadne} and {\sc Lepto}, 
were also used to calculate detector acceptances 
and to correct the data to the hadron level. For this purpose they are used with
the {\sc Djangoh}~1.1~\cite{cpc:81:381} interface and QED radiative effects
are included using the {\sc Heracles}~4.6.1~\cite{cpc:69:155} program.
Both MC~programs use the Lund
string model~\cite{prep:97:31} for hadronisation as implemented in
{\sc Jetset}~7.4~\cite{cpc:46:43,cpc:82:74}. 
Hadrons are assumed stable 
if their lifetime is larger than \mbox{$3\times10^{-11}$~s}; 
their decay products are not considered.
This excludes in particular the 
charged decay products of $K^0$ and $\Lambda$ particles.

All generated events  were
passed through the ZEUS detector- and trigger-simulation programs
which are based on GEANT~3~\cite{tech:cern-dd-ee-84-1}. They were
reconstructed and analysed by the same program chain as the data.

% ----------------------------------------------------------------------------
%       Studies of systematic effects
% ----------------------------------------------------------------------------
\section{Correction procedure and systematic uncertainties}
\label{sec:syscheck}

All measured distributions were corrected to the hadron level.
The correction factors were calculated bin~by~bin using MC~events.
For the scaled momentum spectra, the factors are typically below 1.2 
for $Q^2<5000~\gev^2$, but rise up to 1.5 for higher $Q^2$.
The corrections for charged-particle densities as a
function of the pseudorapidity are of a similar magnitude and approach
1.5 as  $\eta^{\rm Breit}$ increases towards the most positive values
measured.

Cross sections, measured separately for each data-taking period, 
were combined using a standard weighted average~\cite{Amsler:2008zzb}. 
The dependence of the scaled momentum distributions on the variation of
the proton-beam energy in the different data samples was determined using
MC events; the resulting changes were found to be smaller than 0.5\%  and
were neglected. Finally, the results were corrected to the QED Born level
using correction factors obtained from  MC, 
reducing the charged hadron multiplicities by up to 4\%.  

The systematic uncertainties were investigated separately for
data with $Q^2$ above and below $160~\gev^2$. 
The numbers in parentheses correspond 
to the largest variations observed in  the scaled momentum spectra.
The uncertainties in the limiting-fragmentation distributions
are of similar magnitude. 
For data with $Q^2>160~\gev^2$, the systematic uncertainties are due to:
\begin{itemize}
\item[$\bullet$]
imperfections in the simulation affecting the 
determination of the efficiency of
event reconstruction and event selection $(^{+1}_{-2}\%)$.
This was evaluated by modifying the selection cuts within
the experimental resolutions.

\item[$\bullet$]
an uncertainty of 3\% in the 
overall tracking efficiency~$(\pm 3\%)$.

\item[$\bullet$]
track reconstruction uncertainties close to the borders of acceptance 
 $(^{+6}_{-3}\%)$. \  This was 
 evaluated by
\begin{itemize}
\item[$-$]
raising (lowering) the cut on $P_{T}^{\rm track}$ 
to 160~MeV (140~MeV);

\item[$-$]
requiring  $|\eta|< 1.5$  instead of  1.75;  
this effect dominates for $Q^2>10 000 \gev^2$;

\item[$-$]
including tracks not associated to the primary vertex. 
\end{itemize}

\item[$\bullet$]
alignment uncertainties affecting the
calculation of the boost vector 
to the Breit frame $(^{+3}_{-2}\%)$. This was evaluated by
\begin{itemize}
\item[$-$]
varying the polar angle for the scattered electron by $\pm 2$~mrad; 

\item[$-$]
varying the polar angles for the hadrons by $\pm 4$~mrad.
\end{itemize}
 
\item[$\bullet$]
assumptions concerning the details of the simulation 
of the hadronic final state ($-4\%$). This
was estimated by using {\sc Lepto} instead of {\sc Ariadne}.

\end{itemize}
For data with $Q^2<160~\gev^2$,
the systematic uncertainties 
are slightly different and are due to:
\begin{itemize}
\item[$\bullet$]
imperfections in the simulation affecting the 
determination of the efficiency of
event reconstruction and event selection $(^{+3}_{-1}\%)$.

\item[$\bullet$]
track reconstruction uncertainties $(^{+5}_{-0.5}\%)$.

\item[$\bullet$]
assumptions concerning the details of the simulation 
of the hadronic final state ($+7\%$).

\item[$\bullet$]
an uncertainty about the size of the contribution 
of diffractive events ($^{+2}_{-1}$\%). 
This was estimated by varying the $\eta_{\mathrm max}$ cut
by $\pm 0.2$ units. 

\end{itemize}
Further details can be found elsewhere~\cite{Beata}.
All individual uncertainties were added in quadrature.

%%%%%%%%%%%%%%%%%%%%%%% 
\section{Results}
\label{sec-res}
%%%%%%%%%%%%%%%%%%%%%%

\subsection{Scaled momentum spectra}
%%%%%%%%%%%%%%%%%%%%%%%%%%%%%%%%%%%%

\begin{figure}[p]
\vfill
\begin{center}
\includegraphics[width=15cm]{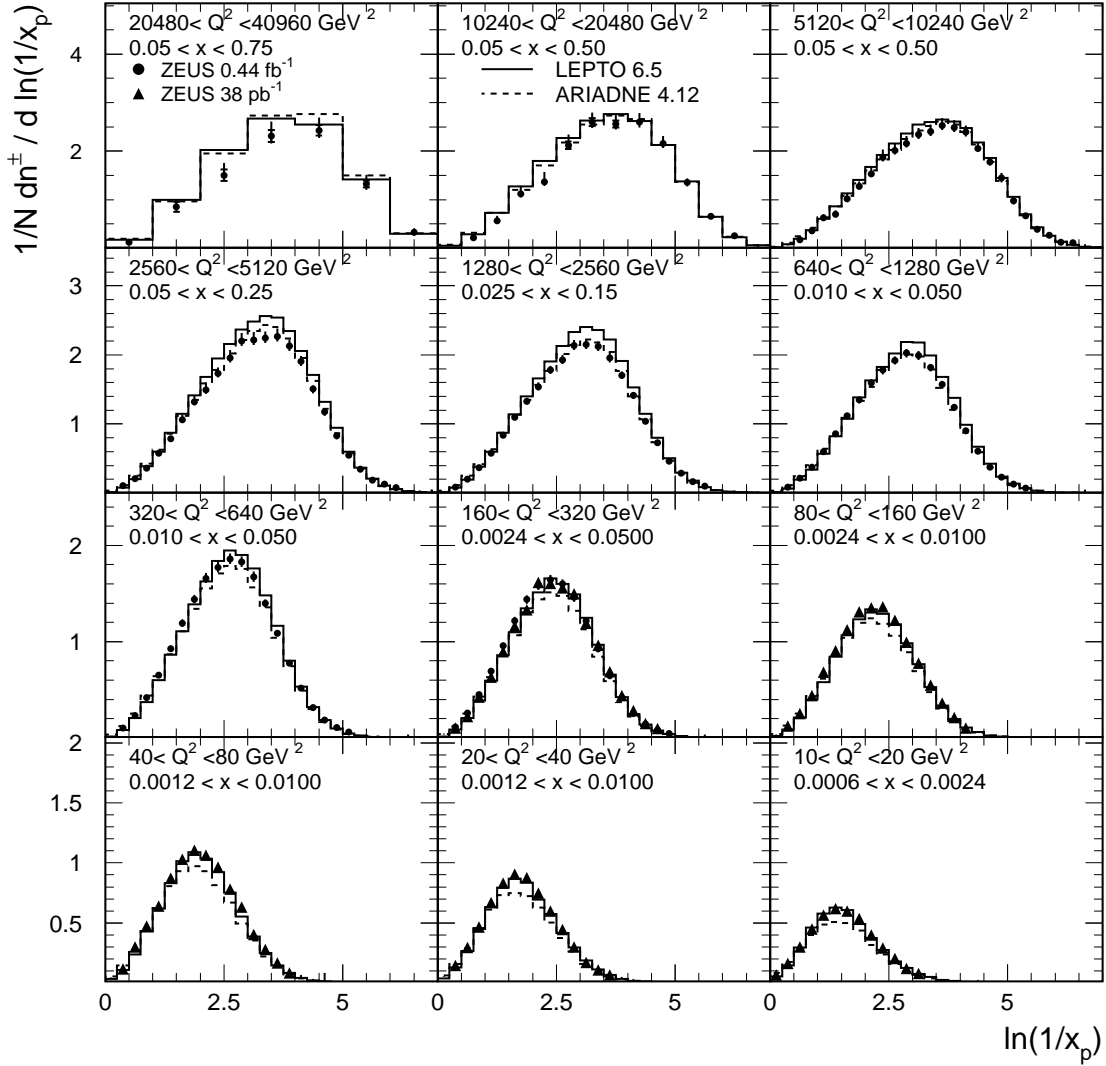}
\end{center}
\caption{
The scaled momentum spectra,
$1/N~ dn^{\pm} /d\ln(1/x_p)$,
for different $(x,Q^2)$ bins. 
The dots represent the new, 
the triangles the previous ZEUS measurement.
The data overlap in the $160 < Q^2 < 320 \,GeV^{\,2} $ bin.
The inner error bars, where visible, indicate statistical uncertainties,
the outer statistical and
systematic uncertainties added in quadrature.
The full and dashed lines represent the {\sc Lepto} 
and the {\sc Ariadne} predictions, respectively.
}
\label{fig-xi}
\vfill
\end{figure}

\begin{figure}[p]
\vfill
\begin{center}
\includegraphics[width=15cm]{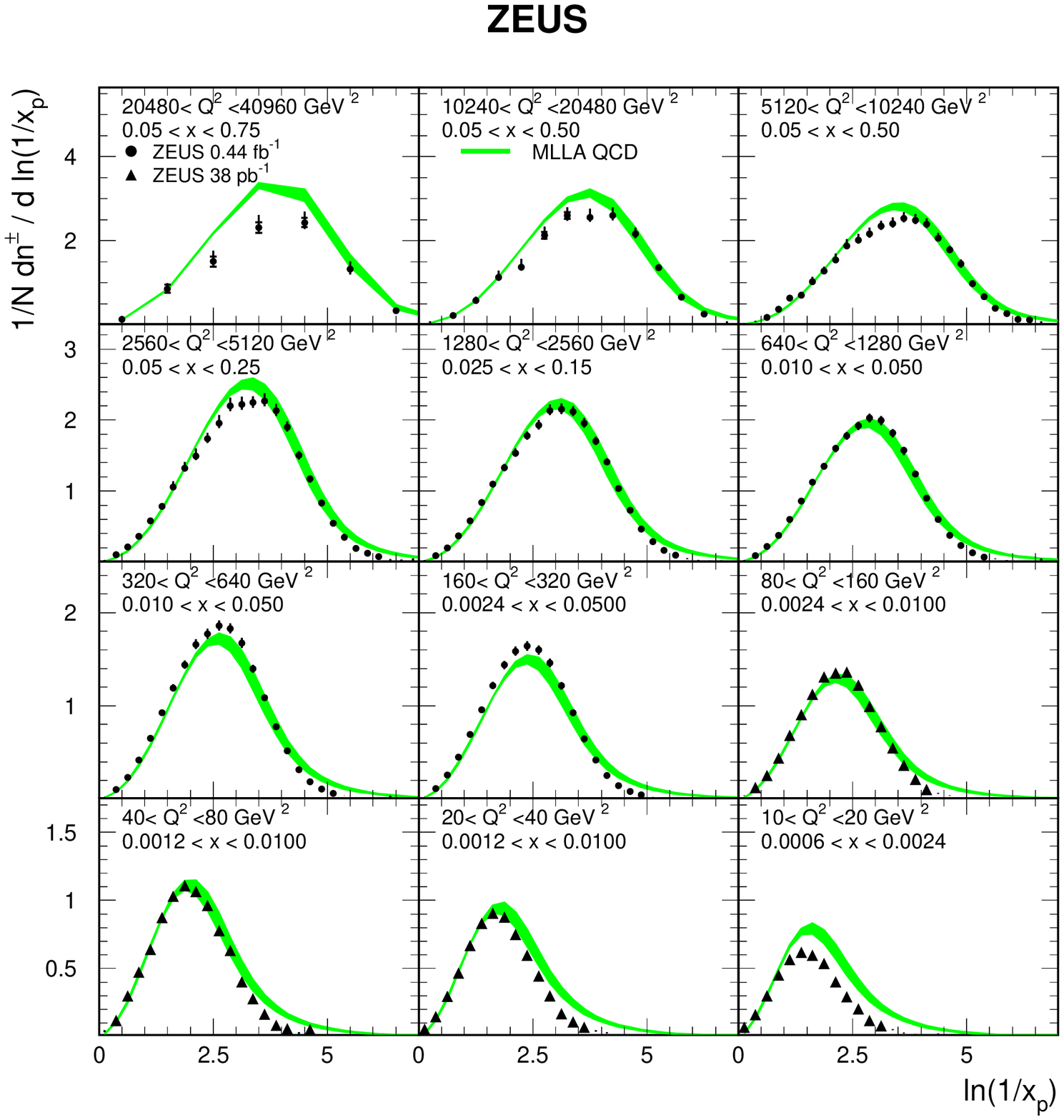}
\end{center}
\caption{
The scaled momentum spectra,
$1/N~ dn^{\pm} /d\ln(1/x_p)$,
for different $(x,Q^2)$ bins. 
The band represents the range of the MLLA+LPHD predictions. 
Other details as in Fig.~\ref{fig-xi}.
% with rather conservative uncertainties.
}
\label{fig-xi-ochs}
\vfill
\end{figure}

\begin{figure}[p]
\vfill
\begin{center}
\includegraphics[width=15cm]{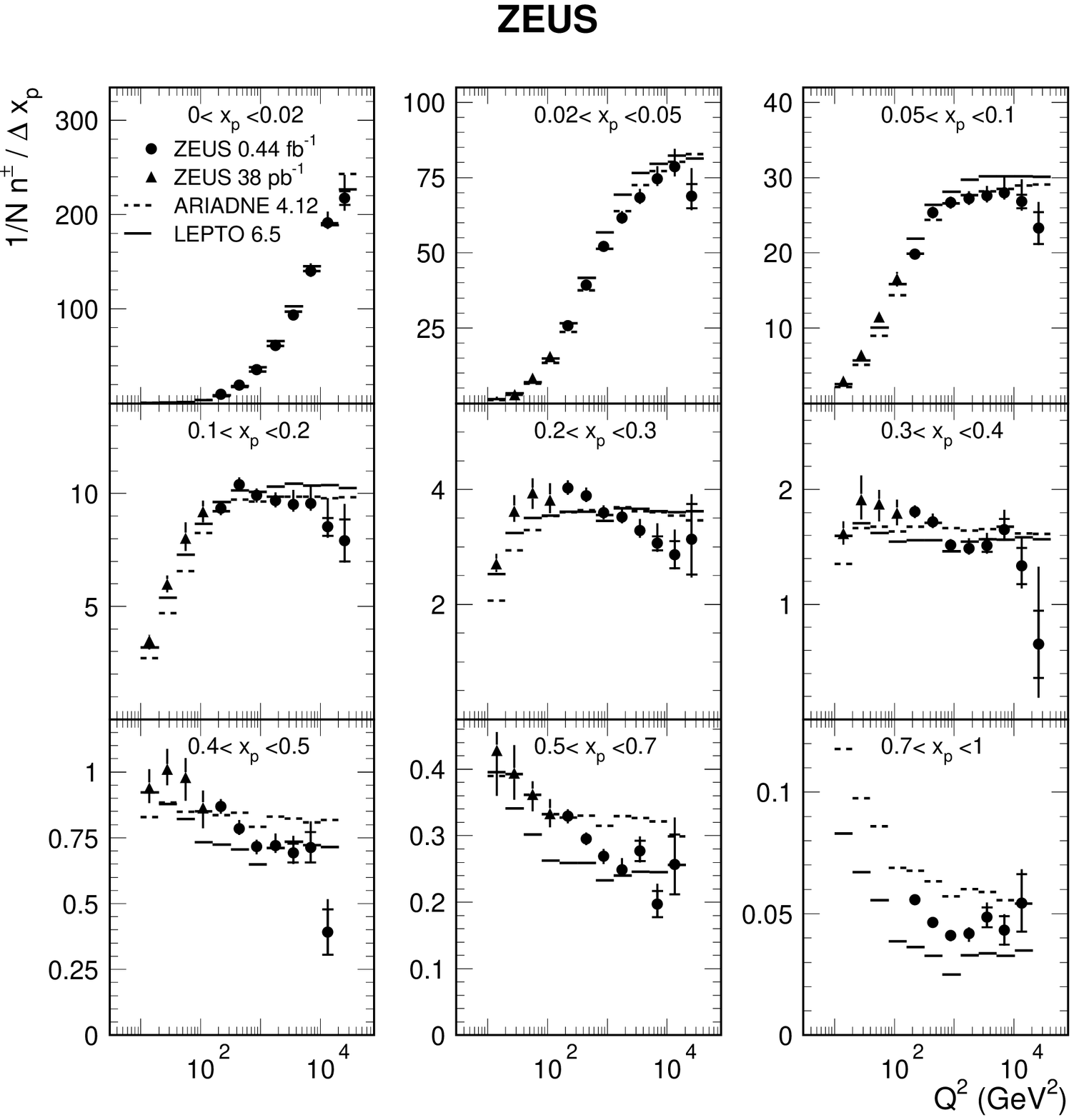}
\end{center}
\caption{
The number of charged particles per event per unit of $x_p$,
$ 1/N~ n^{\pm}/\Delta x_p$,
as a function of $Q^2$ in $x_p$ bins of width $\Delta x_p$.
Other details as in Fig.~\ref{fig-xi}.
}
\label{fig-xp-mc}
\vfill
\end{figure}

\begin{figure}[p]
\vfill
\begin{center}
\includegraphics[width=15cm]{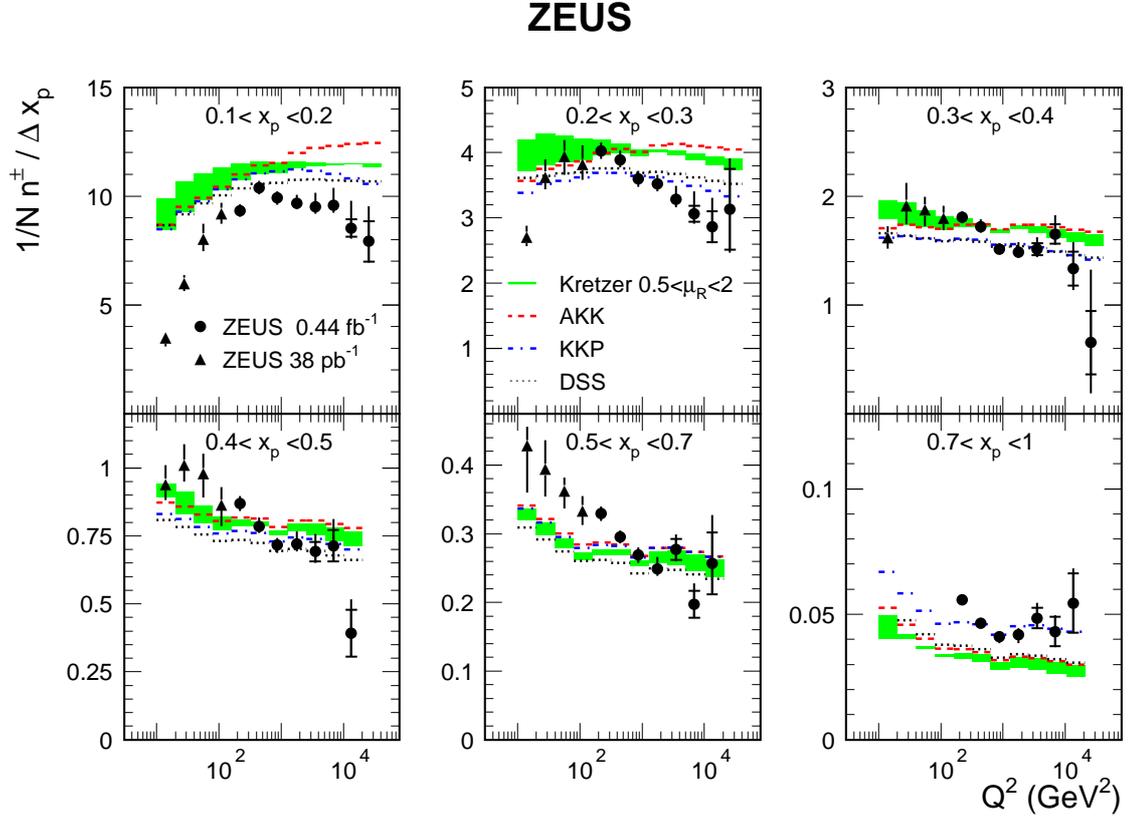}
\end{center}
\caption{
The number of charged particles per event per unit 
of $x_p$, $ 1/N~ n^{\pm}/\Delta x_p$,
as a function of $Q^2$ in $x_p$ bins with width 
$\Delta x_p$ as in  Fig.~\ref{fig-xp-mc}.
The shaded band represents the NLO calculation  by
Kretzer~\protect\cite{Kretzer:2000yf} 
with its renormalisation scale uncertainty.
Additional NLO calculations are shown:  
Kniehl, Kramer, P\"otter~\protect\cite{Kniehl:2000cr}(KKP), 
Albino, Kniehl, Kramer~\protect\cite{Albino:2005me}(AKK) and 
De Florian, Sassot and 
Stratmann~\protect\cite{Florian:2007aj,*Florian:2007hc}(DSS). 
}
\label{fig-xp}
\vfill
\end{figure}

Scaled momentum spectra  were measured in the current region 
in the Breit frame as a function of $Q^2$ in the kinematic range
$160<Q^2<40960\gev^2$  and $0.002 < x < 0.75$. 
Results are presented in Figs.~\ref{fig-xi}--\ref{fig-xp}
and Tables~\ref{table-ksi1}--\ref{table-xp3}.
Also shown in Figs.~\ref{fig-xi}--\ref{fig-xp} are previously
published results for $10<Q^2<160\gev^2$. In addition,
in Fig.~\ref{fig-xi} data from a previous ZEUS  
publication~\cite{epj:c11:251}
are given for $160<Q^2<320\gev^2$. They agree well with the 
measurements presented here.  The same is true  for previously
obtained results up to $Q^2=5120 \gev^2$ which are not shown.

The normalised spectrum, $1/N~dn^{\pm}/d \ln (1/x_p )$, 
with $N$ being the number of events  and
$n^{\pm}$ being the number of charged particles, 
is shown in Figs.~\ref{fig-xi}--\ref{fig-xi-ochs}.
These scaled momentum spectra exhibit a hump-backed form with an 
approximately Gaussian shape around the peak. 
The mean charged multiplicities are given by the integrals of 
the spectra. As $Q^2$ increases, the multiplicity increases and, 
in addition, the peak of the spectrum moves to larger values of $\ln (1/x_p)$.

In Fig.~\ref{fig-xi}, the predictions of {\sc Ariadne} and 
{\sc Lepto} are compared to the data.
They reproduce the main features of the data but do not agree in 
detail. 
For the highest\,$Q^2$ bin,  
both models predict too many charged particles 
at medium and low  values 
of $\ln (1/x_p)$.
{\sc Lepto} also predicts too many particles for medium-$Q^2$ bins
while 
{\sc Ariadne} predicts too few for low-$Q^2$ bins.

In Fig.~\ref{fig-xi-ochs}, 
the MLLA+LPHD  predictions\cite{MLLA,Khoze:1996dn} 
are compared to the data.   Too many particles  are predicted
for the highest- and lowest-$Q^2$ bins, while at medium $Q^2$   
the data is reasonably well described.
At low $Q^2$, the observed particle deficit 
can be interpreted as a significant migration
of particles to the target region of the Breit frame;
this was also previously observed \cite{zfp:c67:93,*pl:b414:428}.
At medium $Q^2$, the agreement is surprising, because
BGF contributes significantly to the cross section and
the observed particles should reflect 
the $q\bar{q}$~final state which is not included in the prediction.
At the highest $Q^2$ available, the failure of the MLLA prediction probably
reflects the fact that the $e^+e^-$ data used to obtain the
input parameters are dominated by $Z^0$ exchange while, in $ep$~collisions,
photon exchange still dominates.   

The MLLA+LPHD calculations predict long tails towards large
values of $\ln (1/x_p)$ over the complete range of~$Q^2$.
These tails are sensitive to the mass correction applied 
in the calculation~\cite{Khoze:1996py,Dixon}.
The data do not show such tails in the predicted size.
A better description of the large $\ln (1/x_p)$
region is obtained  if $m_{\mathrm{eff}}=Q_0= 0.9 \leff$ is taken
instead of $m_{\mathrm{eff}}=Q_0= \leff$.

\subsection{Scaling violation}
%%%%%%%%%%%%%%%%%%%%%%%%%%%%%%

As the energy scale, $Q$, increases, the phase 
space for soft gluon radiation increases,  leading to a rise 
of the number of soft particles with small $x_p$. 
These scaling violations can be seen when the data are plotted in bins of
$x_p$ as a function of $Q^2$.  Figure~\ref{fig-xp-mc} and
Tables~\ref{table-xp1}--\ref{table-xp3} show that 
the number of charged particles increases with $Q^2$ at low $x_p$ 
and decreases with $Q^2$ at high $x_p$.  Neither {\sc Lepto} nor
{\sc Ariadne} provides a good description of this $Q^2$ dependence 
over the whole range of $x_p$.

Figure~\ref{fig-xp} shows the data together with four NLO+FF QCD 
predictions~\cite{Kretzer:2000yf,Kniehl:2000cr,Albino:2005me,*Albino:2008fy,Florian:2007aj,*Florian:2007hc}   
for $x_p>0.1$, where theoretical uncertainties are small 
and the predictions not too strongly affected by hadron-mass effects 
which are not included in the calculations~\cite{albino-2007-75}.
The fragmentation functions (FF) used in all four calculations
were extracted from  $e^+e^-$ data.  The four predictions
are similar in shape and have similar uncertainties.
The uncertainties  are only illustrated for the 
calculation of Kretzer~\cite{Kretzer:2000yf}.
The NLO calculations also do not provide a good description of the data.
Too many particles are predicted at small $x_p$ and too few at large $x_p$.
In general, the scaling violations predicted are not strong enough.

\begin{figure}[p]
\vfill
\begin{center}
\includegraphics[width=14cm]{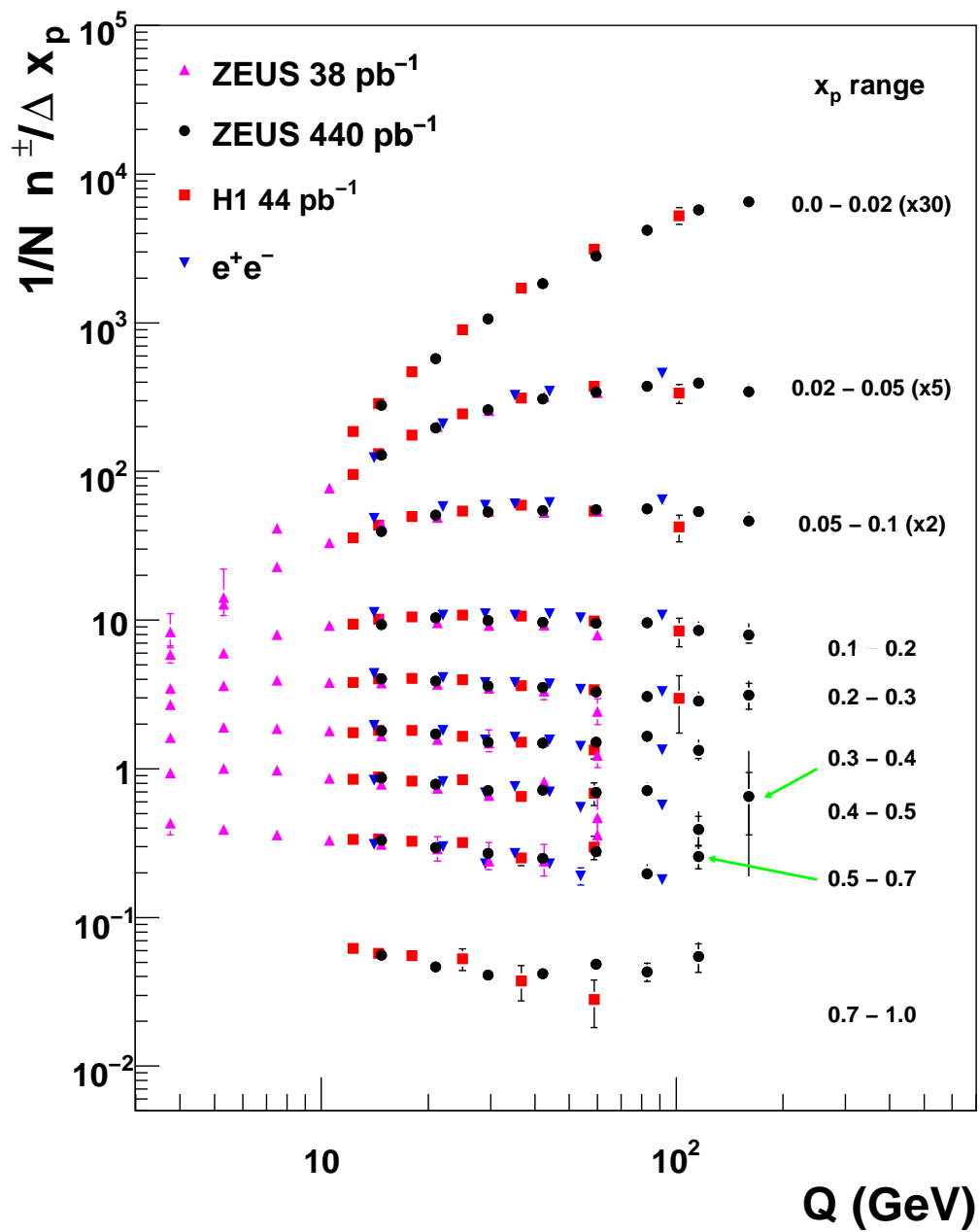}
\end{center}
\caption{
The number of charged particles per event 
per unit of $x_p$, $ 1/N~ n^{\pm}/\Delta x_p$,
as a function of $Q$ in $x_p$ bins with width $\Delta x_p$.
Also shown are data from H1~\protect\cite{h1:frag} and 
$e^+e^-$~\protect\cite{mark2:1988,*Braunschweig:1990yd,
*delphi:1993,*Li:1989sn}.  
The dots (triangles) represent the new (previous) ZEUS measurement, 
the squares the H1 data and the inverted triangles the $e^+e^-$ data.
The inner error bars, where visible, indicate statistical uncertainties,
the outer statistical and systematic uncertainties added in quadrature.
The three lowest $x_p$ bins are scaled by  factors of 30,~5 and~2, 
respectively.
}
\label{fig-xp-h1}
\vfill
\end{figure}

\begin{figure}[p]
\vfill
\begin{center}
\includegraphics[width=15cm]{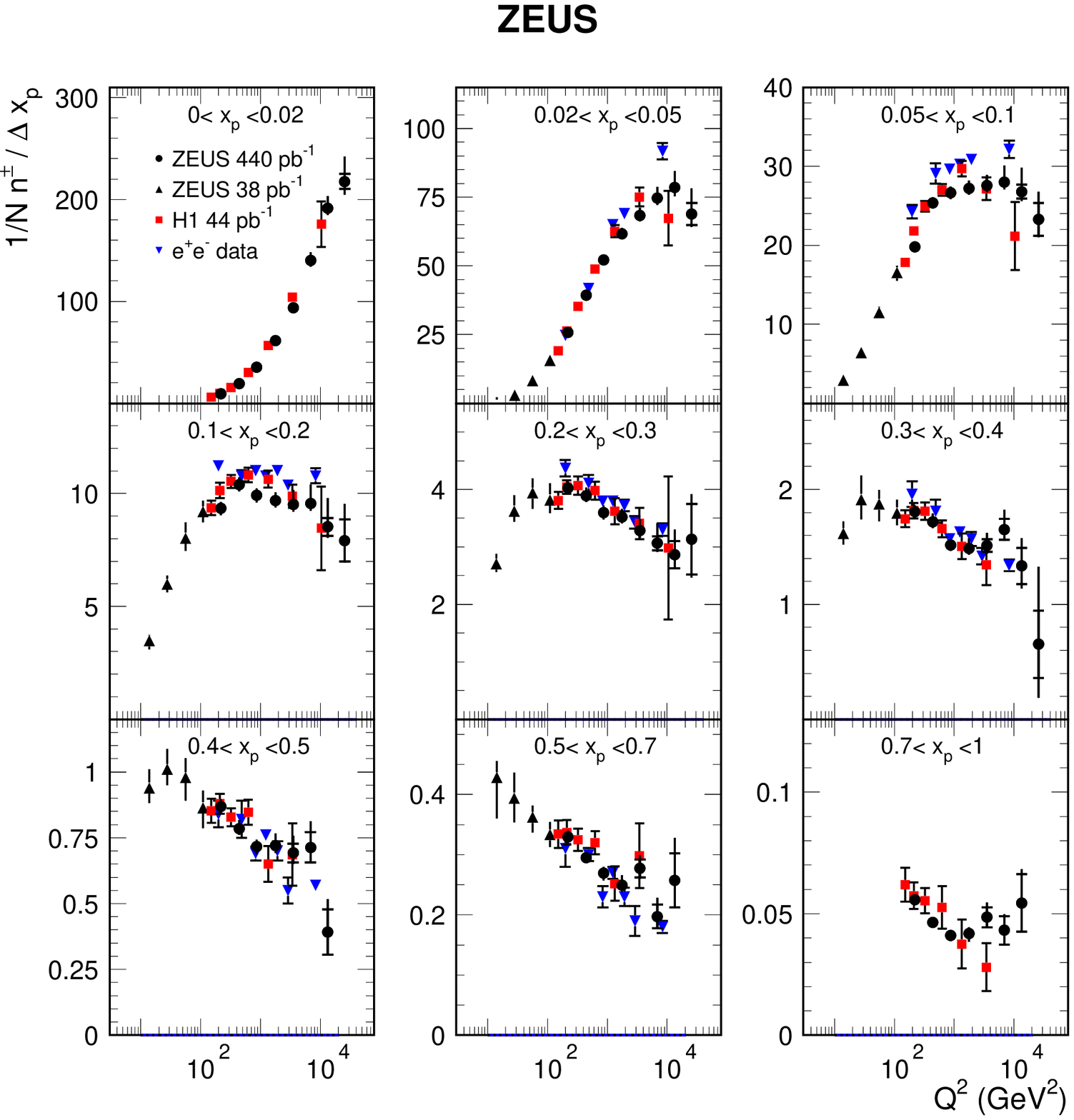}
\end{center}
\caption{
The number of charged particles per event per unit of $x_p$,
$ 1/N~ n^{\pm}/\Delta x_p$,
as a function of $Q^2$ in $x_p$ bins with width $\Delta x_p$.
Other details as in Fig.~\ref{fig-xp-h1}.
}
\label{fig-xp-h1-ee}
\vfill
\end{figure}

Figure~\ref{fig-xp-h1} shows the same data as Fig.~\ref{fig-xp}
together with results from H1~\protect\cite{h1:frag} and from $e^+e^-$ 
experiments~\protect\cite{mark2:1988,*Braunschweig:1990yd,*Li:1989sn,*delphi:1993}.
For a proper comparison, the 
the particle momenta from  $e^+e^-$ data were scaled 
to half of the centre-of-mass energy
as discussed in the introduction  and the scale was set to
$Q=2 ~E_{\rm beam}$, where $E_{\rm beam}$ is the beam energy. 
In addition, corrections for the different treatment 
of $K^0$ and $\Lambda$ decays were applied. The overall agreement between 
the different data sets supports fragmentation universality.
The presentation of the data using a linear scale
as presented in~Fig.~\ref{fig-xp-h1-ee} 
does, however, show some significant differences between $e^+e^-$ and $ep$,
in particular around the $Z^0$ mass at $0.02<x_p<0.2$
and at low $Q^2$ at $0.1<x_p<0.2$.

\subsection{Limiting fragmentation}
%%%%%%%%%%%%%%%%%%%%%%%%%%%%%%%%%%%%%%%%%%

The concept of limiting fragmentation~\cite{limiting} is based on
the assumption that a Lorentz-contracted object passes through
another object at rest,  leaving behind an excited state 
with properties depending neither on the energy nor the identity of 
the passing object.
This excited state fragments into particles in a restricted
window of rapidity, called the limiting-fragmentation region.  
In this region, the limiting-fragmentation hypothesis predicts that
the density of charged particles per unit of rapidity depends only on $W$.

\begin{figure}[p]
\vfill
\begin{center}
\includegraphics[width=15cm]{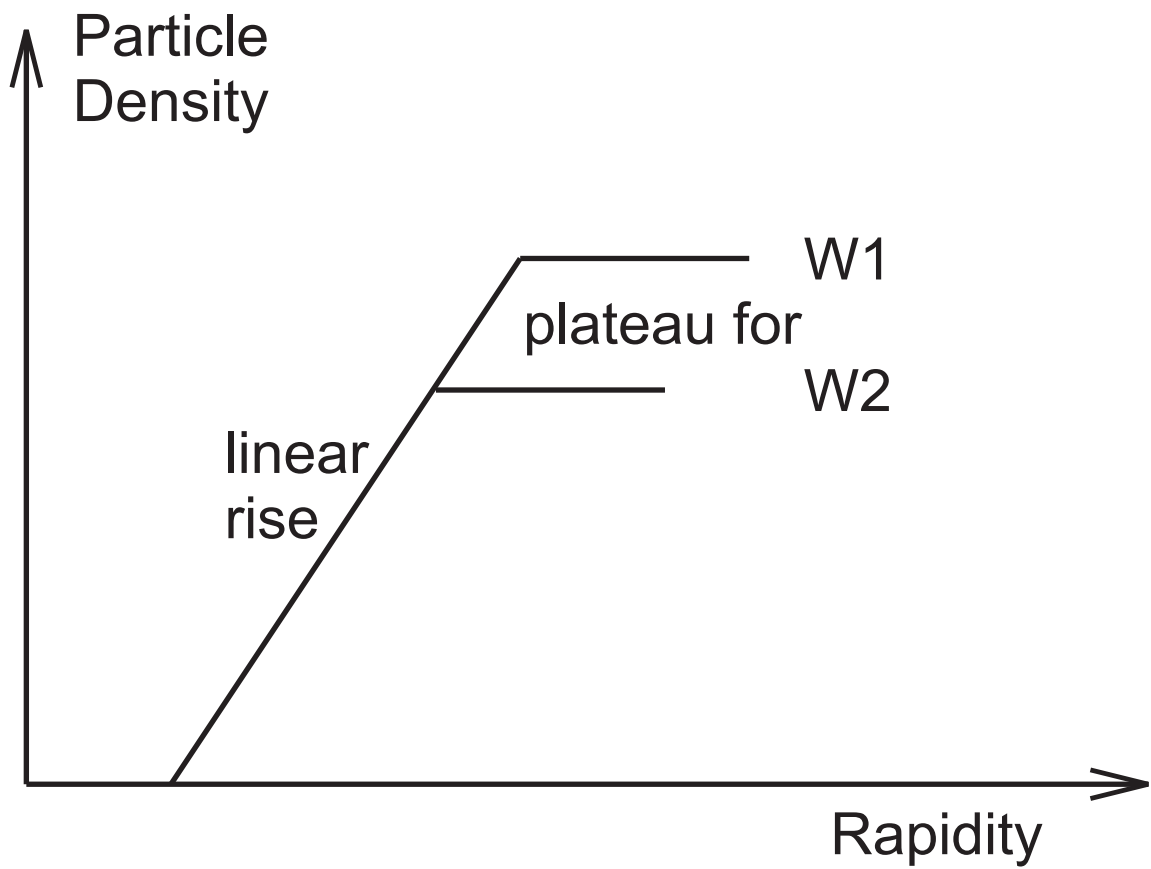}
\end{center}
\caption{
The main features of the prediction of the Bialas-Jezabek model
      based on the limiting-fragmentation hypothesis
      for the dependence of the particles density on the
      rapidity in hadronic collisions for two values of $W$ with
      $W1>W2$.
}
\label{fig-lim}
\vfill
\end{figure}

Limiting fragmentation has been observed in a variety of hadronic 
collisions~\cite{DeinesJones:1999ap,*UA5,Back:2004je}, including nucleus-nucleus
interactions~\cite{Back:2005hs,*Adams:2005cy,*PHOBOS,*Bearden:2001qq,*Bearden:2001xw}.
It was observed that in the region of
limiting fragmentation the particle density
increases linearly with the rapidity before reaching a plateau.
The slope of the increase did not show a $W$~dependence, but the height
of the plateau increased with $W$. These features
are illustrated in Fig.~\ref{fig-lim}.
Bialas and Jezabek~\cite{Bialas} proposed a statistical model to explain
the missing $W$ dependence of the slopes.
In this model, soft particle production in hadronic collisions 
is described in terms
of multiple gluon exchanges between partons of the colliding hadrons
and by the subsequent radiation of hadronic clusters.

The application of the limiting-fragmentation hypothesis to
$e^+e^-$ annihilations is not straight-forward.
However, again a behaviour as illustrated in Fig.~\ref{fig-lim}
was observed, only in this case the
slopes increase with $W$~\cite{Back:2004je,acta}.

In the case of $ep$ collisions, the passing object is the proton 
while the virtual photon exchanged in the interaction is 
the object ``at rest''. It is assumed to be the excited hadron
which fragments~\cite{acta}.

\begin{figure}[p]
\vfill
\begin{center}
\includegraphics[width=15cm]{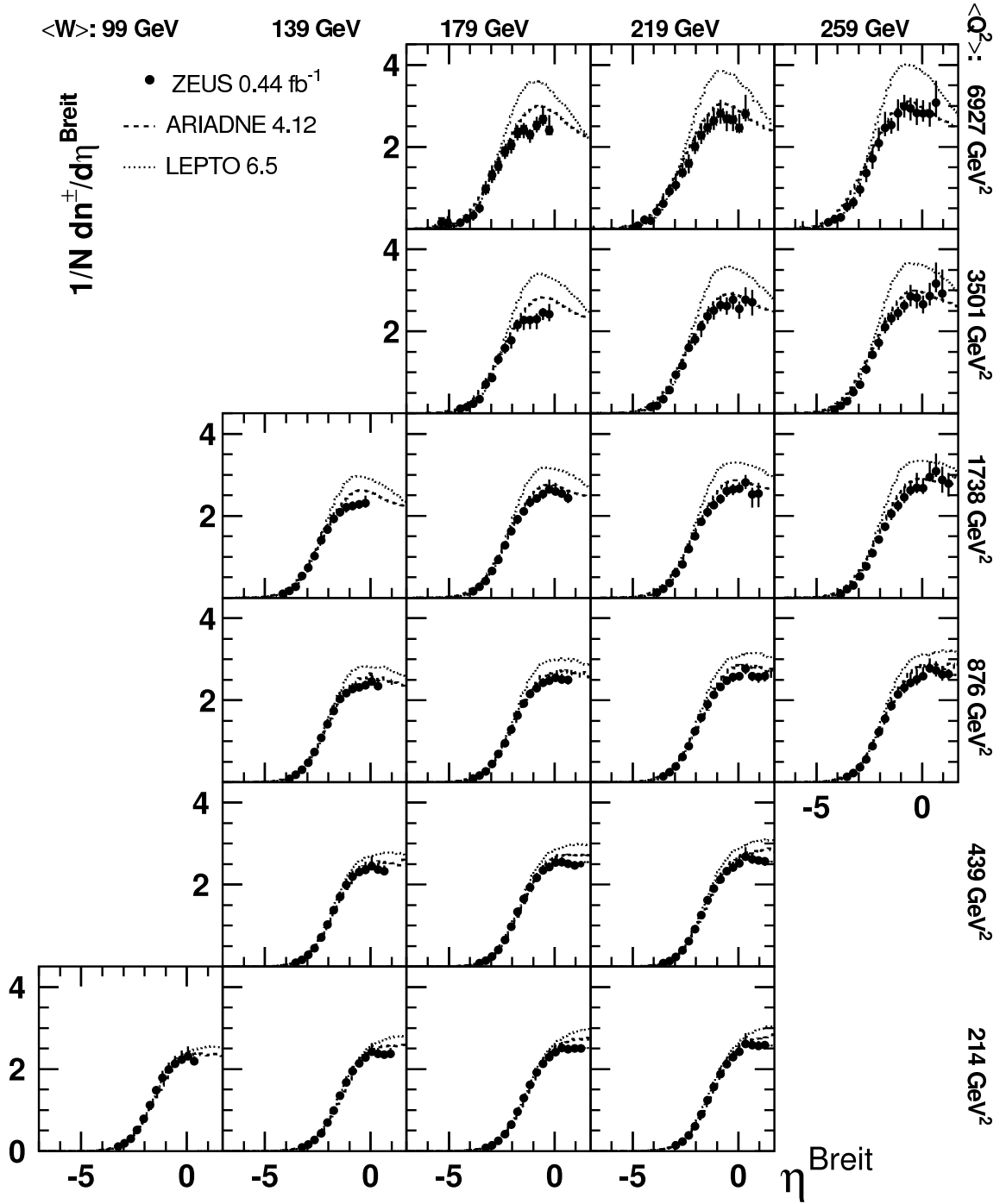}
\end{center}
\caption{
The normalised charged-particle density per unit of $\eta^{\rm Breit}$,
$1/N~ dn^{\pm} /d\eta^{\rm Breit}$, for different $(W,Q^2)$ bins 
for $Q^2>160$~GeV\,$^2$. The dots represent the ZEUS measurement.
The error bars, where visible, indicate 
statistical and
systematic uncertainties added in quadrature.
The dashed and dotted lines represent the {\sc Lepto}  
and {\sc Ariadne} predictions, respectively.
}
\label{fig-limfrag}
\vfill
\end{figure}

\begin{figure}[p]
\vfill
\begin{center}
\includegraphics[width=12cm]{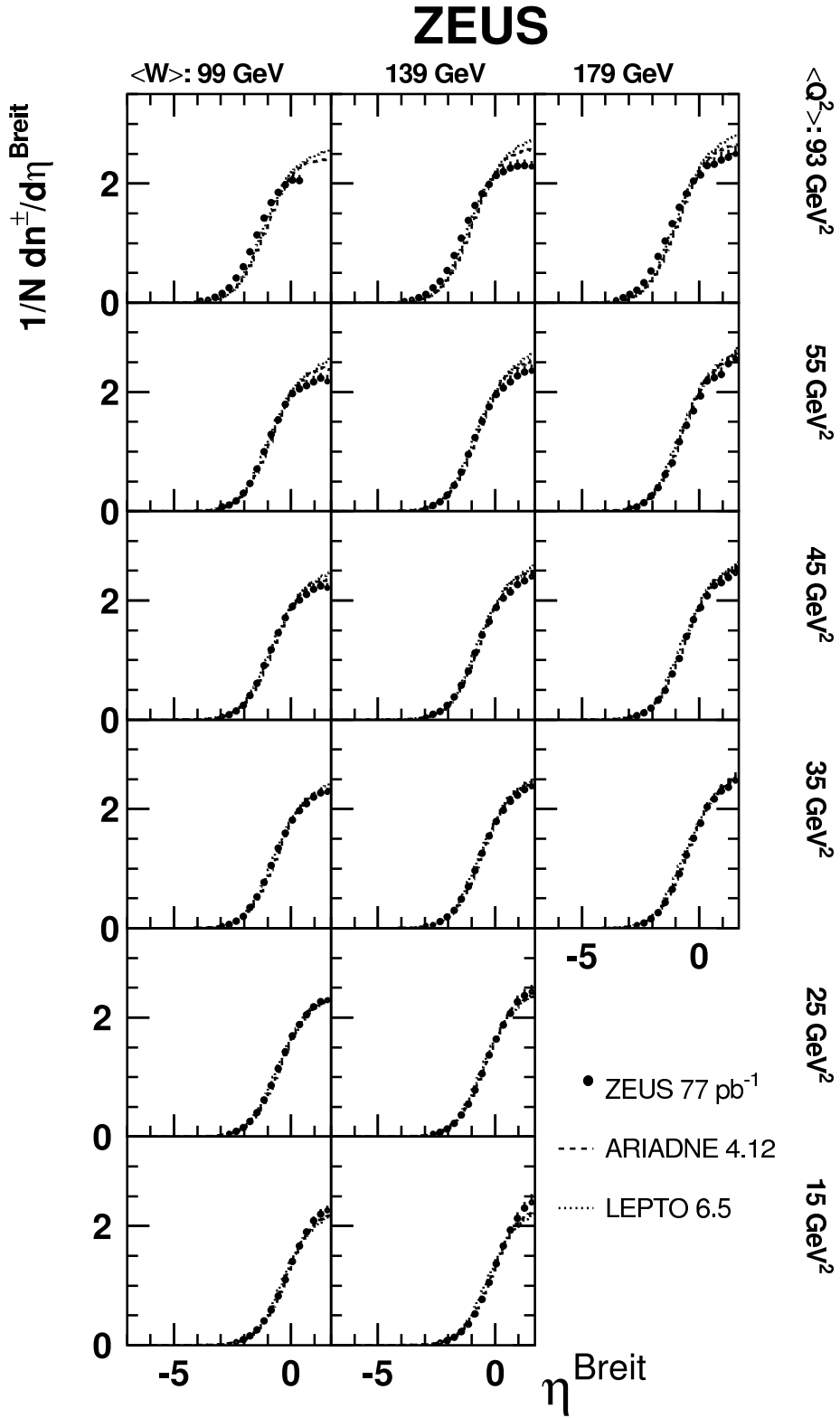}
\end{center}
\vspace*{1.8cm}
\caption{
The normalised charged-particle density per unit of $\eta^{\rm Breit}$, 
$1/N~ dn^{\pm} /d\eta^{\rm Breit}$, 
for different $(W,Q^2)$ bins for $10<Q^2<160$~GeV\,$^2$.
Other details as in Fig.~\ref{fig-limfrag}.
}
\label{fig-limfrag-low}
\vfill
\end{figure}

Figures~\ref{fig-limfrag}--\ref{fig-limfrag-low} 
and Tables~\ref{table-q6}--\ref{table-q1} present
the density of charged particles per unit of pseudorapidity,
$\eta^{\rm Breit}$, for $10<Q^2<10240~{\rm GeV}^2$ 
in bins of $Q^2$ and $W$ as listed in Table~\ref{tab:every}.
A region of linear rise and the onset of a plateau
are observed in all bins.
This supports the applicability of the hypothesis of limiting
fragmentation to the case of $ep$ collisions.
However, for low $Q^2$, the plateau is only reached
in the target region, $\eta^{\rm Breit}>0$.

\begin{figure}[p]
\vfill
\begin{center}
\includegraphics[width=20cm]{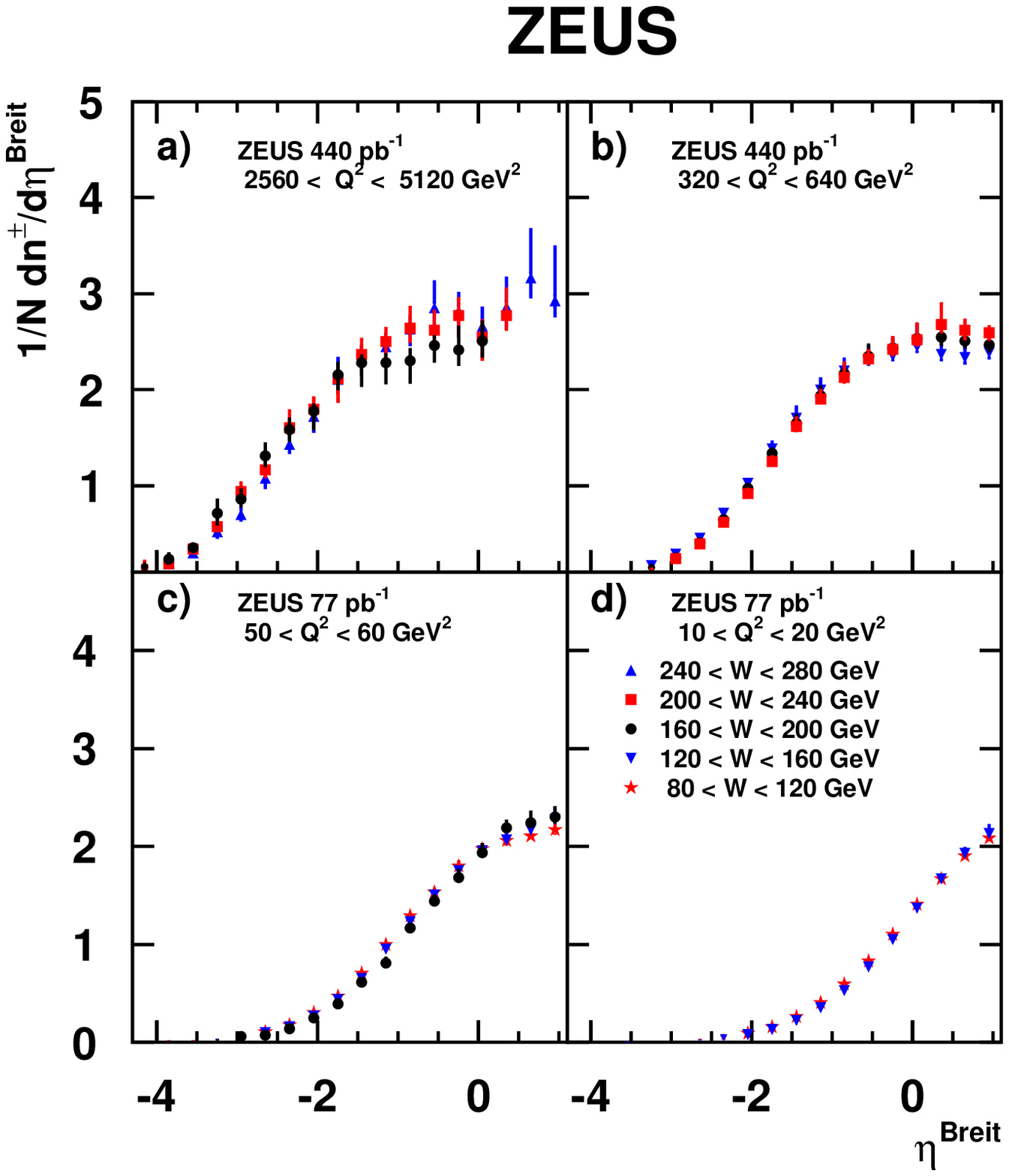}
\end{center}
\caption{
The normalised charged-particle density per unit of $\eta^{\rm Breit}$, 
$1/N~ dn^{\pm} /d\eta^{\rm Breit}$, for 4 bins in $Q^2$  and 5 bins in $W$. 
The error bars, where visible, indicate 
statistical and systematic uncertainties added in quadrature.
}
\label{fig-slope-1}
\vfill
\end{figure}

\begin{figure}[p]
\vfill
\begin{center}
\includegraphics[width=18cm]{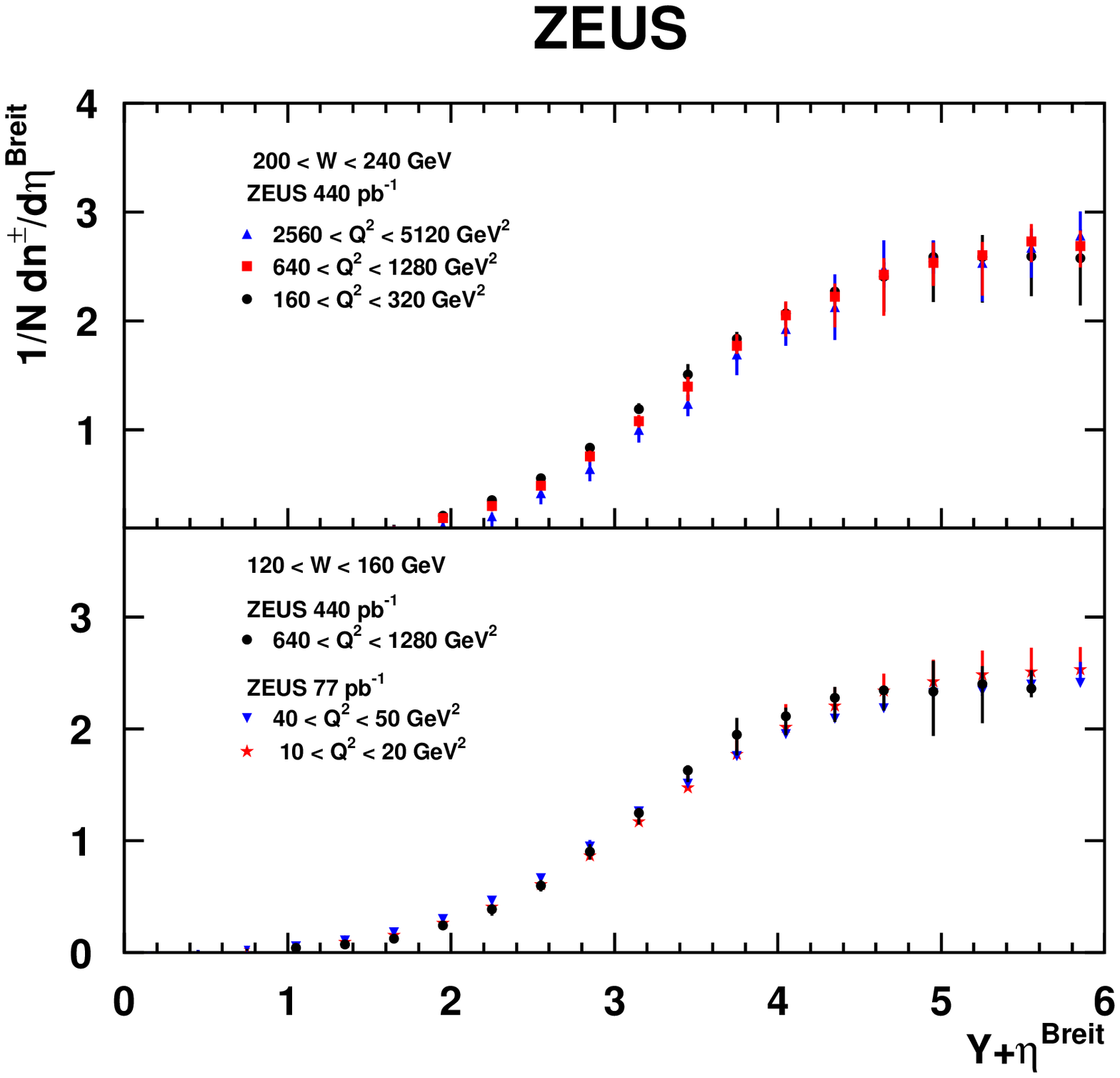}
\end{center}
\caption{
The normalised charged-particle density per unit of $\eta^{\rm Breit}$, 
$1/N~ dn^{\pm} /d\eta^{\rm Breit}$, for 2 bins in $W$ and 5 bins in $Q^2$. 
The distributions were rebinned by shifting the entries for each event 
by Y$=$ln$\,Q/m_{\pi}$ as explained in the text.
The error bars, where visible, indicate 
statistical and systematic uncertainties added in quadrature.
}
\label{fig-slope-3}
\vfill
\end{figure}

The slopes in the region of linear rise do not depend significantly  
on either $Q^2$ or $W$, as also demonstrated in Figs.~\ref{fig-slope-1}
and~\ref{fig-slope-3}.  
The lack of a $W$ dependence indicates that the model
of Bialas and Jezabek is also applicable for $ep$ collisions.

Figures.~\ref{fig-limfrag} and \ref{fig-limfrag-low}
also show 
predictions from {\sc Ariadne} and {\sc Lepto}.
Overall, {\sc Ariadne} provides reasonable predictions 
for the whole range in $Q^2$ and $W$.  {\sc Lepto}, however,
predicts a sizeable increase in the height of the plateau 
with $Q^2$ and $W$
which is not observed in the data.
The predictions in the plateau region are sensitive to the input 
parameters used in the fragmentation functions.
The usage of input parameters derived from SMC~data~\cite{Adeva:2004dh}
in {\sc Lepto} results in the prediction of
a softer spectrum, reflected 
in a charged-particle density of up to 30\,\% too high.

The hypothesis of limiting fragmentation was further tested by
studying the charged-particle densities in the rest frame of 
the fragmenting object, i.e. the virtual photon. 
The $\eta^{\rm Breit}$
distributions were rebinned by shifting event by event all entries 
by ln\,$Q/m_{\pi}$ , thus scaling the available energy to the pion mass.
The resulting distributions
are shown in Fig.~\ref{fig-slope-3}.
The distributions are very similar but 
for $Q^2>160$~GeV$^2$ a slightly larger slope is observed.
This is a region where the BGF contribution is decreasing.
In general, the observations support the hypothesis of 
limiting fragmentation
and the model of Bialas and Jezabek.
This indicates that, even at high $Q^2$, soft processes are involved
in the fragmentation and a statistical approach is justified.

\section{Conclusions}
\label{sec-con}

Scaled momentum spectra have been measured in NC DIS 
for the current region in the Breit frame over the large range
of $Q^2$ from  10~GeV$^2$ to 40960~GeV$^2$.
Large scaling violations are observed.
Comparing the data to $e^+e^-$ results generally supports the concept
of quark-fragmentation universality.
Neither MLLA+LPHD nor  NLO+FF calculations
describe the data well. A better, albeit not prefect description is provided 
by the {\sc Ariadne} program.

The limiting-fragmentation hypothesis 
and the statistical model of Bialas and Jezabek
were tested 
by studying the density of charged particles as a function 
of the pseudorapidity, $\eta^{\rm Breit}$, over the range 
of $10<Q^2<10240$~GeV$^2$. 
A region of linear rise and the onset of a plateau
are observed over the whole range in $Q^2$ and
support the limiting fragmentation hypothesis.
The independence of the slopes on $W$ supports the statistical
approach of Bialas and Jezabek.

\section*{Acknowledgments}

We appreciate the contributions to the construction and maintenance 
of the ZEUS detector of many people who are not listed as authors.  
The HERA machine group and the DESY computing staff are especially 
acknowledged for their success in providing excellent operation of 
the collider and the data analysis environment.  We thank the DESY 
directorate for their strong support and encouragement.
We thank W.~Khoze, W.~Ochs, R.~Sassot, A.~Bia{\l}as and M.~Je{\.z}abek 
for fruitful discussions. 
We especially would like to thank S.~Albino for providing the QCD calculations 
and for instructive discussions.

\vfill\eject

%------------------------------------------------------------------------------
%       Bibliography
%------------------------------------------------------------------------------
%\include{paper-ref}
\providecommand{\etal}{et al.\xspace}
\providecommand{\coll}{Coll.\xspace}
\catcode`\@=11
\def\@bibitem#1{%
\ifmc@bstsupport
  \mc@iftail{#1}%
    {;\newline\ignorespaces}%
    {\ifmc@first\else.\fi\orig@bibitem{#1}}
  \mc@firstfalse
\else
  \mc@iftail{#1}%
    {\ignorespaces}%
    {\orig@bibitem{#1}}%
\fi}%
\catcode`\@=12

%------------------------------------------------------------------------------
%       Tables
%------------------------------------------------------------------------------
%-------------------------------------------------------------------------------
%       Tables
%-------------------------------------------------------------------------------

\begin{table}[p]
\begin{center}
\begin{tabular}{||r l r||l l l |r||r l r|r||}
\hline
\multicolumn{3}{||c||}{$Q^2_{DA}$ (GeV$^2$)} &  \multicolumn{4}{c||}{scaled momenta} &  \multicolumn{4}{c||}{limiting fragmentation} \\ \cline{4-11}
& & & \multicolumn{3}{c|}{$x_{DA}$} & no. of events & \multicolumn{3}{c|}{$W$ (GeV)} & no. of events\\
\hline
$10$ & -- & $30$ &  & --- &  & --- & $80$ & -- & $160$ & 1 142 102 \\
\hline
$30$ & -- & $160$ &  & ---& & --- & $80$ & -- & $200$ & 1 185 843 \\
\hline
$ 160$ & -- & $320 $ & $0.0024$ & -- & $0.0500 $ & 684 077 & $80$ & -- & $240$ & 584 578 \\
\hline
$ 320$ & -- & $640 $ & $0.0100$ & -- & $0.0500 $ & 170 784 & $120$ & -- & $240$ & 152 367\\ 
$ 640$ &  -- & $ 1280 $ & & & &72 268 & $120$ & -- & $280$ & 75 176\\ 
\hline
$ 1280$ & -- & $ 2560 $ & $0.025$ & -- &$ 0.15  $ & 31 608 & $120$ & -- & $280$ & 29 949\\  
\hline
$ 2560$ & -- & $ 5120 $ & $0.05$ & -- & $0.25  $ & 10 858 & $160$ & -- & $280$ & 8 031\\  
\hline
$ 5120$ &  -- & $10240 $ & $0.05$ & -- &$ 0.50  $ & 4 748 & $160$ & -- & $280$ & 2 914\\ 
$ 10240$ & -- &$ 20480 $ & & & & 1 197 & & ---& & --- \\ 
\hline
$ 20480$ & --&$ 40960  $ & $0.05$ & --&$ 0.75 $ & 205 & & ---& & --- \\ 
\hline\hline
\end{tabular}
\caption{{\small Number of accepted events in  ($Q^2$,$x$) bins 
for the scaled momentum analysis and in ($W$,$Q^2$) bins
for limiting-fragmentation studies.  }}
\label{tab:every}
\end{center}
\end{table}

\begin{table}
\centering
   \begin{tabular}{|l|c|c|}
\hline
\hline
$\ln(1/x_p)$  &  $160<Q^2<320$~GeV$^2$  & $320<Q^2<640$~GeV$^2$\\ \hline
\hline
0.00 -- 0.25 &	0.0319	 $\pm$ 0.0005$^{+0.0011 }_{-0.0012 }$ &	0.0259	 $\pm$ 0.0009$^{+0.0009 }_{-0.0010 }$ \\ \hline
0.25 -- 0.50 &	0.1156	 $\pm$ 0.0011$^{+0.0041 }_{-0.0039 }$ &	0.101	 $\pm$ 0.002$^{+0.003 }_{-0.004 }$ \\ \hline
0.50 -- 0.75 &	0.2573	 $\pm$ 0.0016$^{+0.0079 }_{-0.0079 }$ &	0.232	 $\pm$ 0.003$^{+0.010 }_{-0.007 }$ \\ \hline
0.75 -- 1.00 &	0.451	 $\pm$ 0.002$^{+0.014 }_{-0.014 }$ &	0.419	 $\pm$ 0.004$^{+0.015 }_{-0.012 }$ \\ \hline
1.00 -- 1.25  &	0.695	 $\pm$ 0.003$^{+0.022 }_{-0.021 }$ &	0.654	 $\pm$ 0.005$^{+0.026 }_{-0.019 }$ \\ \hline
1.25 -- 1.50 &	0.959	 $\pm$ 0.003$^{+0.030 }_{-0.028 }$ &	0.925	 $\pm$ 0.006$^{+0.031 }_{-0.026 }$ \\ \hline
1.50 -- 1.75 &	1.216	 $\pm$ 0.004$^{+0.039 }_{-0.037 }$ &	1.191	 $\pm$ 0.007$^{+0.039 }_{-0.034 }$ \\ \hline
1.75 -- 2.00 &	1.437	 $\pm$ 0.004$^{+0.044 }_{-0.044 }$ &	1.440	 $\pm$ 0.008$^{+0.044 }_{-0.044 }$ \\ \hline
2.00 -- 2.25 &	1.587	 $\pm$ 0.004$^{+0.051 }_{-0.049 }$ &	1.658	 $\pm$ 0.008$^{+0.054 }_{-0.047 }$ \\ \hline
2.25 -- 2.50 &	 1.643	 $\pm$ 0.004$^{+0.051 }_{-0.051 }$ &	 1.772	 $\pm$ 0.008$^{+0.056 }_{-0.051 }$ \\ \hline
2.50 -- 2.75 &	 1.599	 $\pm$ 0.004$^{+0.050 }_{-0.049 }$ &	 1.861	 $\pm$ 0.008$^{+0.056 }_{-0.053 }$ \\ \hline
2.75 -- 3.00 &	 1.459	 $\pm$ 0.004$^{+0.049 }_{-0.044 }$ &	 1.829	 $\pm$ 0.008$^{+0.055 }_{-0.053 }$ \\ \hline
3.00 -- 3.25 &	 1.215	 $\pm$ 0.003$^{+0.038 }_{-0.038 }$ &	 1.673	 $\pm$ 0.008$^{+0.056 }_{-0.049 }$ \\ \hline
3.25 -- 3.50 &	 0.927	 $\pm$ 0.003$^{+0.028 }_{-0.036 }$ &	 1.398	 $\pm$ 0.007$^{+0.043 }_{-0.042 }$ \\ \hline
3.50 -- 3.75 &	 0.649	 $\pm$ 0.002$^{+0.021 }_{-0.036 }$ &	 1.087	 $\pm$ 0.006$^{+0.033 }_{-0.033 }$ \\ \hline
3.75 -- 4.00 &	 0.4209	 $\pm$ 0.0019$^{+0.0127 }_{-0.0277 }$ &	 0.778	 $\pm$ 0.005$^{+0.026 }_{-0.028 }$ \\ \hline
4.00 -- 4.25 &	 0.2561	 $\pm$ 0.0015$^{+0.0078 }_{-0.0219 }$ &	 0.517	 $\pm$ 0.004$^{+0.018 }_{-0.025 }$ \\ \hline
4.25 -- 4.50 &	 0.1473	 $\pm$ 0.0011$^{+0.0047 }_{-0.0141 }$ &	 0.317	 $\pm$ 0.003$^{+0.010 }_{-0.017 }$ \\ \hline
4.50 -- 4.75 &	 0.0842	 $\pm$ 0.0009$^{+0.0032 }_{-0.0081 }$ &	 0.184	 $\pm$ 0.002$^{+0.006 }_{-0.013 }$ \\ \hline
4.75 -- 5.00 &	 0.0474	 $\pm$ 0.0007$^{+0.0016 }_{-0.0045 }$ &	 0.1079	 $\pm$ 0.0019$^{+0.0032 }_{-0.0061 }$ \\ \hline
5.00 -- 5.25 &	 0.0266	 $\pm$ 0.0006$^{+0.0008 }_{-0.0022 }$ &	 0.0598	 $\pm$ 0.0015$^{+0.0024 }_{-0.0046 }$ \\ \hline
5.25 -- 5.50 &	 0.0167	 $\pm$ 0.0006$^{+0.0007 }_{-0.0025 }$ &	 0.0356	 $\pm$ 0.0013$^{+0.0011 }_{-0.0021 }$ \\ \hline
5.50 -- 5.75 &	 0.0102	 $\pm$ 0.0006$^{+0.0014 }_{-0.0017 }$ &	 0.0208	 $\pm$ 0.0012$^{+0.0011 }_{-0.0019 }$ \\ \hline
5.75 -- 6.00 &	 0.0069	 $\pm$ 0.0008$^{+0.0008 }_{-0.0040 }$ &	 0.0126	 $\pm$ 0.0012$^{+0.0011 }_{-0.0041 }$ \\ \hline
6.00 -- 6.25 &	 0.0040	 $\pm$ 0.0011$^{+0.0015 }_{-0.0051 }$ &	 0.0092	 $\pm$ 0.0016$^{+0.0033 }_{-0.0012 }$ \\ \hline
6.25 -- 6.50 &	 0.008	 $\pm$ 0.007$^{+0.002 }_{-0.004 }$ &	 0.004	 $\pm$ 0.002$^{+0.031 }_{-0.008 }$ \\ \hline
\hline
\end{tabular}

\caption{
The bin-averaged scaled momentum spectra, $1/N~ dn^{\pm} 
/d\ln(1/x_p)$, for $160<Q^2<640$~GeV $^2$. 
The first uncertainty is statistical, the second systematic.
}
\label{table-ksi1}
\end{table}

\begin{table}
\centering
   \begin{tabular}{|l|c|c|}
\hline
\hline
$\ln(1/x_p)$  &  $640<Q^2<1280$~GeV$^2$  & $1280<Q^2<2560$~GeV$^2$\\ \hline
\hline
0.00 -- 0.25 &	0.0228	 $\pm$ 0.0013$^{+0.0017 }_{-0.0014 }$ &	0.0236	 $\pm$ 0.0016$^{+0.0014 }_{-0.0013 }$ \\ \hline
0.25 -- 0.50 &	0.086	 $\pm$ 0.003$^{+0.006 }_{-0.006 }$ &	0.087	 $\pm$ 0.004$^{+0.003 }_{-0.004 }$ \\ \hline
0.50 -- 0.75 &	0.218	 $\pm$ 0.004$^{+0.007 }_{-0.011 }$ &	0.199	 $\pm$ 0.006$^{+0.011 }_{-0.006 }$ \\ \hline
0.75 -- 1.00 &	0.371	 $\pm$ 0.006$^{+0.013 }_{-0.011 }$ &	0.368	 $\pm$ 0.009$^{+0.023 }_{-0.011 }$ \\ \hline
1.00 -- 1.25 &	0.597	 $\pm$ 0.007$^{+0.018 }_{-0.022 }$ &	0.577	 $\pm$ 0.011$^{+0.032 }_{-0.016 }$ \\ \hline
1.25 -- 1.50 &	0.857	 $\pm$ 0.009$^{+0.027 }_{-0.031 }$ &	0.840	 $\pm$ 0.014$^{+0.035 }_{-0.024 }$ \\ \hline
1.50 -- 1.75 &	1.121	 $\pm$ 0.010$^{+0.034 }_{-0.037 }$ &	1.098	 $\pm$ 0.015$^{+0.039 }_{-0.032 }$ \\ \hline
1.75 -- 2.00 &	1.346	 $\pm$ 0.011$^{+0.041 }_{-0.044 }$ &	1.325	 $\pm$ 0.017$^{+0.052 }_{-0.039 }$ \\ \hline
2.00 -- 2.25 &	1.598	 $\pm$ 0.012$^{+0.048 }_{-0.050 }$ &	1.535	 $\pm$ 0.017$^{+0.063 }_{-0.044 }$ \\ \hline
2.25 -- 2.50 &	 1.780	 $\pm$ 0.013$^{+0.053 }_{-0.058 }$ &	 1.779	 $\pm$ 0.018$^{+0.060 }_{-0.051 }$ \\ \hline
2.50 -- 2.75 &	 1.920	 $\pm$ 0.013$^{+0.056 }_{-0.058 }$ &	 1.928	 $\pm$ 0.019$^{+0.069 }_{-0.056 }$ \\ \hline
2.75 -- 3.00 &	 2.027	 $\pm$ 0.013$^{+0.061 }_{-0.060 }$ &	 2.135	 $\pm$ 0.019$^{+0.082 }_{-0.061 }$ \\ \hline
3.00 -- 3.25 &	 1.995	 $\pm$ 0.013$^{+0.060 }_{-0.066 }$ &	 2.152	 $\pm$ 0.019$^{+0.077 }_{-0.063 }$ \\ \hline
3.25 -- 3.50 &	 1.816	 $\pm$ 0.012$^{+0.054 }_{-0.055 }$ &	 2.121	 $\pm$ 0.018$^{+0.073 }_{-0.063 }$ \\ \hline
3.50 -- 3.75 &	 1.575	 $\pm$ 0.011$^{+0.047 }_{-0.068 }$ &	 1.953	 $\pm$ 0.017$^{+0.078 }_{-0.058 }$ \\ \hline
3.75 -- 4.00 &	 1.241	 $\pm$ 0.009$^{+0.040 }_{-0.037 }$ &	 1.705	 $\pm$ 0.016$^{+0.054 }_{-0.052 }$ \\ \hline
4.00 -- 4.25 &	 0.901	 $\pm$ 0.008$^{+0.028 }_{-0.037 }$ &	 1.412	 $\pm$ 0.014$^{+0.046 }_{-0.042 }$ \\ \hline
4.25 -- 4.50 &	 0.603	 $\pm$ 0.006$^{+0.020 }_{-0.022 }$ &	 1.038	 $\pm$ 0.012$^{+0.038 }_{-0.033 }$ \\ \hline
4.50 -- 4.75 &	 0.376	 $\pm$ 0.005$^{+0.012 }_{-0.028 }$ &	 0.731	 $\pm$ 0.010$^{+0.027 }_{-0.033 }$ \\ \hline
4.75 -- 5.00 &	 0.232	 $\pm$ 0.004$^{+0.007 }_{-0.012 }$ &	 0.463	 $\pm$ 0.007$^{+0.019 }_{-0.026 }$ \\ \hline
5.00 -- 5.25 &	 0.133	 $\pm$ 0.003$^{+0.004 }_{-0.009 }$ &	 0.288	 $\pm$ 0.006$^{+0.009 }_{-0.018 }$ \\ \hline
5.25 -- 5.50 &	 0.070	 $\pm$ 0.002$^{+0.002 }_{-0.004 }$ &	 0.163	 $\pm$ 0.005$^{+0.005 }_{-0.009 }$ \\ \hline
5.50 -- 5.75 &	 0.047	 $\pm$ 0.002$^{+0.002 }_{-0.002 }$ &	 0.097	 $\pm$ 0.004$^{+0.006 }_{-0.006 }$ \\ \hline
5.75 -- 6.00 &	 0.023	 $\pm$ 0.002$^{+0.002 }_{-0.003 }$ &	 0.049	 $\pm$ 0.003$^{+0.002 }_{-0.006 }$ \\ \hline
6.00 -- 6.25 &	 0.013	 $\pm$ 0.002$^{+0.003 }_{-0.002 }$ &	 0.026	 $\pm$ 0.003$^{+0.004 }_{-0.007 }$ \\ \hline
6.25 -- 6.50 &	 0.007	 $\pm$ 0.003$^{+0.006 }_{-0.016 }$ &	 0.017	 $\pm$ 0.004$^{+0.005 }_{-0.006 }$ \\ \hline
6.50 -- 6.75 &	 0.024	 $\pm$ 0.022$^{+0.006 }_{-0.001 }$ &	 0.011	 $\pm$ 0.006$^{+0.010 }_{-0.013 }$ \\ \hline
\hline
\end{tabular}

\caption{The bin-averaged scaled momentum spectra, $1/N~ dn^{\pm} 
/d\ln(1/x_p)$, for $640<Q^2<2560$~GeV $^2$. 
The first uncertainty is statistical, the second systematic.
}
\label{table-ksi2}
\end{table}

\begin{table}
\centering
   \begin{tabular}{|l|c|c|}
\hline
\hline
$\ln(1/x_p)$  &  $2560<Q^2<5120$~GeV$^2$  & $5120<Q^2<10240$~GeV$^2$\\ \hline
\hline
0.00 -- 0.25 &	0.0246	 $\pm$ 0.0028$^{+0.0036 }_{-0.0011 }$ &	0.0261	 $\pm$ 0.0044$^{+0.0033 }_{-0.0019 }$ \\ \hline
0.25 -- 0.50 &	0.105	 $\pm$ 0.008$^{+0.005 }_{-0.003 }$ &	0.075	 $\pm$ 0.010$^{+0.008 }_{-0.003 }$ \\ \hline
0.50 -- 0.75 &	0.210	 $\pm$ 0.012$^{+0.014 }_{-0.006 }$ &	0.178	 $\pm$ 0.017$^{+0.018 }_{-0.004 }$ \\ \hline
0.75 -- 1.00 &	0.362	 $\pm$ 0.016$^{+0.031 }_{-0.011 }$ &	0.362	 $\pm$ 0.026$^{+0.044 }_{-0.010 }$ \\ \hline
1.00 -- 1.25 &	0.58	 $\pm$ 0.02$^{+0.03 }_{-0.02 }$ &	0.630	 $\pm$ 0.035$^{+0.055 }_{-0.018 }$ \\ \hline
1.25 -- 1.50 &	0.78	 $\pm$ 0.02$^{+0.03 }_{-0.02 }$ &	0.700	 $\pm$ 0.037$^{+0.069 }_{-0.019 }$ \\ \hline
1.50 -- 1.75 &	1.06	 $\pm$ 0.03$^{+0.07 }_{-0.03 }$ &	1.02	 $\pm$ 0.04$^{+0.11 }_{-0.03 }$ \\ \hline
1.75 -- 2.00 &	1.32	 $\pm$ 0.03$^{+0.08 }_{-0.04 }$ &	1.28	 $\pm$ 0.05$^{+0.09 }_{-0.03 }$ \\ \hline
2.00 -- 2.25 &	1.49	 $\pm$ 0.03$^{+0.10 }_{-0.04 }$ &	1.54	 $\pm$ 0.05$^{+0.14 }_{-0.04 }$ \\ \hline
2.25 -- 2.50 &	 1.73	 $\pm$ 0.03$^{+0.08 }_{-0.05 }$ &	 1.87	 $\pm$ 0.06$^{+0.15 }_{-0.06 }$ \\ \hline
2.50 -- 2.75 &	 1.95	 $\pm$ 0.03$^{+0.11 }_{-0.06 }$ &	 2.01	 $\pm$ 0.05$^{+0.14 }_{-0.06 }$ \\ \hline
2.75 -- 3.00 &	 2.20	 $\pm$ 0.04$^{+0.11 }_{-0.06 }$ &	 2.16	 $\pm$ 0.05$^{+0.14 }_{-0.06 }$ \\ \hline
3.00 -- 3.25 &	 2.22	 $\pm$ 0.03$^{+0.11 }_{-0.06 }$ &	 2.35	 $\pm$ 0.05$^{+0.12 }_{-0.07 }$ \\ \hline
3.25 -- 3.50 &	 2.25	 $\pm$ 0.03$^{+0.09 }_{-0.07 }$ &	 2.40	 $\pm$ 0.05$^{+0.13 }_{-0.07 }$ \\ \hline
3.50 -- 3.75 &	 2.27	 $\pm$ 0.03$^{+0.11 }_{-0.07 }$ &	 2.53	 $\pm$ 0.05$^{+0.14 }_{-0.08 }$ \\ \hline
3.75 -- 4.00 &	 2.13	 $\pm$ 0.03$^{+0.09 }_{-0.07 }$ &	 2.48	 $\pm$ 0.05$^{+0.15 }_{-0.08 }$ \\ \hline
4.00 -- 4.25 &	 1.90	 $\pm$ 0.03$^{+0.07 }_{-0.06 }$ &	 2.39	 $\pm$ 0.05$^{+0.13 }_{-0.08 }$ \\ \hline
4.25 -- 4.50 &	 1.51	 $\pm$ 0.02$^{+0.05 }_{-0.06 }$ &	 2.06	 $\pm$ 0.04$^{+0.12 }_{-0.06 }$ \\ \hline
4.50 -- 4.75 &	 1.17	 $\pm$ 0.02$^{+0.04 }_{-0.05 }$ &	 1.78	 $\pm$ 0.04$^{+0.08 }_{-0.06 }$ \\ \hline
4.75 -- 5.00 &	 0.828	 $\pm$ 0.017$^{+0.030 }_{-0.043 }$ &	 1.45	 $\pm$ 0.04$^{+0.10 }_{-0.08 }$ \\ \hline
5.00 -- 5.25 &	 0.547	 $\pm$ 0.014$^{+0.019 }_{-0.035 }$ &	 0.97	 $\pm$ 0.03$^{+0.06 }_{-0.07 }$ \\ \hline
5.25 -- 5.50 &	 0.345	 $\pm$ 0.011$^{+0.012 }_{-0.040 }$ &	 0.67	 $\pm$ 0.02$^{+0.05 }_{-0.05 }$ \\ \hline
5.50 -- 5.75 &	 0.188	 $\pm$ 0.009$^{+0.008 }_{-0.020 }$ &	 0.388	 $\pm$ 0.018$^{+0.041 }_{-0.051 }$ \\ \hline
5.75 -- 6.00 &	 0.126	 $\pm$ 0.008$^{+0.005 }_{-0.023 }$ &	 0.269	 $\pm$ 0.017$^{+0.022 }_{-0.027 }$ \\ \hline
6.00 -- 6.25 &	 0.077	 $\pm$ 0.008$^{+0.004 }_{-0.005 }$ &	 0.126	 $\pm$ 0.015$^{+0.024 }_{-0.022 }$ \\ \hline
6.25 -- 6.50 &	 0.033	 $\pm$ 0.007$^{+0.004 }_{-0.009 }$ &	 0.112	 $\pm$ 0.018$^{+0.022 }_{-0.013 }$ \\ \hline
6.50 -- 6.75 &	 0.027	 $\pm$ 0.010$^{+0.001 }_{-0.017 }$ &	 0.058	 $\pm$ 0.019$^{+0.016 }_{-0.011 }$ \\ \hline
\hline
\end{tabular}

\caption{The bin-averaged scaled momentum spectra, $1/N~ dn^{\pm} 
/d\ln(1/x_p)$, for $2560<Q^2<10240$~GeV $^2$.
The first uncertainty is statistical, the second systematic.
}
\label{table-ksi3}
\end{table}

\begin{table}
\centering
   \begin{tabular}{|l|c|}
\hline
\hline
$\ln(1/x_p)$  &  $10240<Q^2<20480$~GeV$^2$ \\ \hline
\hline
0.0 -- 0.5 &	0.062	 $\pm$ 0.011$^{+0.007 }_{-0.001 }$ \\ \hline
0.5 -- 1.0 &	0.220	 $\pm$ 0.028$^{+0.039 }_{-0.006 }$ \\ \hline
1.0 -- 1.5 &	0.570	 $\pm$ 0.047$^{+0.087 }_{-0.017 }$ \\ \hline
1.5 -- 2.0 &	1.12	 $\pm$ 0.07$^{+0.15 }_{-0.04 }$ \\ \hline
2.0 -- 2.5 &	1.37	 $\pm$ 0.07$^{+0.19 }_{-0.04 }$ \\ \hline
2.5 -- 3.0 &	2.12	 $\pm$ 0.08$^{+0.20 }_{-0.06 }$ \\ \hline
3.0 -- 3.5 &	2.60	 $\pm$ 0.08$^{+0.18 }_{-0.10 }$ \\ \hline
3.5 -- 4.0 &	2.56	 $\pm$ 0.07$^{+0.19 }_{-0.08 }$ \\ \hline
4.0 -- 4.5 &	2.60	 $\pm$ 0.07$^{+0.18 }_{-0.09 }$ \\ \hline
4.5 -- 5.0 &	 2.16	 $\pm$ 0.06$^{+0.14 }_{-0.08 }$ \\ \hline
5.0 -- 5.5 &	 1.36	 $\pm$ 0.05$^{+0.07 }_{-0.06 }$ \\ \hline
5.5 -- 6.0 &	 0.66	 $\pm$ 0.03$^{+0.04 }_{-0.04 }$ \\ \hline
6.0 -- 6.5 &	 0.25	 $\pm$ 0.03$^{+0.03 }_{-0.02 }$ \\ \hline
6.5 -- 7.0 &	 0.04	 $\pm$ 0.02$^{+0.04 }_{-0.02 }$ \\ \hline
\hline
\end{tabular}

\caption{
The bin-averaged scaled momentum spectra, $1/N~ dn^{\pm} 
/d\ln(1/x_p)$, for $10240<Q^2<20480$~GeV $^2$.
The first uncertainty is statistical, the second systematic.
}
\label{table-ksi4}
\end{table}

\begin{table}
\centering
   \begin{tabular}{|l|c|}
\hline
\hline
$\ln(1/x_p)$  &  $20480<Q^2<40960$~GeV$^2$ \\ \hline
\hline
0.0 -- 1.0 &	0.13	 $\pm$ 0.04$^{+0.03 }_{-0.03 }$ \\ \hline
1.0 -- 2.0 &	0.85	 $\pm$ 0.10$^{+0.10 }_{-0.03 }$ \\ \hline
2.0 -- 3.0 &	1.51	 $\pm$ 0.12$^{+0.22 }_{-0.05 }$ \\ \hline
3.0 -- 4.0 &	2.31	 $\pm$ 0.13$^{+0.27 }_{-0.08 }$ \\ \hline
4.0 -- 5.0 &	2.43	 $\pm$ 0.11$^{+0.24 }_{-0.11 }$ \\ \hline
5.0 -- 6.0 &	1.32	 $\pm$ 0.07$^{+0.17 }_{-0.10 }$ \\ \hline
6.0 -- 7.0 &	0.33	 $\pm$ 0.05$^{+0.07 }_{-0.05 }$ \\ \hline
\hline
\end{tabular}

\caption{
The bin-averaged scaled momentum spectra, $1/N~ dn^{\pm} 
/d\ln(1/x_p)$, for $20480<Q^2<40960$~GeV $^2$. 
The first uncertainty is statistical, the second systematic.
}
\label{table-ksi5}
\end{table}

\begin{table}
\centering
   \begin{tabular}{|r|c|}
\hline
 \hline
 $<Q^2>$,  GeV$^2$  &  $0<x_p<0.02   $ \\ \hline
 \hline
 $    218$ & $   9.32\pm    0.03^{ +   0.29}_{ -   0.75}$ \\ \hline
 $    440$ & $  19.10\pm    0.09^{ +   0.59}_{ -   0.86}$ \\ \hline
 $    871$ & $  35.53\pm    0.18^{ +   1.08}_{ -   1.47}$ \\ \hline
 $   1767$ & $  61.3\pm    0.3^{ +   2.1}_{ -   2.0}$ \\ \hline
 $   3530$ & $  93.6\pm    0.7^{ +   3.2}_{ -   3.7}$ \\ \hline
 $   6870$ & $ 140.1\pm    1.3^{ +   8.1}_{ -   5.9}$ \\ \hline
 $  13380$ & $ 191\pm    3^{ +  11}_{ -   6}$ \\ \hline
 $  25700$ & $ 218\pm    7^{ +  23}_{ -  11}$ \\ \hline
 \hline
 $<Q^2>$,  GeV$^2$  &  $0.02<x_p<0.05$ \\ \hline
 \hline
 $    218$ & $  25.86\pm    0.04^{ +   0.80}_{ -   0.94}$ \\ \hline
 $    440$ & $  39.34\pm    0.11^{ +   1.23}_{ -   1.18}$ \\ \hline
 $    871$ & $  52.15\pm    0.18^{ +   1.55}_{ -   1.63}$ \\ \hline
 $   1767$ & $  61.6\pm    0.3^{ +   2.1}_{ -   1.8}$ \\ \hline
 $   3530$ & $  68.3\pm    0.5^{ +   2.9}_{ -   2.0}$ \\ \hline
 $   6870$ & $  74.6\pm    0.8^{ +   4.0}_{ -   2.3}$ \\ \hline
 $  13380$ & $  78.6\pm    1.7^{ +   5.6}_{ -   2.7}$ \\ \hline
 $  25700$ & $  69\pm    4^{ +   8}_{ -   2}$ \\ \hline
 \hline
 $<Q^2>$,  GeV$^2$  &  $0.05<x_p<0.1 $ \\ \hline
 \hline
 $    218$ & $  19.79\pm    0.03^{ +   0.63}_{ -   0.61}$ \\ \hline
 $    440$ & $  25.34\pm    0.07^{ +   0.78}_{ -   0.74}$ \\ \hline
 $    871$ & $  26.67\pm    0.11^{ +   0.78}_{ -   0.79}$ \\ \hline
 $   1767$ & $  27.20\pm    0.16^{ +   0.98}_{ -   0.79}$ \\ \hline
 $   3530$ & $  27.6\pm    0.3^{ +   1.3}_{ -   0.8}$ \\ \hline
 $   6870$ & $  28.0\pm    0.5^{ +   2.0}_{ -   0.8}$ \\ \hline
 $  13380$ & $  26.8\pm    0.9^{ +   2.9}_{ -   0.7}$ \\ \hline
 $  25700$ & $  23.3\pm    2.1^{ +   2.8}_{ -   0.8}$ \\ \hline
\hline
\end{tabular}

\caption{
The number of charged particles per event and unit of $x_p$, $1/N~n^{\pm} /\Delta x_p$,  as a function of $Q^2$ in bins of $x_p$ with widths $\Delta x_p$. 
The first uncertainty is statistical, the second systematic.
}
\label{table-xp1}
\end{table}

\begin{table}
\centering
   \begin{tabular}{|r|c|}
\hline
 \hline
 $<Q^2>$,  GeV$^2$  &  $0.1<x_p<0.2  $ \\ \hline
 \hline
 $    218$ & $   9.338\pm    0.015^{ +   0.30}_{ -   0.29}$ \\ \hline
 %$    218$ & $   9.338\pm    0.015^{ +   0.295}_{ -   0.288}$ \\ \hline
 $    440$ & $  10.39\pm    0.03^{ +   0.32}_{ -   0.30}$ \\ \hline
 $    871$ & $   9.92\pm    0.05^{ +   0.30}_{ -   0.32}$ \\ \hline
 $   1767$ & $   9.68\pm    0.07^{ +   0.38}_{ -   0.28}$ \\ \hline
 $   3530$ & $   9.51\pm    0.13^{ +   0.63}_{ -   0.28}$ \\ \hline
 $   6870$ & $   9.6\pm    0.2^{ +   0.8}_{ -   0.3}$ \\ \hline
 $  13380$ & $   8.5\pm    0.4^{ +   1.2}_{ -   0.3}$ \\ \hline
 $  25700$ & $   7.9\pm    0.9^{ +   1.3}_{ -   0.3}$ \\ \hline
 \hline
 $<Q^2>$,  GeV$^2$  &  $0.2<x_p<0.3  $ \\ \hline
 \hline
 $    218$ & $   4.03\pm    0.01^{ +   0.13}_{ -   0.12}$ \\ \hline
 $    440$ & $   3.89\pm    0.02^{ +   0.14}_{ -   0.11}$ \\ \hline
 $    871$ & $   3.60\pm    0.03^{ +   0.11}_{ -   0.12}$ \\ \hline
 $   1767$ & $   3.52\pm    0.04^{ +   0.13}_{ -   0.10}$ \\ \hline
 $   3530$ & $   3.29\pm    0.08^{ +   0.19}_{ -   0.10}$ \\ \hline
 $   6870$ & $   3.06\pm    0.12^{ +   0.33}_{ -   0.09}$ \\ \hline
 $  13380$ & $   2.87\pm    0.24^{ +   0.37}_{ -   0.08}$ \\ \hline
 $  25700$ & $   3.1\pm    0.6^{ +   0.5}_{ -   0.3}$ \\ \hline
 \hline
 $<Q^2>$,  GeV$^2$  &  $0.3<x_p<0.4  $ \\ \hline
 \hline
 $    218$ & $   1.806\pm    0.007^{ +   0.057}_{ -   0.053}$ \\ \hline
 $    440$ & $   1.717\pm    0.013^{ +   0.070}_{ -   0.050}$ \\ \hline
 $    871$ & $   1.515\pm    0.019^{ +   0.047}_{ -   0.046}$ \\ \hline
 $   1767$ & $   1.49\pm    0.03^{ +   0.08}_{ -   0.04}$ \\ \hline
 $   3530$ & $   1.51\pm    0.05^{ +   0.09}_{ -   0.05}$ \\ \hline
 $   6870$ & $   1.65\pm    0.09^{ +   0.14}_{ -   0.04}$ \\ \hline
 $  13380$ & $   1.33\pm    0.16^{ +   0.18}_{ -   0.12}$ \\ \hline
 $  25700$ & $   0.7\pm    0.3^{ +   0.6}_{ -   0.4}$ \\ \hline
\hline
\end{tabular}

\caption{
The number of charged particles per event and unit of $x_p$, $1/N~n^{\pm} /\Delta x_p$,
  as a function of $Q^2$ in bins of $x_p$ with widths $\Delta x_p$. 
The first uncertainty is statistical, the second systematic.
}
\label{table-xp2}
\end{table}

\begin{table}
\centering
   \begin{tabular}{|r|c|}
\hline
 \hline
 $<Q^2>$,  GeV$^2$  &  $0.4<x_p<0.5  $ \\ \hline
 \hline
 $    218$ & $   0.868\pm    0.005^{ +   0.028}_{ -   0.027}$ \\ \hline
 $    440$ & $   0.785\pm    0.009^{ +   0.031}_{ -   0.023}$ \\ \hline
 $    871$ & $   0.716\pm    0.013^{ +   0.022}_{ -   0.025}$ \\ \hline
 $   1767$ & $   0.721\pm    0.019^{ +   0.040}_{ -   0.021}$ \\ \hline
 $   3530$ & $   0.692\pm    0.036^{ +   0.053}_{ -   0.021}$ \\ \hline
 $   6870$ & $   0.714\pm    0.057^{ +   0.081}_{ -   0.018}$ \\ \hline
 $  13380$ & $   0.392\pm    0.086^{ +   0.089}_{ -   0.013}$ \\ \hline
 $  25700$ & $   0.68\pm    0.34^{ +   0.19}_{ -   0.16}$ \\ \hline
 \hline
 $<Q^2>$,  GeV$^2$  &  $0.5<x_p<0.7  $ \\ \hline
 \hline
 $    218$ & $   0.329\pm    0.002^{ +   0.010}_{ -   0.010}$ \\ \hline
 $    440$ & $   0.295\pm    0.004^{ +   0.009}_{ -   0.008}$ \\ \hline
 $    871$ & $   0.269\pm    0.006^{ +   0.009}_{ -   0.010}$ \\ \hline
 $   1767$ & $   0.249\pm    0.008^{ +   0.015}_{ -   0.007}$ \\ \hline
 $   3530$ & $   0.277\pm    0.015^{ +   0.016}_{ -   0.008}$ \\ \hline
 $   6870$ & $   0.197\pm    0.020^{ +   0.024}_{ -   0.006}$ \\ \hline
 $  13380$ & $   0.257\pm    0.045^{ +   0.054}_{ -   0.006}$ \\ \hline
 $  25700$ & $   0.33\pm    0.15^{ +   0.11}_{ -   0.07}$ \\ \hline
 \hline
 $<Q^2>$,  GeV$^2$  &  $0.7<x_p<1.0  $ \\ \hline
 \hline
 $    218$ & $   0.056\pm    0.001^{ +   0.002}_{ -   0.002}$ \\ \hline
 $    440$ & $   0.046\pm    0.001^{ +   0.001}_{ -   0.001}$ \\ \hline
 $    871$ & $   0.041\pm    0.002^{ +   0.001}_{ -   0.002}$ \\ \hline
 $   1767$ & $   0.042\pm    0.002^{ +   0.001}_{ -   0.003}$ \\ \hline
 $   3530$ & $   0.049\pm    0.004^{ +   0.004}_{ -   0.001}$ \\ \hline
 $   6870$ & $   0.043\pm    0.006^{ +   0.003}_{ -   0.002}$ \\ \hline
 $  13380$ & $   0.054\pm    0.012^{ +   0.007}_{ -   0.002}$ \\ \hline
 $  25700$ & $   0.013\pm    0.013^{ +   0.002}_{ -   0.021}$ \\ \hline
\hline
\end{tabular}

\caption{
The number of charged particles per event and unit of $x_p$, $1/N~n^{\pm} /\Delta x_p$,
  as a function of $Q^2$ in bins of $x_p$ with widths $\Delta x_p$. 
The first uncertainty is statistical, the second systematic.
}
\label{table-xp3}
\end{table}

%-------------------------------------------------------------------------------
%       end of tables  
%-------------------------------------------------------------------------------

%-------------------------------------------------------------------------------
%       Tables
%-------------------------------------------------------------------------------

%-------------------------------------------------------------------------------
%       eta_Breit as F(Q2,W)  for Q2>160 GeV2
%-------------------------------------------------------------------------------

\begin{table}
\centering
   \begin{tabular}{|r l r|c|c|c|}
\hline
\hline
\multicolumn{3}{|c|}{$\eta^{\rm Breit}$}  &  $160<W<200$~GeV  & $200<W<240$~GeV & $240<W<280$~GeV\\ \hline
\hline
-5.5 & --& -5.2 &	0.15	 $\pm$ 0.11	$^{+0.06}	_{-0.06}$ &	0.054	 $\pm$ 0.027	$^{+0.030}	_{-0.006}$ &	0.017	 $\pm$ 0.012	$^{+0.015}	_{-0.014}$ \\ \hline
-5.2& --& -4.9 &	0.14	 $\pm$ 0.08	$^{+0.14}	_{-0.05}$ &	0.04	 $\pm$ 0.02	$^{+0.04}	_{-0.03}$ &	0.027	 $\pm$ 0.016	$^{+0.017}	_{-0.053}$ \\ \hline
-4.9& --& -4.6 &	0.07	 $\pm$ 0.03	$^{+0.04}	_{-0.06}$ &	0.08	 $\pm$ 0.03 	$^{+0.04}	_{-0.02}$ &	0.068	 $\pm$ 0.024	$^{+0.045}	_{-0.018}$ \\ \hline
-4.6& --& -4.3 &	0.15	 $\pm$ 0.04	$^{+0.05}	_{-0.04}$ &	0.22	 $\pm$ 0.04	$^{+0.11}	_{-0.07}$ &	0.16	 $\pm$ 0.03	$^{+0.09}	_{-0.08}$ \\ \hline
-4.3& -- &-4.0 &	 0.25	 $\pm$ 0.05	$^{+0.07}	_{-0.09}$ &	 0.20	 $\pm$ 0.04	$^{+0.07}	_{-0.04}$ &	 0.23	 $\pm$ 0.04	$^{+0.06}	_{-0.08}$ \\ \hline
-4.0& --& -3.7 &	 0.33	 $\pm$ 0.06	$^{+0.01}	_{-0.12}$ &	 0.42	 $\pm$ 0.06	$^{+0.17}	_{-0.04}$ &	 0.28	 $\pm$ 0.04	$^{+0.09}	_{-0.10}$ \\ \hline
-3.7& --& -3.4 &	 0.51	 $\pm$ 0.07	$^{+0.14}	_{-0.11}$ &	 0.61	 $\pm$ 0.06	$^{+0.24}	_{-0.08}$ &	 0.54	 $\pm$ 0.05	$^{+0.14}	_{-0.11}$ \\ \hline
-3.4& -- &-3.1 &	 0.97	 $\pm$ 0.09	$^{+0.17}	_{-0.12}$ &	 0.92	 $\pm$ 0.08	$^{+0.15}	_{-0.10}$ &	 0.64	 $\pm$ 0.06	$^{+0.10}	_{-0.14}$ \\ \hline
-3.1& --& -2.8 &	 1.31	 $\pm$ 0.10	$^{+0.22}	_{-0.16}$ &	 1.07	 $\pm$ 0.08	$^{+0.27}	_{-0.12}$ &	 0.96	 $\pm$ 0.07	$^{+0.15}	_{-0.15}$ \\ \hline
-2.8 &-- &-2.5 &	 1.54	 $\pm$ 0.10	$^{+0.21}	_{-0.17}$ &	 1.37	 $\pm$ 0.08	$^{+0.22}	_{-0.10}$ &	 1.36	 $\pm$ 0.08	$^{+0.16}	_{-0.20}$ \\ \hline
-2.5& --& -2.2 &	 1.88	 $\pm$ 0.10	$^{+0.09}	_{-0.15}$ &	 1.60	 $\pm$ 0.09	$^{+0.20}	_{-0.23}$ &	 1.72	 $\pm$ 0.09	$^{+0.24}	_{-0.24}$ \\ \hline
-2.2& --& -1.9 &	 2.05	 $\pm$ 0.10	$^{+0.11}	_{-0.19}$ &	 2.01	 $\pm$ 0.09	$^{+0.25}	_{-0.16}$ &	 2.09	 $\pm$ 0.09	$^{+0.37}	_{-0.17}$ \\ \hline
-1.9& --& -1.6 &	 2.34	 $\pm$ 0.10	$^{+0.16}	_{-0.14}$ &	 2.28	 $\pm$ 0.09	$^{+0.27}	_{-0.19}$ &	 2.46	 $\pm$ 0.10	$^{+0.38}	_{-0.17}$ \\ \hline
-1.6& --& -1.3 &	 2.42	 $\pm$ 0.10	$^{+0.13}	_{-0.19}$ &	 2.47	 $\pm$ 0.09	$^{+0.32}	_{-0.14}$ &	 2.54	 $\pm$ 0.10	$^{+0.13}	_{-0.19}$ \\ \hline
-1.3& --& 1.0 &	 2.30	 $\pm$ 0.09	$^{+0.10}	_{-0.18}$ &	 2.63	 $\pm$ 0.09	$^{+0.32}	_{-0.20}$ &	 2.81	 $\pm$ 0.10	$^{+0.33}	_{-0.41}$ \\ \hline
-1.0& --& -0.7 &	 2.52	 $\pm$ 0.09	$^{+0.20}	_{-0.14}$ &	 2.82	 $\pm$ 0.09	$^{+0.32}	_{-0.24}$ &	 2.98	 $\pm$ 0.11	$^{+0.27}	_{-0.25}$ \\ \hline
-0.7& --& -0.4 &	 2.67	 $\pm$ 0.09	$^{+0.29}	_{-0.12}$ &	 2.68	 $\pm$ 0.09	$^{+0.18}	_{-0.29}$ &	 2.94	 $\pm$ 0.11	$^{+0.23}	_{-0.28}$ \\ \hline
-0.4& -- &-0.1 &	 2.41	 $\pm$ 0.09	$^{+0.36}	_{-0.09}$ &	 2.66	 $\pm$ 0.09	$^{+0.29}	_{-0.27}$ &	 2.82	 $\pm$ 0.10	$^{+0.34}	_{-0.28}$ \\ \hline
-0.1& --& 0.2 &	  &	 2.46	 $\pm$ 0.08	$^{+0.34}	_{-0.09}$ &	 2.81	 $\pm$ 0.10	$^{+0.32}	_{-0.18}$ \\ \hline
0.2& --& 0.5 &	  &	 2.81	 $\pm$ 0.10	$^{+0.46}	_{-0.15}$ &	 2.80	 $\pm$ 0.10	$^{+0.26}	_{-0.29}$ \\ \hline
0.5& --& 0.8 &	  &	  &	 3.08	 $\pm$ 0.11	$^{+0.53}	_{-0.33}$ \\ \hline
\hline
\end{tabular}

\caption{
The charged-particle density, $1/N~dn^{\pm} /d\eta^{\rm Breit}$, 
for $5120<Q^2<10240$~GeV~$^2$ and 3 bins in $W$.
The first uncertainty is statistical, the second systematic.
}
\label{table-q6}
\end{table}

\begin{table}
\centering
   \begin{tabular}{|r l r|c|c|c|}
\hline
\hline
\multicolumn{3}{|c|}{$\eta^{\rm Breit}$}  &  $160<W<200$~GeV  & $200<W<240$~GeV & $240<W<280$~GeV\\ \hline
\hline
-4.6& --& -4.3 &	0.112	 $\pm$ 0.019	$^{+0.042}	_{-0.012}$ &	0.074	 $\pm$ 0.015	$^{+0.019}	_{-0.028}$ &	0.069	 $\pm$ 0.014	$^{+0.050}	_{-0.031}$ \\ \hline
-4.3& --& -4.0 &	 0.16	 $\pm$ 0.02	$^{+0.04}	_{-0.03}$ &	 0.15	 $\pm$ 0.02	$^{+0.07}	_{-0.01}$ &	 0.095	 $\pm$ 0.015	$^{+0.041}	_{-0.024}$ \\ \hline
-4.0& --& -3.7 &	 0.24	 $\pm$ 0.03	$^{+0.07}	_{-0.02}$ &	 0.19	 $\pm$ 0.02	$^{+0.08}	_{-0.03}$ &	 0.19	 $\pm$ 0.02	$^{+0.05}	_{-0.03}$ \\ \hline
-3.7& -- &-3.4 &	 0.35	 $\pm$ 0.03	$^{+0.05}	_{-0.04}$ &	 0.34	 $\pm$ 0.03	$^{+0.05}	_{-0.02}$ &	 0.30	 $\pm$ 0.03	$^{+0.08}	_{-0.03}$ \\ \hline
-3.4 &-- &-3.1 &	 0.71	 $\pm$ 0.04	$^{+0.14}	_{-0.12}$ &	 0.57	 $\pm$ 0.04	$^{+0.08}	_{-0.05}$ &	 0.52	 $\pm$ 0.03	$^{+0.07}	_{-0.07}$ \\ \hline
-3.1& --& -2.8 &	 0.86	 $\pm$ 0.04	$^{+0.11}	_{-0.07}$ &	 0.94	 $\pm$ 0.04	$^{+0.10}	_{-0.11}$ &	 0.70	 $\pm$ 0.04	$^{+0.08}	_{-0.06}$ \\ \hline
-2.8 &--& -2.5 &	 1.31	 $\pm$ 0.05	$^{+0.13}	_{-0.11}$ &	 1.16	 $\pm$ 0.05	$^{+0.16}	_{-0.07}$ &	 1.08	 $\pm$ 0.05	$^{+0.13}	_{-0.10}$ \\ \hline
-2.5& -- &-2.2 &	 1.59	 $\pm$ 0.05	$^{+0.11}	_{-0.12}$ &	 1.60	 $\pm$ 0.05	$^{+0.18}	_{-0.07}$ &	 1.43	 $\pm$ 0.05	$^{+0.14}	_{-0.09}$ \\ \hline
-2.2 &-- &-1.9 &	 1.78	 $\pm$ 0.05	$^{+0.06}	_{-0.20}$ &	 1.80	 $\pm$ 0.06	$^{+0.12}	_{-0.13}$ &	 1.72	 $\pm$ 0.05	$^{+0.13}	_{-0.17}$ \\ \hline
-1.9& -- &-1.6 &	 2.15	 $\pm$ 0.06	$^{+0.12}	_{-0.15}$ &	 2.11	 $\pm$ 0.06	$^{+0.11}	_{-0.24}$ &	 2.11	 $\pm$ 0.06	$^{+0.22}	_{-0.15}$ \\ \hline
-1.6 &-- &-1.3 &	 2.28	 $\pm$ 0.06	$^{+0.07}	_{-0.24}$ &	 2.37	 $\pm$ 0.06	$^{+0.16}	_{-0.18}$ &	 2.32	 $\pm$ 0.06	$^{+0.13}	_{-0.16}$ \\ \hline
-1.3& --& -1.0 &	 2.28	 $\pm$ 0.05	$^{+0.09}	_{-0.22}$ &	 2.50	 $\pm$ 0.06	$^{+0.13}	_{-0.21}$ &	 2.45	 $\pm$ 0.06	$^{+0.13}	_{-0.17}$ \\ \hline
-1.0 &--& -0.7 &	 2.30	 $\pm$ 0.05	$^{+0.13}	_{-0.24}$ &	 2.64	 $\pm$ 0.06	$^{+0.23}	_{-0.14}$ &	 2.63	 $\pm$ 0.06	$^{+0.12}	_{-0.17}$ \\ \hline
-0.7& --& -0.4 &	 2.46	 $\pm$ 0.05	$^{+0.10}	_{-0.17}$ &	 2.62	 $\pm$ 0.06	$^{+0.21}	_{-0.20}$ &	 2.85	 $\pm$ 0.07	$^{+0.27}	_{-0.23}$ \\ \hline
-0.4 &--& -0.1 &	 2.41	 $\pm$ 0.05	$^{+0.25}	_{-0.16}$ &	 2.77	 $\pm$ 0.06	$^{+0.19}	_{-0.23}$ &	 2.82	 $\pm$ 0.06	$^{+0.19}	_{-0.26}$ \\ \hline
-0.1& --& 0.2 &	  &	 2.55	 $\pm$ 0.06	$^{+0.17}	_{-0.24}$ &	 2.66	 $\pm$ 0.06	$^{+0.20}	_{-0.21}$ \\ \hline
0.2 &--& 0.5 &	  &	 2.77	 $\pm$ 0.06	$^{+0.29}	_{-0.15}$ &	 2.86	 $\pm$ 0.06	$^{+0.31}	_{-0.23}$ \\ \hline
0.5& --& 0.8 &	  &	 2.71	 $\pm$ 0.07	$^{+0.30}	_{-0.31}$ &	 3.16	 $\pm$ 0.07	$^{+0.52}	_{-0.20}$ \\ \hline
0.8& --& 1.1 &	  &	  &	 2.92	 $\pm$ 0.07	$^{+0.58}	_{-0.15}$ \\ \hline
\hline
\end{tabular}

\caption{
The charged-particle density, $1/N~dn^{\pm} /d\eta^{\rm Breit}$,  for $2560<Q^2<5120$~GeV~$^2$ and 3 bins in $W$.
The first uncertainty is statistical, the second systematic.
}
\label{table-q5}
\end{table}

\begin{sidewaystable}
\centering
   \begin{tabular}{|r l r|c|c|c|c|}
\hline
\hline
\multicolumn{3}{|c|}{$\eta^{\rm Breit}$}  &  $120<W<160$~GeV & $160<W<200$~GeV & $200<W<240$~GeV & $240<W<280$~GeV \\ \hline
\hline
-4.3& --& -4.0 &	 0.103	 $\pm$ 0.010	$^{+0.052}	_{-0.014}$ &	 0.071	 $\pm$ 0.008	$^{+0.013}	_{-0.016}$ &	 0.071	 $\pm$ 0.008	$^{+0.013}	_{-0.018}$ &	 0.057	 $\pm$ 0.007	$^{+0.022}	_{-0.011}$ \\ \hline
-4.0& --& -3.7 &	 0.170	 $\pm$ 0.013	$^{+0.010}	_{-0.036}$ &	 0.161	 $\pm$ 0.012	$^{+0.022}	_{-0.025}$ &	 0.127	 $\pm$ 0.010	$^{+0.025}	_{-0.027}$ &	 0.088	 $\pm$ 0.008	$^{+0.025}	_{-0.010}$ \\ \hline
-3.7& --& -3.4 &	 0.268	 $\pm$ 0.016	$^{+0.043}	_{-0.034}$ &	 0.264	 $\pm$ 0.015	$^{+0.046}	_{-0.034}$ &	 0.218	 $\pm$ 0.013	$^{+0.030}	_{-0.036}$ &	 0.203	 $\pm$ 0.012	$^{+0.028}	_{-0.011}$ \\ \hline
-3.4& --& -3.1 &	 0.53	 $\pm$ 0.02	$^{+0.10}	_{-0.05}$ &	 0.414	 $\pm$ 0.018	$^{+0.070}	_{-0.032}$ &	 0.359	 $\pm$ 0.017	$^{+0.032}	_{-0.035}$ &	 0.300	 $\pm$ 0.014	$^{+0.053}	_{-0.035}$ \\ \hline
-3.1& -- &-2.8 &	 0.73	 $\pm$ 0.02	$^{+0.07}	_{-0.03}$ &	 0.65	 $\pm$ 0.02	$^{+0.07}	_{-0.04}$ &	 0.61	 $\pm$ 0.02	$^{+0.10}	_{-0.04}$ &	 0.524	 $\pm$ 0.019	$^{+0.055}	_{-0.029}$ \\ \hline
-2.8& -- &-2.5 &	 1.02	 $\pm$ 0.03	$^{+0.12}	_{-0.06}$ &	 0.93	 $\pm$ 0.03	$^{+0.08}	_{-0.05}$ &	 0.81	 $\pm$ 0.02	$^{+0.06}	_{-0.08}$ &	 0.77	 $\pm$ 0.02	$^{+0.03}	_{-0.07}$ \\ \hline
-2.5& --& -2.2 &	 1.41	 $\pm$ 0.03	$^{+0.20}	_{-0.09}$ &	 1.28	 $\pm$ 0.03	$^{+0.05}	_{-0.07}$ &	 1.19	 $\pm$ 0.03	$^{+0.10}	_{-0.06}$ &	 1.09	 $\pm$ 0.03	$^{+0.07}	_{-0.09}$ \\ \hline
-2.2 &--& -1.9 &	 1.67	 $\pm$ 0.03	$^{+0.13}	_{-0.09}$ &	 1.62	 $\pm$ 0.03	$^{+0.12}	_{-0.06}$ &	 1.50	 $\pm$ 0.03	$^{+0.09}	_{-0.09}$ &	 1.42	 $\pm$ 0.03	$^{+0.08}	_{-0.08}$ \\ \hline
-1.9 &-- &-1.6 &	 1.93	 $\pm$ 0.03	$^{+0.15}	_{-0.09}$ &	 1.92	 $\pm$ 0.03	$^{+0.08}	_{-0.12}$ &	 1.85	 $\pm$ 0.03	$^{+0.10}	_{-0.08}$ &	 1.74	 $\pm$ 0.03	$^{+0.09}	_{-0.10}$ \\ \hline
-1.6& -- &-1.3 &	 2.09	 $\pm$ 0.03	$^{+0.10}	_{-0.11}$ &	 2.11	 $\pm$ 0.03	$^{+0.09}	_{-0.08}$ &	 2.08	 $\pm$ 0.03	$^{+0.16}	_{-0.11}$ &	 2.04	 $\pm$ 0.03	$^{+0.13}	_{-0.15}$ \\ \hline
-1.3 &-- &-1.0 &	 2.21	 $\pm$ 0.03	$^{+0.09}	_{-0.13}$ &	 2.33	 $\pm$ 0.03	$^{+0.14}	_{-0.11}$ &	 2.26	 $\pm$ 0.04	$^{+0.12}	_{-0.07}$ &	 2.25	 $\pm$ 0.04	$^{+0.13}	_{-0.13}$ \\ \hline
-1.0 &-- &-0.7 &	 2.24	 $\pm$ 0.03	$^{+0.11}	_{-0.09}$ &	 2.42	 $\pm$ 0.03	$^{+0.10}	_{-0.09}$ &	 2.41	 $\pm$ 0.04	$^{+0.10}	_{-0.12}$ &	 2.45	 $\pm$ 0.04	$^{+0.24}	_{-0.14}$ \\ \hline
-0.7 &-- &-0.4 &	 2.28	 $\pm$ 0.03	$^{+0.10}	_{-0.10}$ &	 2.53	 $\pm$ 0.03	$^{+0.09}	_{-0.11}$ &	 2.60	 $\pm$ 0.04	$^{+0.19}	_{-0.13}$ &	 2.62	 $\pm$ 0.04	$^{+0.23}	_{-0.13}$ \\ \hline
-0.4& -- &-0.1 &	 2.31	 $\pm$ 0.03	$^{+0.18}	_{-0.11}$ &	 2.65	 $\pm$ 0.03	$^{+0.24}	_{-0.09}$ &	 2.64	 $\pm$ 0.04	$^{+0.13}	_{-0.14}$ &	 2.68	 $\pm$ 0.04	$^{+0.36}	_{-0.12}$ \\ \hline
-0.1 &--& 0.2 &	  &	 2.60	 $\pm$ 0.03	$^{+0.21}	_{-0.09}$ &	 2.66	 $\pm$ 0.04	$^{+0.15}	_{-0.09}$ &	 2.68	 $\pm$ 0.04	$^{+0.23}	_{-0.14}$ \\ \hline
0.2& --& 0.5 &	  &	 2.54	 $\pm$ 0.03	$^{+0.16}	_{-0.10}$ &	 2.81	 $\pm$ 0.04	$^{+0.18}	_{-0.16}$ &	 2.95	 $\pm$ 0.04	$^{+0.38}	_{-0.11}$ \\ \hline
0.5 &--& 0.8 &	  &	 2.43	 $\pm$ 0.04	$^{+0.18}	_{-0.12}$ &	 2.52	 $\pm$ 0.04	$^{+0.16}	_{-0.32}$ &	 3.08	 $\pm$ 0.04	$^{+0.44}	_{-0.13}$ \\ \hline
0.8& --& 1.1 &	  &	  &	 2.54	 $\pm$ 0.04	$^{+0.25}	_{-0.33}$ &	 2.88	 $\pm$ 0.04	$^{+0.32}	_{-0.32}$ \\ \hline
1.1 &-- &1.3 &	  &	  &	  &	 2.79	 $\pm$ 0.04	$^{+0.26}	_{-0.32}$ \\ \hline
\hline
\end{tabular}

\caption{
The charged-particle density, $1/N~dn^{\pm} /d\eta^{\rm Breit}$, for $1280<Q^2<2560$~GeV~$^2$ and 4 bins in $W$.
The first uncertainty is statistical, the second systematic.
}
\label{table-q4}
\end{sidewaystable}

\begin{sidewaystable}
\centering
   \begin{tabular}{|r l r|c|c|c|c|}
\hline
\hline
\multicolumn{3}{|c|}{$\eta^{\rm Breit}$}  &  $120<W<160$~GeV & $160<W<200$~GeV & $200<W<240$~GeV & $240<W<280$~GeV \\ \hline
\hline
-4.0 &-- &-3.7 &	 0.086	 $\pm$ 0.006	$^{+0.014}	_{-0.016}$ &	 0.087	 $\pm$ 0.006	$^{+0.008}	_{-0.014}$ &	 0.073	 $\pm$ 0.005	$^{+0.014}	_{-0.008}$ &	 0.069	 $\pm$ 0.005	$^{+0.022}	_{-0.006}$ \\ \hline
-3.7 &-- &-3.4 &	 0.185	 $\pm$ 0.008	$^{+0.032}	_{-0.018}$ &	 0.159	 $\pm$ 0.007	$^{+0.020}	_{-0.007}$ &	 0.138	 $\pm$ 0.007	$^{+0.011}	_{-0.017}$ &	 0.130	 $\pm$ 0.007	$^{+0.021}	_{-0.031}$ \\ \hline
-3.4 &-- &-3.1 &	 0.301	 $\pm$ 0.010	$^{+0.036}	_{-0.019}$ &	 0.263	 $\pm$ 0.009	$^{+0.038}	_{-0.033}$ &	 0.238	 $\pm$ 0.009	$^{+0.040}	_{-0.008}$ &	 0.224	 $\pm$ 0.009	$^{+0.020}	_{-0.015}$ \\ \hline
-3.1& -- &-2.8 &	 0.476	 $\pm$ 0.013	$^{+0.036}	_{-0.034}$ &	 0.444	 $\pm$ 0.012	$^{+0.059}	_{-0.018}$ &	 0.387	 $\pm$ 0.011	$^{+0.027}	_{-0.026}$ &	 0.364	 $\pm$ 0.012	$^{+0.049}	_{-0.016}$ \\ \hline
-2.8 &-- &-2.5 &	 0.735	 $\pm$ 0.015	$^{+0.040}	_{-0.062}$ &	 0.683	 $\pm$ 0.015	$^{+0.096}	_{-0.029}$ &	 0.617	 $\pm$ 0.014	$^{+0.035}	_{-0.027}$ &	 0.559	 $\pm$ 0.014	$^{+0.038}	_{-0.039}$ \\ \hline
-2.5& -- &-2.2 &	 1.080	 $\pm$ 0.018	$^{+0.086}	_{-0.052}$ &	 0.950	 $\pm$ 0.017	$^{+0.085}	_{-0.048}$ &	 0.881	 $\pm$ 0.017	$^{+0.056}	_{-0.048}$ &	 0.874	 $\pm$ 0.018	$^{+0.097}	_{-0.035}$ \\ \hline
-2.2 &-- &-1.9 &	 1.418	 $\pm$ 0.019	$^{+0.128}	_{-0.058}$ &	 1.283	 $\pm$ 0.019	$^{+0.048}	_{-0.041}$ &	 1.25	 $\pm$ 0.02	$^{+0.10}	_{-0.05}$ &	 1.22	 $\pm$ 0.02	$^{+0.14}	_{-0.04}$ \\ \hline
-1.9& -- &-1.6 &	 1.74	 $\pm$ 0.02	$^{+0.16}	_{-0.06}$ &	 1.62	 $\pm$ 0.02	$^{+0.10}	_{-0.07}$ &	 1.58	 $\pm$ 0.02	$^{+0.19}	_{-0.06}$ &	 1.54	 $\pm$ 0.02	$^{+0.12}	_{-0.05}$ \\ \hline
-1.6 &--& -1.3 &	 2.03	 $\pm$ 0.02	$^{+0.18}	_{-0.09}$ &	 1.92	 $\pm$ 0.02	$^{+0.11}	_{-0.06}$ &	 1.90	 $\pm$ 0.02	$^{+0.16}	_{-0.11}$ &	 1.86	 $\pm$ 0.02	$^{+0.18}	_{-0.06}$ \\ \hline
-1.3& -- &-1.0 &	 2.17	 $\pm$ 0.02	$^{+0.08}	_{-0.07}$ &	 2.15	 $\pm$ 0.02	$^{+0.11}	_{-0.08}$ &	 2.13	 $\pm$ 0.02	$^{+0.19}	_{-0.07}$ &	 2.15	 $\pm$ 0.03	$^{+0.19}	_{-0.06}$ \\ \hline
-1.0 &--& -0.7 &	 2.27	 $\pm$ 0.02	$^{+0.14}	_{-0.09}$ &	 2.29	 $\pm$ 0.02	$^{+0.08}	_{-0.10}$ &	 2.33	 $\pm$ 0.03	$^{+0.17}	_{-0.08}$ &	 2.31	 $\pm$ 0.03	$^{+0.15}	_{-0.13}$ \\ \hline
-0.7& -- &-0.4 &	 2.31	 $\pm$ 0.02	$^{+0.11}	_{-0.08}$ &	 2.43	 $\pm$ 0.02	$^{+0.11}	_{-0.08}$ &	 2.48	 $\pm$ 0.03	$^{+0.17}	_{-0.08}$ &	 2.43	 $\pm$ 0.03	$^{+0.24}	_{-0.11}$ \\ \hline
-0.4 &--& -0.1 &	 2.37	 $\pm$ 0.02	$^{+0.11}	_{-0.09}$ &	 2.48	 $\pm$ 0.02	$^{+0.14}	_{-0.07}$ &	 2.56	 $\pm$ 0.03	$^{+0.24}	_{-0.12}$ &	 2.50	 $\pm$ 0.03	$^{+0.25}	_{-0.14}$ \\ \hline
-0.1& -- &0.2 &	 2.45	 $\pm$ 0.02	$^{+0.21}	_{-0.07}$ &	 2.55	 $\pm$ 0.02	$^{+0.11}	_{-0.07}$ &	 2.59	 $\pm$ 0.03	$^{+0.21}	_{-0.11}$ &	 2.59	 $\pm$ 0.03	$^{+0.22}	_{-0.10}$ \\ \hline
0.2 &--& 0.5 &	 2.35	 $\pm$ 0.02	$^{+0.18}	_{-0.08}$ &	 2.51	 $\pm$ 0.02	$^{+0.14}	_{-0.10}$ &	 2.77	 $\pm$ 0.03	$^{+0.25}	_{-0.13}$ &	 2.78	 $\pm$ 0.03	$^{+0.25}	_{-0.10}$ \\ \hline
0.5& -- &0.8 &	 &	 2.49	 $\pm$ 0.02	$^{+0.10}	_{-0.11}$ &	 2.58	 $\pm$ 0.02	$^{+0.14}	_{-0.11}$ &	 2.73	 $\pm$ 0.03	$^{+0.22}	_{-0.15}$ \\ \hline
0.8 &--& 1.1 &	  &	  &	 2.57	 $\pm$ 0.03	$^{+0.11}	_{-0.14}$ &	 2.63	 $\pm$ 0.03	$^{+0.21}	_{-0.16}$ \\ \hline
1.1& -- &1.4 &	  &	  &	 2.59	 $\pm$ 0.03	$^{+0.15}	_{-0.13}$ &	 2.64	 $\pm$ 0.03	$^{+0.11}	_{-0.14}$ \\ \hline
\hline
\end{tabular}

\caption{
The charged-particle density, $1/N~dn^{\pm} /d\eta^{\rm Breit}$, 
for $640<Q^2<1280$~GeV~$^2$ and 4 bins in $W$.
The first uncertainty is statistical, the second systematic.
}
\label{table-q3}
\end{sidewaystable}

\begin{table}
\centering
   \begin{tabular}{|r l r|c|c|c|}
\hline
\hline
\multicolumn{3}{|c|}{$\eta^{\rm Breit}$}  &  $120<W<160$~GeV  & $160<W<200$~GeV & $200<W<240$~GeV\\ \hline
\hline
-3.7& --& -3.4 &	 0.094	 $\pm$ 0.003	$^{+0.013}	_{-0.011}$ &	 0.083	 $\pm$ 0.003	$^{+0.008}	_{-0.008}$ &	 0.082	 $\pm$ 0.004	$^{+0.010}	_{-0.013}$ \\ \hline
-3.4& -- &-3.1 &	 0.168	 $\pm$ 0.005	$^{+0.009}	_{-0.017}$ &	 0.145	 $\pm$ 0.004	$^{+0.013}	_{-0.012}$ &	 0.134	 $\pm$ 0.004	$^{+0.009}	_{-0.012}$ \\ \hline
-3.1 &-- &-2.8 &	 0.285	 $\pm$ 0.006	$^{+0.019}	_{-0.018}$ &	 0.246	 $\pm$ 0.006	$^{+0.013}	_{-0.023}$ &	 0.241	 $\pm$ 0.006	$^{+0.014}	_{-0.013}$ \\ \hline
-2.8& -- &-2.5 &	 0.451	 $\pm$ 0.007	$^{+0.016}	_{-0.032}$ &	 0.404	 $\pm$ 0.007	$^{+0.015}	_{-0.019}$ &	 0.395	 $\pm$ 0.007	$^{+0.031}	_{-0.028}$ \\ \hline
-2.5& -- &-2.2 &	 0.714	 $\pm$ 0.009	$^{+0.039}	_{-0.033}$ &	 0.646	 $\pm$ 0.009	$^{+0.037}	_{-0.025}$ &	 0.621	 $\pm$ 0.009	$^{+0.024}	_{-0.029}$ \\ \hline
-2.2& -- &-1.9 &	 1.022	 $\pm$ 0.010	$^{+0.049}	_{-0.042}$ &	 0.969	 $\pm$ 0.011	$^{+0.052}	_{-0.040}$ &	 0.915	 $\pm$ 0.011	$^{+0.060}	_{-0.033}$ \\ \hline
-1.9 &-- &-1.6 &	 1.384	 $\pm$ 0.012	$^{+0.084}	_{-0.049}$ &	 1.335	 $\pm$ 0.012	$^{+0.058}	_{-0.045}$ &	 1.253	 $\pm$ 0.013	$^{+0.065}	_{-0.051}$ \\ \hline
-1.6& -- &-1.3 &	 1.702	 $\pm$ 0.013	$^{+0.133}	_{-0.069}$ &	 1.652	 $\pm$ 0.013	$^{+0.099}	_{-0.077}$ &	 1.618	 $\pm$ 0.014	$^{+0.103}	_{-0.057}$ \\ \hline
-1.3 &--& -1.0 &	 1.998	 $\pm$ 0.013	$^{+0.133}	_{-0.051}$ &	 1.936	 $\pm$ 0.014	$^{+0.105}	_{-0.067}$ &	 1.903	 $\pm$ 0.015	$^{+0.102}	_{-0.052}$ \\ \hline
-1.0& -- &-0.7 &	 2.192	 $\pm$ 0.013	$^{+0.140}	_{-0.050}$ &	 2.166	 $\pm$ 0.015	$^{+0.121}	_{-0.066}$ &	 2.129	 $\pm$ 0.016	$^{+0.165}	_{-0.064}$ \\ \hline
-0.7 &-- &-0.4 &	 2.305	 $\pm$ 0.013	$^{+0.158}	_{-0.054}$ &	 2.350	 $\pm$ 0.015	$^{+0.126}	_{-0.074}$ &	 2.323	 $\pm$ 0.017	$^{+0.095}	_{-0.068}$ \\ \hline
-0.4& -- &-0.1 &	 2.356	 $\pm$ 0.013	$^{+0.203}	_{-0.061}$ &	 2.434	 $\pm$ 0.016	$^{+0.101}	_{-0.071}$ &	 2.422	 $\pm$ 0.018	$^{+0.136}	_{-0.069}$ \\ \hline
-0.1 &--& 0.2 &	 2.447	 $\pm$ 0.013	$^{+0.241}	_{-0.065}$ &	 2.536	 $\pm$ 0.015	$^{+0.160}	_{-0.068}$ &	 2.522	 $\pm$ 0.017	$^{+0.175}	_{-0.073}$ \\ \hline
0.2& -- &0.5 &	 2.364	 $\pm$ 0.013	$^{+0.169}	_{-0.071}$ &	 2.549	 $\pm$ 0.015	$^{+0.124}	_{-0.095}$ &	 2.680	 $\pm$ 0.017	$^{+0.230}	_{-0.092}$ \\ \hline
0.5 &-- &0.8 &	 2.332	 $\pm$ 0.014	$^{+0.139}	_{-0.069}$ &	 2.505	 $\pm$ 0.015	$^{+0.093}	_{-0.080}$ &	 2.617	 $\pm$ 0.017	$^{+0.120}	_{-0.102}$ \\ \hline
0.8& -- &1.1 &	  &	 2.468	 $\pm$ 0.015	$^{+0.097}	_{-0.081}$ &	 2.591	 $\pm$ 0.016	$^{+0.081}	_{-0.080}$ \\ \hline
1.1& -- &1.4 &	  &	 2.500	 $\pm$ 0.016	$^{+0.101}	_{-0.091}$ &	 2.569	 $\pm$ 0.017	$^{+0.071}	_{-0.105}$ \\ \hline
\hline
\end{tabular}

\caption{
The charged-particle density, $1/N~dn^{\pm} /d\eta^{\rm Breit}$, 
for $320<Q^2<640$~GeV~$^2$ and 3 bins in $W$.
The first uncertainty is statistical, the second systematic.
}
\label{table-q2}
\end{table}

\begin{sidewaystable}[p]
\centering
   \begin{tabular}{|r l r|c|c|c|c|}
\hline
\hline
\multicolumn{3}{|c|}{$\eta^{\rm Breit}$}  &  $80<W<120$~GeV & $120<W<160$~GeV & $160<W<200$~GeV & $200<W=240$~GeV \\ \hline
\hline
-3.4& --& -3.1 &	 0.107	 $\pm$ 0.002	$^{+0.007}	_{-0.007}$ &	 0.092	 $\pm$ 0.002	$^{+0.004}	_{-0.005}$ &	 0.082	 $\pm$ 0.002	$^{+0.003}	_{-0.005}$ &	 0.079	 $\pm$ 0.002	$^{+0.005}	_{-0.005}$ \\ \hline
-3.1& -- &-2.8 &	 0.186	 $\pm$ 0.003	$^{+0.006}	_{-0.010}$ &	 0.155	 $\pm$ 0.002	$^{+0.004}	_{-0.015}$ &	 0.144	 $\pm$ 0.003	$^{+0.008}	_{-0.012}$ &	 0.136	 $\pm$ 0.003	$^{+0.004}	_{-0.012}$ \\ \hline
-2.8 &-- &-2.5 &	 0.304	 $\pm$ 0.004	$^{+0.010}	_{-0.011}$ &	 0.270	 $\pm$ 0.003	$^{+0.012}	_{-0.016}$ &	 0.248	 $\pm$ 0.003	$^{+0.011}	_{-0.015}$ &	 0.233	 $\pm$ 0.004	$^{+0.012}	_{-0.010}$ \\ \hline
-2.5& --& -2.2 &	 0.512	 $\pm$ 0.005	$^{+0.017}	_{-0.020}$ &	 0.436	 $\pm$ 0.004	$^{+0.015}	_{-0.018}$ &	 0.411	 $\pm$ 0.004	$^{+0.011}	_{-0.017}$ &	 0.386	 $\pm$ 0.005	$^{+0.016}	_{-0.031}$ \\ \hline
-2.2 &-- &-1.9 &	 0.790	 $\pm$ 0.006	$^{+0.023}	_{-0.030}$ &	 0.695	 $\pm$ 0.005	$^{+0.018}	_{-0.022}$ &	 0.646	 $\pm$ 0.005	$^{+0.020}	_{-0.020}$ &	 0.609	 $\pm$ 0.006	$^{+0.020}	_{-0.037}$ \\ \hline
-1.9& -- &-1.6 &	 1.119	 $\pm$ 0.006	$^{+0.045}	_{-0.042}$ &	 0.992	 $\pm$ 0.006	$^{+0.033}	_{-0.020}$ &	 0.959	 $\pm$ 0.006	$^{+0.034}	_{-0.032}$ &	 0.902	 $\pm$ 0.007	$^{+0.035}	_{-0.038}$ \\ \hline
-1.6 &-- &-1.3 &	 1.479	 $\pm$ 0.007	$^{+0.108}	_{-0.058}$ &	 1.349	 $\pm$ 0.007	$^{+0.080}	_{-0.040}$ &	 1.288	 $\pm$ 0.007	$^{+0.049}	_{-0.034}$ &	 1.242	 $\pm$ 0.008	$^{+0.072}	_{-0.040}$ \\ \hline
-1.3& -- &-1.0 &	 1.783	 $\pm$ 0.007	$^{+0.173}	_{-0.056}$ &	 1.679	 $\pm$ 0.007	$^{+0.143}	_{-0.041}$ &	 1.619	 $\pm$ 0.008	$^{+0.078}	_{-0.048}$ &	 1.568	 $\pm$ 0.009	$^{+0.082}	_{-0.057}$ \\ \hline
-1.0 &-- &-0.7 &	 1.989	 $\pm$ 0.007	$^{+0.167}	_{-0.054}$ &	 1.947	 $\pm$ 0.008	$^{+0.132}	_{-0.052}$ &	 1.918	 $\pm$ 0.009	$^{+0.104}	_{-0.056}$ &	 1.880	 $\pm$ 0.010	$^{+0.165}	_{-0.079}$ \\ \hline
-0.7& -- &-0.4 &	 2.126	 $\pm$ 0.007	$^{+0.197}	_{-0.059}$ &	 2.140	 $\pm$ 0.008	$^{+0.164}	_{-0.061}$ &	 2.134	 $\pm$ 0.009	$^{+0.145}	_{-0.060}$ &	 2.120	 $\pm$ 0.011	$^{+0.159}	_{-0.073}$ \\ \hline
-0.4 &--& -0.1 &	 2.233	 $\pm$ 0.007	$^{+0.223}	_{-0.066}$ &	 2.277	 $\pm$ 0.008	$^{+0.143}	_{-0.060}$ &	 2.294	 $\pm$ 0.010	$^{+0.183}	_{-0.065}$ &	 2.288	 $\pm$ 0.011	$^{+0.165}	_{-0.061}$ \\ \hline
-0.1& -- &0.2 &	 2.309	 $\pm$ 0.007	$^{+0.242}	_{-0.066}$ &	 2.421	 $\pm$ 0.008	$^{+0.211}	_{-0.067}$ &	 2.417	 $\pm$ 0.010	$^{+0.173}	_{-0.052}$ &	 2.424	 $\pm$ 0.011	$^{+0.204}	_{-0.070}$ \\ \hline
0.2 &-- &0.5 &	 2.190	 $\pm$ 0.008	$^{+0.127}	_{-0.067}$ &	 2.376	 $\pm$ 0.008	$^{+0.140}	_{-0.056}$ &	 2.513	 $\pm$ 0.010	$^{+0.151}	_{-0.064}$ &	 2.617	 $\pm$ 0.011	$^{+0.258}	_{-0.072}$ \\ \hline
0.5& -- &0.8 &	  &	 2.354	 $\pm$ 0.008	$^{+0.098}	_{-0.068}$ &	 2.485	 $\pm$ 0.009	$^{+0.105}	_{-0.069}$ &	 2.587	 $\pm$ 0.011	$^{+0.151}	_{-0.066}$ \\ \hline
0.8& --& 1.1 &	  &	 2.377	 $\pm$ 0.009	$^{+0.135}	_{-0.070}$ &	 2.501	 $\pm$ 0.009	$^{+0.075}	_{-0.056}$ &	 2.569	 $\pm$ 0.011	$^{+0.114}	_{-0.074}$ \\ \hline
1.1 &-- &1.4 &	  &	  &	 2.500	 $\pm$ 0.010	$^{+0.070}	_{-0.070}$ &	 2.587	 $\pm$ 0.011	$^{+0.094}	_{-0.075}$ \\ \hline
\hline
\end{tabular}

\caption{
The charged-particle density, $1/N~dn^{\pm} /d\eta^{\rm Breit}$,  
for $160<Q^2<320$~GeV~$^2$ and 4 bins in $W$.
The first uncertainty is statistical, the second systematic.
}
\label{table-q1}
\end{sidewaystable}

%-------------------------------------------------------------------------------
%       eta_Breit as F(Q2,W)  for Q2<160 GeV2
%-------------------------------------------------------------------------------

\begin{table}
\centering
   \begin{tabular}{|r l r|c|c|c|}
\hline
\hline
\multicolumn{3}{|c|}{$\eta^{\rm Breit}$}  &  $80<W<120$~GeV  & $120<W<160$~GeV & $160<W<200$~GeV\\ \hline
\hline
-4.0& --& -3.7 &	 0.034	 $\pm$ 0.002$^{+0.001 }_{-0.002 }$ &	 0.033	 $\pm$ 0.002$^{+0.002  }_{-0.002  }$ &	 0.027	 $\pm$ 0.002$^{+0.002  }_{-0.001  }$ \\ \hline
-3.7& --& -3.4 &	 0.057	 $\pm$ 0.002$^{+0.003 }_{-0.001  }$ &	 0.053	 $\pm$ 0.003$^{+0.002  }_{-0.003 }$ &	 0.049	 $\pm$ 0.003$^{+0.002  }_{-0.002 }$ \\ \hline
-3.4& -- &-3.1 &	 0.093	 $\pm$ 0.003$^{+0.003 }_{-0.001  }$ &	 0.086	 $\pm$ 0.003$^{+0.002 }_{-0.001  }$ &	 0.084	 $\pm$ 0.004$^{+0.004 }_{-0.004 }$ \\ \hline
-3.1& -- &-2.8 &	 0.158	 $\pm$ 0.004$^{+0.005 }_{-0.004 }$ &	 0.141	 $\pm$ 0.004$^{+0.007 }_{-0.002 }$ &	 0.138	 $\pm$ 0.004$^{+0.004 }_{-0.004 }$ \\ \hline
-2.8 &-- &-2.5 &	 0.254	 $\pm$ 0.005$^{+0.006 }_{-0.005 }$ &	 0.248	 $\pm$ 0.005$^{+0.006 }_{-0.004 }$ &	 0.214	 $\pm$ 0.005$^{+0.004 }_{-0.004 }$ \\ \hline
-2.5& -- &-2.2 &	 0.413	 $\pm$ 0.006$^{+0.012 }_{-0.008 }$ &	 0.361	 $\pm$ 0.006$^{+0.011 }_{-0.005 }$ &	 0.339	 $\pm$ 0.006$^{+0.007 }_{-0.006 }$ \\ \hline
-2.2 &-- &-1.9 &	 0.600	 $\pm$ 0.007$^{+0.019 }_{-0.010 }$ &	 0.542	 $\pm$ 0.007$^{+0.012 }_{-0.008 }$ &	 0.530	 $\pm$ 0.008$^{+0.007 }_{-0.011 }$ \\ \hline
-1.9& -- &-1.6 &	 0.856	 $\pm$ 0.008$^{+0.023 }_{-0.013 }$ &	 0.796	 $\pm$ 0.009$^{+0.012 }_{-0.008 }$ &	 0.773	 $\pm$ 0.009$^{+0.009 }_{-0.012 }$ \\ \hline
-1.6 &-- &-1.3 &	 1.140	 $\pm$ 0.009$^{+0.025 }_{-0.010 }$ &	 1.084	 $\pm$ 0.010$^{+0.008 }_{-0.011 }$ &	 1.034	 $\pm$ 0.011$^{+0.016 }_{-0.011 }$ \\ \hline
-1.3& -- &-1.0 &	 1.424	 $\pm$ 0.010$^{+0.032 }_{-0.011 }$ &	 1.383	 $\pm$ 0.011$^{+0.016 }_{-0.013 }$ &	 1.324	 $\pm$ 0.012$^{+0.020 }_{-0.014 }$ \\ \hline
-1.0& -- &-0.7 &	 1.682	 $\pm$ 0.011$^{+0.027 }_{-0.015 }$ &	 1.634	 $\pm$ 0.012$^{+0.025 }_{-0.009 }$ &	 1.600	 $\pm$ 0.013$^{+0.022 }_{-0.015 }$ \\ \hline
-0.7& -- &-0.4 &	 1.855	 $\pm$ 0.011$^{+0.032 }_{-0.017 }$ &	 1.833	 $\pm$ 0.012$^{+0.031 }_{-0.013 }$ &	 1.827	 $\pm$ 0.014$^{+0.045 }_{-0.015 }$ \\ \hline
-0.4 &--& -0.1 &	 1.979	 $\pm$ 0.012$^{+0.063 }_{-0.021 }$ &	 1.986	 $\pm$ 0.013$^{+0.065 }_{-0.025 }$ &	 2.039	 $\pm$ 0.014$^{+0.042 }_{-0.012 }$ \\ \hline
-0.1& -- &0.2 &	 2.060	 $\pm$ 0.012$^{+0.075 }_{-0.020 }$ &	 2.134	 $\pm$ 0.013$^{+0.069 }_{-0.022 }$ &	 2.142	 $\pm$ 0.014$^{+0.081 }_{-0.020 }$ \\ \hline
0.2 &--& 0.5 &	 2.051	 $\pm$ 0.012$^{+0.093 }_{-0.044 }$ &	 2.201	 $\pm$ 0.013$^{+0.093 }_{-0.031 }$ &	 2.301	 $\pm$ 0.014$^{+0.092 }_{-0.032 }$ \\ \hline
0.5& -- &0.8 &	 2.082	 $\pm$ 0.012$^{+0.100 }_{-0.025 }$ &	 2.263	 $\pm$ 0.013$^{+0.107 }_{-0.017 }$ &	 2.327	 $\pm$ 0.014$^{+0.116 }_{-0.033 }$ \\ \hline
0.8 &-- &1.1 &	  &	 2.294	 $\pm$ 0.013$^{+0.111 }_{-0.038 }$ &	 2.395	 $\pm$ 0.014$^{+0.139 }_{-0.030 }$ \\ \hline
1.1& -- &1.4 &	  &	 2.302	 $\pm$ 0.013$^{+0.104 }_{-0.025 }$ &	 2.440	 $\pm$ 0.015$^{+0.126 }_{-0.033 }$ \\ \hline
1.4 &-- &1.7 &	  &	 2.287	 $\pm$ 0.014$^{+0.092 }_{-0.026 }$ &	 2.497	 $\pm$ 0.015$^{+0.139 }_{-0.018 }$ \\ \hline
\hline
\end{tabular}

\caption{
The charged-particle density, $1/N~dn^{\pm} /d\eta^{\rm Breit}$,  
for $60<Q^2<160$~GeV~$^2$ and 3 bins in $W$.
The first uncertainty is statistical, the second systematic.
}
\label{tab-q6}
\end{table}

\begin{table}
\centering
   \begin{tabular}{|r l r|c|c|c|}
\hline
\hline
\multicolumn{3}{|c|}{$\eta^{\rm Breit}$}  &  $80<W<120$~GeV  & $120<W<160$~GeV & $160<W<200$~GeV\\ \hline
\hline
-3.1& -- &-2.8 &	 0.070	 $\pm$ 0.005$^{+0.010 }_{-0.008 }$ &	 0.051	 $\pm$ 0.004$^{+0.008 }_{-0.003 }$ &	 0.057	 $\pm$ 0.005$^{+0.009 }_{-0.005 }$ \\ \hline
-2.8& -- &-2.5 &	 0.105	 $\pm$ 0.005$^{+0.005 }_{-0.005 }$ &	 0.096	 $\pm$ 0.006$^{+0.013 }_{-0.006 }$ &	 0.073	 $\pm$ 0.005$^{+0.005 }_{-0.004 }$ \\ \hline
-2.5 &-- &-2.2 &	 0.177	 $\pm$ 0.007$^{+0.014 }_{-0.007 }$ &	 0.157	 $\pm$ 0.007$^{+0.009 }_{-0.009 }$ &	 0.139	 $\pm$ 0.008$^{+0.010 }_{-0.007 }$ \\ \hline
-2.2& -- &-1.9 &	 0.298	 $\pm$ 0.009$^{+0.015 }_{-0.012 }$ &	 0.271	 $\pm$ 0.010$^{+0.016 }_{-0.010 }$ &	 0.250	 $\pm$ 0.010$^{+0.011 }_{-0.014 }$ \\ \hline
-1.9 &-- &-1.6 &	 0.469	 $\pm$ 0.011$^{+0.016 }_{-0.015 }$ &	 0.436	 $\pm$ 0.012$^{+0.012 }_{-0.013 }$ &	 0.391	 $\pm$ 0.012$^{+0.024 }_{-0.015 }$ \\ \hline
-1.6& -- &-1.3 &	 0.706	 $\pm$ 0.014$^{+0.028 }_{-0.009 }$ &	 0.654	 $\pm$ 0.014$^{+0.021 }_{-0.018 }$ &	 0.611	 $\pm$ 0.016$^{+0.036 }_{-0.019 }$ \\ \hline
-1.3 &-- &-1.0 &	 0.999	 $\pm$ 0.016$^{+0.021 }_{-0.026 }$ &	 0.951	 $\pm$ 0.017$^{+0.021 }_{-0.019 }$ &	 0.809	 $\pm$ 0.017$^{+0.062 }_{-0.016 }$ \\ \hline
-1.0& -- &-0.7 &	 1.293	 $\pm$ 0.018$^{+0.035 }_{-0.020 }$ &	 1.233	 $\pm$ 0.019$^{+0.023 }_{-0.027 }$ &	 1.17	 $\pm$ 0.02$^{+0.05 }_{-0.03 }$ \\ \hline
-0.7 &-- &-0.4 &	 1.534	 $\pm$ 0.019$^{+0.036 }_{-0.039 }$ &	 1.51	 $\pm$ 0.02$^{+0.03 }_{-0.02 }$ &	 1.44	 $\pm$ 0.02$^{+0.06 }_{-0.02 }$ \\ \hline
-0.4& -- &-0.1 &	 1.80	 $\pm$ 0.02$^{+0.05 }_{-0.02  }$ &	 1.75	 $\pm$ 0.02$^{+0.05 }_{-0.02 }$ &	 1.68	 $\pm$ 0.03$^{+0.09 }_{-0.03 }$ \\ \hline
-0.1 &--& 0.2 &	 1.97	 $\pm$ 0.02$^{+0.06 }_{-0.02 }$ &	 1.96	 $\pm$ 0.02$^{+0.05 }_{-0.02 }$ &	 1.93	 $\pm$ 0.03$^{+0.10 }_{-0.03 }$ \\ \hline
0.2& -- &0.5 &	 2.06	 $\pm$ 0.02$^{+0.08 }_{-0.05 }$ &	 2.07	 $\pm$ 0.02$^{+0.09 }_{-0.03 }$ &	 2.19	 $\pm$ 0.03$^{+0.08 }_{-0.05 }$ \\ \hline
0.5 &-- &0.8 &	 2.11	 $\pm$ 0.02$^{+0.10 }_{-0.04 }$ &	 2.17	 $\pm$ 0.02$^{+0.11 }_{-0.02 }$ &	 2.24	 $\pm$ 0.03$^{+0.12 }_{-0.03 }$ \\ \hline
0.8& -- &1.1 &	 2.17	 $\pm$ 0.02$^{+0.10 }_{-0.04 }$ &	 2.28	 $\pm$ 0.02$^{+0.13 }_{-0.03 }$ &	 2.30	 $\pm$ 0.03$^{+0.11 }_{-0.06 }$ \\ \hline
1.1 &-- &1.4 &	 2.24	 $\pm$ 0.03$^{+0.08 }_{-0.03}$ &	 2.33	 $\pm$ 0.03$^{+0.10 }_{-0.04 }$ &	 2.48	 $\pm$ 0.03$^{+0.12 }_{-0.06 }$ \\ \hline
1.4& -- &1.7 &	 2.18	 $\pm$ 0.03$^{+0.10 }_{-0.03 }$ &	 2.36	 $\pm$ 0.03$^{+0.10 }_{-0.05 }$ &	 2.56	 $\pm$ 0.03$^{+0.16 }_{-0.07 }$ \\ \hline
\hline
\end{tabular}

\caption{
The charged-particle density, $1/N~dn^{\pm} /d\eta^{\rm Breit}$,   
for $50<Q^2<60$~GeV~$^2$ and 3 bins in $W$.
The first uncertainty is statistical, the second systematic.
}
\label{tab-q5}
\end{table}

\begin{table}
\centering
   \begin{tabular}{|r l r|c|c|c|}
\hline
\hline
\multicolumn{3}{|c|}{$\eta^{\rm Breit}$}  &  $80<W<120$~GeV  & $120<W<160$~GeV & $160<W<200$~GeV\\ \hline
\hline
-3.1& --& -2.8 &	 0.049	 $\pm$ 0.003$^{+0.005 }_{-0.004 }$ &	 0.044	 $\pm$ 0.003$^{+0.004 }_{-0.004 }$ &	 0.039	 $\pm$ 0.003$^{+0.004 }_{-0.005 }$ \\ \hline
-2.8& -- &-2.5 &	 0.083	 $\pm$ 0.004$^{+0.008 }_{-0.003 }$ &	 0.089	 $\pm$ 0.005$^{+0.007 }_{-0.010 }$ &	 0.064	 $\pm$ 0.004$^{+0.008 }_{-0.004 }$ \\ \hline
-2.5 &-- &-2.2 &	 0.149	 $\pm$ 0.005$^{+0.010 }_{-0.006 }$ &	 0.136	 $\pm$ 0.006$^{+0.009 }_{-0.005 }$ &	 0.118	 $\pm$ 0.006$^{+0.007 }_{-0.006 }$ \\ \hline
-2.2& -- &-1.9 &	 0.242	 $\pm$ 0.006$^{+0.008 }_{-0.004 }$ &	 0.242	 $\pm$ 0.008$^{+0.009 }_{-0.012 }$ &	 0.194	 $\pm$ 0.007$^{+0.013 }_{-0.007 }$ \\ \hline
-1.9& -- &-1.6 &	 0.405	 $\pm$ 0.008$^{+0.017 }_{-0.010 }$ &	 0.384	 $\pm$ 0.009$^{+0.017 }_{-0.011 }$ &	 0.330	 $\pm$ 0.009$^{+0.014 }_{-0.008 }$ \\ \hline
-1.6& -- &-1.3 &	 0.607	 $\pm$ 0.010$^{+0.021 }_{-0.008 }$ &	 0.576	 $\pm$ 0.011$^{+0.006 }_{-0.018 }$ &	 0.495	 $\pm$ 0.011$^{+0.021 }_{-0.008 }$ \\ \hline
-1.3 &-- &-1.0 &	 0.912	 $\pm$ 0.013$^{+0.017 }_{-0.013 }$ &	 0.815	 $\pm$ 0.013$^{+0.025 }_{-0.014 }$ &	 0.766	 $\pm$ 0.014$^{+0.024 }_{-0.014 }$ \\ \hline
-1.0& -- &-0.7 &	 1.176	 $\pm$ 0.014$^{+0.029 }_{-0.007 }$ &	 1.121	 $\pm$ 0.015$^{+0.018 }_{-0.022 }$ &	 1.029	 $\pm$ 0.016$^{+0.031 }_{-0.011 }$ \\ \hline
-0.7 &-- &-0.4 &	 1.461	 $\pm$ 0.015$^{+0.036 }_{-0.014 }$ &	 1.416	 $\pm$ 0.017$^{+0.016 }_{-0.037 }$ &	 1.39	 $\pm$ 0.02$^{+0.03 }_{-0.02 }$ \\ \hline
-0.4& -- &-0.1 &	 1.710	 $\pm$ 0.017$^{+0.050 }_{-0.017 }$ &	 1.649	 $\pm$ 0.018$^{+0.035 }_{-0.017 }$ &	 1.67	 $\pm$ 0.02$^{+0.06 }_{-0.01  }$ \\ \hline
-0.1 &--& 0.2 &	 1.901	 $\pm$ 0.018$^{+0.057 }_{-0.026 }$ &	 1.883	 $\pm$ 0.019$^{+0.056 }_{-0.026 }$ &	 1.88	 $\pm$ 0.02$^{+0.07 }_{-0.03 }$ \\ \hline
0.2& -- &0.5 &	 2.009	 $\pm$ 0.018$^{+0.067 }_{-0.038 }$ &	 2.044	 $\pm$ 0.019$^{+0.079 }_{-0.028 }$ &	 2.08	 $\pm$ 0.02$^{+0.08 }_{-0.03 }$ \\ \hline
0.5 &--& 0.8 &	 2.106	 $\pm$ 0.018$^{+0.095 }_{-0.051 }$ &	 2.14	 $\pm$ 0.02$^{+0.08 }_{-0.04 }$ &	 2.25	 $\pm$ 0.02$^{+0.10 }_{-0.04 }$ \\ \hline
0.8& --& 1.1 &	 2.187	 $\pm$ 0.018$^{+0.092 }_{-0.032 }$ &	 2.26	 $\pm$ 0.02$^{+0.11 }_{-0.03 }$ &	 2.30	 $\pm$ 0.02$^{+0.12 }_{-0.04 }$ \\ \hline
1.1 &-- &1.4 &	 2.24	 $\pm$ 0.02$^{+0.08 }_{-0.04 }$ &	 2.33	 $\pm$ 0.02$^{+0.12 }_{-0.02 }$ &	 2.39	 $\pm$ 0.02$^{+0.16 }_{-0.04 }$ \\ \hline
1.4 &-- &1.7 &	 2.21	 $\pm$ 0.02$^{+0.08 }_{-0.03 }$ &	 2.40	 $\pm$ 0.02$^{+0.09 }_{-0.04 }$ &	 2.46	 $\pm$ 0.02$^{+0.15 }_{-0.05 }$ \\ \hline
\hline
\end{tabular}

\caption{
The charged-particle density, $1/N~dn^{\pm} /d\eta^{\rm Breit}$,  
for $40<Q^2<50$~GeV~$^2$ and 3 bins in $W$.
The first uncertainty is statistical, the second systematic.
}
\label{tab-q4}
\end{table}

\begin{table}
\centering
   \begin{tabular}{|r l r|c|c|c|}
\hline
\hline
\multicolumn{3}{|c|}{$\eta^{\rm Breit}$}  &  $80<W<120$~GeV  & $120<W<160$~GeV & $160<W<200$~GeV\\ \hline
\hline
-3.1& -- &-2.8 &	 0.036	 $\pm$ 0.002$^{+0.003 }_{-0.002 }$ &	 0.035	 $\pm$ 0.002$^{+0.004 }_{-0.002 }$ &	 0.029	 $\pm$ 0.002$^{+0.004 }_{-0.003 }$ \\ \hline
-2.8& -- &-2.5 &	 0.066	 $\pm$ 0.003$^{+0.003 }_{-0.003 }$ &	 0.055	 $\pm$ 0.003$^{+0.005 }_{-0.002 }$ &	 0.056	 $\pm$ 0.003$^{+0.004 }_{-0.004 }$ \\ \hline
-2.5 &-- &-2.2 &	 0.122	 $\pm$ 0.004$^{+0.003 }_{-0.004 }$ &	 0.108	 $\pm$ 0.004$^{+0.007 }_{-0.004 }$ &	 0.090	 $\pm$ 0.004$^{+0.008 }_{-0.003 }$ \\ \hline
-2.2& -- &-1.9 &	 0.196	 $\pm$ 0.005$^{+0.007 }_{-0.004 }$ &	 0.185	 $\pm$ 0.005$^{+0.006 }_{-0.007 }$ &	 0.160	 $\pm$ 0.005$^{+0.008 }_{-0.006 }$ \\ \hline
-1.9 &-- &-1.6 &	 0.344	 $\pm$ 0.006$^{+0.008 }_{-0.008 }$ &	 0.300	 $\pm$ 0.006$^{+0.006 }_{-0.004 }$ &	 0.260	 $\pm$ 0.006$^{+0.011 }_{-0.006 }$ \\ \hline
-1.6& -- &-1.3 &	 0.529	 $\pm$ 0.007$^{+0.011 }_{-0.008 }$ &	 0.488	 $\pm$ 0.008$^{+0.010 }_{-0.011 }$ &	 0.431	 $\pm$ 0.008$^{+0.024 }_{-0.009 }$ \\ \hline
-1.3 &-- &-1.0 &	 0.767	 $\pm$ 0.009$^{+0.006 }_{-0.010 }$ &	 0.696	 $\pm$ 0.009$^{+0.016 }_{-0.009 }$ &	 0.653	 $\pm$ 0.010$^{+0.020 }_{-0.010 }$ \\ \hline
-1.0 &-- &-0.7 &	 1.050	 $\pm$ 0.010$^{+0.018 }_{-0.011 }$ &	 0.964	 $\pm$ 0.011$^{+0.014 }_{-0.008 }$ &	 0.913	 $\pm$ 0.012$^{+0.045 }_{-0.016 }$ \\ \hline
-0.7 &-- &-0.4 &	 1.342	 $\pm$ 0.011$^{+0.031 }_{-0.008 }$ &	 1.260	 $\pm$ 0.012$^{+0.019 }_{-0.011 }$ &	 1.228	 $\pm$ 0.014$^{+0.045 }_{-0.014 }$ \\ \hline
-0.4 &-- &-0.1 &	 1.597	 $\pm$ 0.012$^{+0.032 }_{-0.017 }$ &	 1.548	 $\pm$ 0.014$^{+0.024 }_{-0.022 }$ &	 1.508	 $\pm$ 0.016$^{+0.052 }_{-0.012 }$ \\ \hline
-0.1& --& 0.2 &	 1.816	 $\pm$ 0.013$^{+0.047 }_{-0.023 }$ &	 1.795	 $\pm$ 0.014$^{+0.053 }_{-0.026 }$ &	 1.757	 $\pm$ 0.016$^{+0.067 }_{-0.013 }$ \\ \hline
0.2& -- &0.5 &	 1.973	 $\pm$ 0.014$^{+0.055 }_{-0.019 }$ &	 1.980	 $\pm$ 0.015$^{+0.086 }_{-0.014 }$ &	 2.035	 $\pm$ 0.017$^{+0.076 }_{-0.015 }$ \\ \hline
0.5 &-- &0.8 &	 2.089	 $\pm$ 0.014$^{+0.091 }_{-0.024 }$ &	 2.133	 $\pm$ 0.015$^{+0.087 }_{-0.026 }$ &	 2.168	 $\pm$ 0.017$^{+0.097 }_{-0.026 }$ \\ \hline
0.8& -- &1.1 &	 2.200	 $\pm$ 0.014$^{+0.078 }_{-0.035 }$ &	 2.229	 $\pm$ 0.015$^{+0.126 }_{-0.019 }$ &	 2.303	 $\pm$ 0.018$^{+0.117 }_{-0.032 }$ \\ \hline
1.1 &-- &1.4 &	 2.268	 $\pm$ 0.015$^{+0.079 }_{-0.031 }$ &	 2.328	 $\pm$ 0.016$^{+0.129 }_{-0.012 }$ &	 2.367	 $\pm$ 0.018$^{+0.145 }_{-0.051 }$ \\ \hline
1.4& -- &1.7 &	 2.285	 $\pm$ 0.017$^{+0.069 }_{-0.031 }$ &	 2.379	 $\pm$ 0.016$^{+0.126 }_{-0.025 }$ &	 2.473	 $\pm$ 0.019$^{+0.140 }_{-0.032 }$ \\ \hline
\hline
\end{tabular}

\caption{
The charged-particle density, $1/N~dn^{\pm} /d\eta^{\rm Breit}$, 
for $30<Q^2<40$~GeV~$^2$ and 3 bins in $W$.
The first uncertainty is statistical, the second systematic.
}
\label{tab-q3}
\end{table}

\begin{table}
\centering
   \begin{tabular}{|r l r|c|c|}
\hline
\hline
\multicolumn{3}{|c|}{$\eta^{\rm Breit}$}  &  $80<W<120$~GeV  & $120<W<160$~GeV \\ \hline
\hline
-3.1& -- &-2.8 &	 0.0268	 $\pm$ 0.0012$^{+0.0016 }_{-0.0018 }$ &	 0.0233	 $\pm$ 0.0012$^{+0.0017 }_{-0.0012 }$  \\ \hline
-2.8& -- &-2.5 &	 0.0475	 $\pm$ 0.0016$^{+0.0031 }_{-0.0012 }$ &	 0.0470	 $\pm$ 0.0019$^{+0.0022 }_{-0.0020 }$  \\ \hline
-2.5 &-- &-2.2 &	 0.091	 $\pm$ 0.002$^{+0.003 }_{-0.003 }$ &	 0.075	 $\pm$ 0.002$^{+0.003 }_{-0.002  }$  \\ \hline
-2.2& -- &-1.9 &	 0.155	 $\pm$ 0.003$^{+0.004 }_{-0.004 }$ &	 0.130	 $\pm$ 0.003$^{+0.005 }_{-0.003 }$  \\ \hline
-1.9 &-- &-1.6 &	 0.256	 $\pm$ 0.004$^{+0.006 }_{-0.005 }$ &	 0.220	 $\pm$ 0.004$^{+0.006 }_{-0.003 }$  \\ \hline
-1.6& -- &-1.3 &	 0.403	 $\pm$ 0.005$^{+0.007 }_{-0.009 }$ &	 0.364	 $\pm$ 0.005$^{+0.010 }_{-0.004 }$  \\ \hline
-1.3 &-- &-1.0 &	 0.620	 $\pm$ 0.006$^{+0.007 }_{-0.007 }$ &	 0.543	 $\pm$ 0.006$^{+0.012 }_{-0.004 }$  \\ \hline
-1.0& -- &-0.7 &	 0.863	 $\pm$ 0.007$^{+0.008 }_{-0.012 }$ &	 0.783	 $\pm$ 0.007$^{+0.009 }_{-0.007 }$  \\ \hline
-0.7 &-- &-0.4 &	 1.142	 $\pm$ 0.008$^{+0.010 }_{-0.012 }$ &	 1.055	 $\pm$ 0.008$^{+0.022 }_{-0.007 }$  \\ \hline
-0.4& -- &-0.1 &	 1.427	 $\pm$ 0.009$^{+0.016 }_{-0.013 }$ &	 1.370	 $\pm$ 0.009$^{+0.019 }_{-0.007 }$  \\ \hline
-0.1 &-- &0.2 &	 1.692	 $\pm$ 0.009$^{+0.027 }_{-0.018 }$ &	 1.634	 $\pm$ 0.010$^{+0.037 }_{-0.007 }$  \\ \hline
0.2& -- &0.5 &	 1.882	 $\pm$ 0.010$^{+0.049 }_{-0.017 }$ &	 1.871	 $\pm$ 0.010$^{+0.064 }_{-0.012 }$  \\ \hline
0.5 &-- &0.8 &	 2.049	 $\pm$ 0.010$^{+0.076 }_{-0.018 }$ &	 2.072	 $\pm$ 0.011$^{+0.074 }_{-0.017 }$  \\ \hline
0.8& -- &1.1 &	 2.177	 $\pm$ 0.010$^{+0.091 }_{-0.021 }$ &	 2.265	 $\pm$ 0.012$^{+0.105 }_{-0.023 }$  \\ \hline
1.1 &-- &1.4 &	 2.269	 $\pm$ 0.011$^{+0.077 }_{-0.032 }$ &	 2.365	 $\pm$ 0.012$^{+0.115 }_{-0.015 }$  \\ \hline
1.4 &-- &1.7 &	 2.290	 $\pm$ 0.011$^{+0.067 }_{-0.017 }$ &	 2.429	 $\pm$ 0.012$^{+0.120 }_{-0.028 }$  \\ \hline
\hline
\end{tabular}

\caption{
The charged-particle density, $1/N~dn^{\pm} /d\eta^{\rm Breit}$, 
for $20<Q^2<30$~GeV~$^2$ and 2 bins in $W$.
The first uncertainty is statistical, the second systematic.
}
\label{tab-q2}
\end{table}

\begin{table}
\centering
   \begin{tabular}{|r l r|c|c|}
\hline
\hline
\multicolumn{3}{|c|}{$\eta^{\rm Breit}$}  &  $80<W<120$~GeV  & $120<W<160$~GeV \\ \hline
\hline
-3.1& -- &-2.8 &	 0.0156	 $\pm$ 0.0006 $^{+0.0002 }_{-0.0007 }$ &	 0.0138	 $\pm$ 0.0006$^{+0.0006 }_{-0.0007 }$  \\ \hline
-2.8& -- &-2.5 &	 0.0300	 $\pm$ 0.0008$^{+0.0007 }_{-0.0013 }$ &	 0.0252	 $\pm$ 0.0008$^{+0.0012 }_{-0.0014 }$  \\ \hline
-2.5& -- &-2.2 &	 0.0548	 $\pm$ 0.0011$^{+0.0023 }_{-0.0020 }$ &	 0.0458	 $\pm$ 0.0011$^{+0.0019 }_{-0.0019 }$  \\ \hline
-2.2& -- &-1.9 &	 0.0915	 $\pm$ 0.0014$^{+0.0015 }_{-0.0024 }$ &	 0.0785	 $\pm$ 0.0014$^{+0.0020 }_{-0.0027 }$  \\ \hline
-1.9 &-- &-1.6 &	 0.1564	 $\pm$ 0.0019$^{+0.0026 }_{-0.0034 }$ &	 0.1327	 $\pm$ 0.0018$^{+0.0024 }_{-0.0032 }$  \\ \hline
-1.6& -- &-1.3 &	 0.260	 $\pm$ 0.002$^{+0.003 }_{-0.005 }$ &	 0.223	 $\pm$ 0.002$^{+0.005 }_{-0.007 }$  \\ \hline
-1.3 &-- &-1.0 &	 0.404	 $\pm$ 0.003$^{+0.005 }_{-0.009 }$ &	 0.354	 $\pm$ 0.003$^{+0.006 }_{-0.009 }$  \\ \hline
-1.0& -- &-0.7 &	 0.591	 $\pm$ 0.003$^{+0.005 }_{-0.010 }$ &	 0.529	 $\pm$ 0.004$^{+0.011 }_{-0.012 }$  \\ \hline
-0.7 &-- &-0.4 &	 0.827	 $\pm$ 0.004$^{+0.007 }_{-0.013 }$ &	 0.769	 $\pm$ 0.004$^{+0.013 }_{-0.022 }$  \\ \hline
-0.4& -- &-0.1 &	 1.100	 $\pm$ 0.005$^{+0.015 }_{-0.016 }$ &	 1.051	 $\pm$ 0.005$^{+0.017 }_{-0.024 }$  \\ \hline
-0.1 &-- &0.2 &	 1.409	 $\pm$ 0.005$^{+0.013 }_{-0.016 }$ &	 1.369	 $\pm$ 0.006$^{+0.024 }_{-0.027 }$  \\ \hline
0.2& -- &0.5 &	 1.665	 $\pm$ 0.006$^{+0.032 }_{-0.018 }$ &	 1.667	 $\pm$ 0.006$^{+0.042 }_{-0.010 }$  \\ \hline
0.5 &-- &0.8 &	 1.900	 $\pm$ 0.006$^{+0.050 }_{-0.016 }$ &	 1.929	 $\pm$ 0.007$^{+0.066 }_{-0.009 }$  \\ \hline
0.8& -- &1.1 &	 2.086	 $\pm$ 0.007$^{+0.069 }_{-0.014 }$ &	 2.132	 $\pm$ 0.007$^{+0.099 }_{-0.011 }$  \\ \hline
1.1 &-- &1.4 &	 2.202	 $\pm$ 0.007$^{+0.086 }_{-0.015 }$ &	 2.304	 $\pm$ 0.008$^{+0.101 }_{-0.016 }$  \\ \hline
1.4 &-- &1.7 &	 2.259	 $\pm$ 0.007$^{+0.081 }_{-0.021 }$ &	 2.395	 $\pm$ 0.008$^{+0.138 }_{-0.013 }$  \\ \hline
\hline
\end{tabular}

\caption{
The charged-particle density, $1/N~dn^{\pm} /d\eta^{\rm Breit}$,  
for $10<Q^2<20$~GeV~$^2$ and 2 bins in $W$.
The first uncertainty is statistical, the second systematic.}
\label{tab-q1}
\end{table}

%-------------------------------------------------------------------------------
%       end of tables  
%-------------------------------------------------------------------------------

%------------------------------------------------------------------------------
%       Figures
%------------------------------------------------------------------------------
%-------------------------------------------------------------------------------
%       Results
%-------------------------------------------------------------------------------
%1

%2

%3

%4

%5

%6

%7

%8

%
%       ... that's it
%

\begin{mcbibliography}{10}

\bibitem{zfp:c67:93}
ZEUS \coll, M.~Derrick \etal,
\newblock Z.\ Phys.{} {\bf C~67},~93~(1995)\relax
\relax
\bibitem{pl:b414:428}
ZEUS \coll, J.~Breitweg \etal,
\newblock Phys.\ Lett.{} {\bf B~414},~428~(1997)\relax
\relax
\bibitem{epj:c11:251}
ZEUS \coll, J.~Breitweg \etal,
\newblock Eur.\ Phys.\ J.{} {\bf C~11},~251~(1999)\relax
\relax
\bibitem{h1:frag}
H1 Coll., F.D. Aaron \etal,
\newblock Phys.~Lett.{} {\bf B~654},~148~(2007)\relax
\relax
\bibitem{alt79}
G.~Altarelli \etal,
\newblock Nucl.\ Phys.{} {\bf B~160},~301~(1979)\relax
\relax
\bibitem{webnas}
P.~Nason and B.R.~Webber,
\newblock Nucl.\ Phys.{} {\bf B~421},~473~(1994)\relax
\relax
\bibitem{Fong:1989qy}
C.P.~Fong and B.R.~Webber,
\newblock Phys. Lett.{} {\bf B~229},~289~(1989)\relax
\relax
\bibitem{Fong:1990nt}
C.P.~Fong and B.R.~Webber,
\newblock Nucl. Phys.{} {\bf B~355},~54~(1991)\relax
\relax
\bibitem{Dokshitzer:1991ej}
Y.L.~Dokshitzer, V.A.~Khoze and S.I.~Troian,
\newblock Int.~J.~Mod.~Phys.{} {\bf A~7},~1875~(1992)\relax
\relax
\bibitem{MLLA}
Y.I.~Dokshitzer \etal,
\newblock {\em Basics of Perturbative QCD}.
\newblock Editions Fronti\'eres, Gif-sur-Yvette, France,
p.169-196, 1991\relax
\relax
\bibitem{Azimov:1984np}
Y.I.~Azimov \etal,
\newblock Z.~Phys.{} {\bf C~27},~65~(1985)\relax
\relax
\bibitem{Kretzer:2000yf}
S.~Kretzer,
\newblock Phys. Rev.{} {\bf D~62},~054001~(2000)\relax
\relax
\bibitem{Kniehl:2000cr}
B.A.~Kniehl, G.~Kramer and B.~P\"otter,
\newblock Phys. Rev. Lett.{} {\bf 85},~5288~(2000)\relax
\relax
\bibitem{Albino:2005me}
S.~Albino, B.A.~Kniehl and G.~Kramer,
\newblock Nucl. Phys.{} {\bf B~725},~181~(2005)\relax
\relax
\bibitem{Albino:2008fy}
S.~Albino \etal,
\newblock Nucl. Phys.{} {\bf B~803},~42~(2008)\relax
\relax
\bibitem{Florian:2007aj}
D.~De Florian, R.~Sassot and M.~Stratmann,
\newblock Phys.~Rev.{} {\bf D~75},~114010~(2007)\relax
\relax
\bibitem{Florian:2007hc}
D.~De Florian, R.~Sassot and M.~Stratmann,
\newblock Phys. Rev.{} {\bf D 76},~074033~(2007)\relax
\relax
\bibitem{Khoze:1996dn}
V.A.~Khoze and W.~Ochs,
\newblock Int. J. Mod. Phys.{} {\bf A~12},~2949~(1997)\relax
\relax
\bibitem{Khoze:1996py}
V.A.~Khoze, S.~Lupia and W.~Ochs,
\newblock Phys.~Lett.{} {\bf B~386},~451~(1996)\relax
\relax
\bibitem{mark2:1988}
MARK II \coll, A.~Petersen \etal,
\newblock Phys.~Rev.{} {\bf D~37},~1~(1988)\relax
\relax
\bibitem{Braunschweig:1990yd}
TASSO \coll, W.~Braunschweig \etal,
\newblock Z.~Phys.{} {\bf C~47},~187~(1990)\relax
\relax
\bibitem{Li:1989sn}
AMY \coll, Y.K.~Li \etal,
\newblock Phys. Rev.{} {\bf D 41},~2675~(1990)\relax
\relax
\bibitem{delphi:1993}
DELPHI \coll, P.~Abreu \etal,
\newblock Phys.~Lett.{} {\bf B~311},~408~(1993)\relax
\relax
\bibitem{:2008hy}
ZEUS \coll, S.~Chekanov \etal,
\newblock JHEP{} {\bf 06},~061~(2008)\relax
\relax
\bibitem{limiting}
J.~Benecke \etal,
\newblock Phys.\ Rev.\ Lett.{} {\bf 188},~2159~(1969)\relax
\relax
\bibitem{Back:2005hs}
PHOBOS \coll, B.B.~Back \etal,
\newblock Phys. Rev.{} {\bf C~74},~021901~(2006)\relax
\relax
\bibitem{Adams:2005cy}
STAR \coll, J.~Adams \etal,
\newblock Phys. Rev.{} {\bf C~73},~034906~(2006)\relax
\relax
\bibitem{PHOBOS}
PHOBOS \coll, B.B.~Back \etal,
\newblock Phys.\ Rev.\ Lett.{} {\bf 91},~052303~(2003)\relax
\relax
\bibitem{Bearden:2001qq}
BRAHMS \coll, I.G.~Bearden \etal,
\newblock Phys. Rev. Lett.{} {\bf 88},~202301~(2002)\relax
\relax
\bibitem{Bearden:2001xw}
BRAHMS \coll, I.G.~Bearden \etal,
\newblock Phys. Lett.{} {\bf B~523},~227~(2001)\relax
\relax
\bibitem{zeus:1993:bluebook}
ZEUS \coll, U.~Holm~(ed.),
\newblock {\em The {ZEUS} Detector}.
\newblock Status Report (unpublished), DESY (1993),
\newblock available on
  \texttt{http://www-zeus.desy.de/bluebook/bluebook.html}\relax
\relax
\bibitem{nim:a279:290}
N.~Harnew \etal,
\newblock Nucl.\ Inst.\ Meth.{} {\bf A~279},~290~(1989)\relax
\relax
\bibitem{npps:b32:181}
B.~Foster \etal,
\newblock Nucl.\ Phys.\ Proc.\ Suppl.{} {\bf B~32},~181~(1993)\relax
\relax
\bibitem{nim:a338:254}
B.~Foster \etal,
\newblock Nucl.\ Inst.\ Meth.{} {\bf A~338},~254~(1994)\relax
\relax
\bibitem{nim:a581:656}
A.~Polini \etal,
\newblock Nucl.\ Inst.\ Meth.{} {\bf A~581},~656~(2007)\relax
\relax
\bibitem{nim:a309:77}
M.~Derrick \etal,
\newblock Nucl.\ Inst.\ Meth.{} {\bf A~309},~77~(1991)\relax
\relax
\bibitem{nim:a309:101}
A.~Andresen \etal,
\newblock Nucl.\ Inst.\ Meth.{} {\bf A~309},~101~(1991)\relax
\relax
\bibitem{nim:a321:356}
A.~Caldwell \etal,
\newblock Nucl.\ Inst.\ Meth.{} {\bf A~321},~356~(1992)\relax
\relax
\bibitem{nim:a336:23}
A.~Bernstein \etal,
\newblock Nucl.\ Inst.\ Meth.{} {\bf A~336},~23~(1993)\relax
\relax
\bibitem{nim:a401:63}
A.~Bamberger \etal,
\newblock Nucl.\ Inst.\ Meth.{} {\bf A~401},~63~(1997)\relax
\relax
\bibitem{nim:a382:419}
A.~Bamberger \etal,
\newblock Nucl.\ Inst.\ Meth.{} {\bf A~382},~419~(1996)\relax
\relax
\bibitem{desy-92-066}
J.~Andruszk\'ow \etal,
\newblock Preprint \mbox{DESY-92-066}, DESY, 1992\relax
\relax
\bibitem{zfp:c63:391}
ZEUS \coll, M.~Derrick \etal,
\newblock Z.\ Phys.{} {\bf C~63},~391~(1994)\relax
\relax
\bibitem{acpp:b32:2025}
J.~Andruszk\'ow \etal,
\newblock Acta Phys.\ Pol.{} {\bf B~32},~2025~(2001)\relax
\relax
\bibitem{nim:a565:572}
M.~Helbich \etal,
\newblock Nucl.\ Inst.\ Meth.{} {\bf A 565},~572~(2006)\relax
\relax
\bibitem{Smith:1992im}
W.H.~Smith, K.~ Tokushuku and L.W.~Wiggers, {\it The ZEUS trigger system},
\newblock Proceedings  Computing in High
  Energy Physics (CHEP 92), Annecy, France, 21-25 Sept 1992,
C. Verkerk and W. Wojcik
(eds.), CERN, Geneva Switzerland (1992), p.222, DESY-92-150B\relax
\relax
\bibitem{gtt}
P.D.~Allfrey \etal,
\newblock Nucl.\ Inst.\ Meth.{} {\bf A 580},~1257~(2007)\relax
\relax
\bibitem{nim:a365:508}
H.~Abramowicz, A.~Caldwell and R.~Sinkus,
\newblock Nucl.\ Inst.\ Meth.{} {\bf A~365},~508~(1995)\relax
\relax
\bibitem{proc:hera:1991:23}
S.~Bentvelsen, J.~Engelen and P.~Kooijman,
\newblock {\em Proc.\ Workshop on Physics at {HERA}}, W.~Buchm\"uller and
  G.~Ingelman~(eds.), Vol.~1, p.~23.
\newblock Hamburg, Germany, DESY (1992)\relax
\relax
\bibitem{Beata}
B.~Brzozowska,
\newblock Thesis, University of Warsaw{},~to be published\relax
\relax
\bibitem{proc:epfacility:1979:391}
F.~Jacquet and A.~Blondel,
\newblock {\em Proceedings of the Study for an $ep$ Facility for {Europe}},
  U.~Amaldi~(ed.), p.~391.
\newblock Hamburg, Germany (1979).
\newblock Also in preprint \mbox{DESY 79/48}\relax
\relax
\bibitem{albino-2007-75}
S.~Albino \etal,
\newblock Phys.~Rev.~D{} {\bf 75},~034018~(2007)\relax
\relax
\bibitem{Bassetto:1982ma}
A.~Bassetto \etal,
\newblock Nucl. Phys.{} {\bf B~207},~189~(1982)\relax
\relax
\bibitem{Mueller:1982cq}
A.H.~Mueller,
\newblock Nucl. Phys.{} {\bf B~213},~85~(1983)\relax
\relax
\bibitem{Webber:1983if}
B.R.~Webber,
\newblock Nucl. Phys.{} {\bf B~238},~492~(1984)\relax
\relax
\bibitem{Chekanov:2009kb}
ZEUS \coll, S.~Chekanov \etal,
\newblock JHEP{} {\bf 08},~077~(2009)\relax
\relax
\bibitem{cpc:71:15}
L.~L\"onnblad,
\newblock Comp.\ Phys.\ Comm.{} {\bf 71},~15~(1992)\relax
\relax
\bibitem{CDM}
G.~Gustafson and U.~Petterson,
\newblock Nucl. Phys.{} {\bf B~306},~746~(1988)\relax
\relax
\bibitem{cpc:101:108}
G.~Ingelman, A.~Edin and J.~Rathsman,
\newblock Comp.\ Phys.\ Comm.{} {\bf 101},~108~(1997)\relax
\relax
\bibitem{cpc:81:381}
K.~Charchula, G.A.~Schuler and H.~Spiesberger,
\newblock Comp.\ Phys.\ Comm.{} {\bf 81},~381~(1994)\relax
\relax
\bibitem{cpc:69:155}
A.~Kwiatkowski, H.~Spiesberger and H.-J.~M\"ohring,
\newblock Comp.\ Phys.\ Comm.{} {\bf 69},~155~(1992)\relax
\relax
\bibitem{prep:97:31}
B.~Andersson \etal,
\newblock Phys.\ Rep.{} {\bf 97},~31~(1983)\relax
\relax
\bibitem{cpc:46:43}
M.~Bengtsson and T.~Sj\"ostrand,
\newblock Comp.\ Phys.\ Comm.{} {\bf 46},~43~(1987)\relax
\relax
\bibitem{cpc:82:74}
T.~Sj\"ostrand,
\newblock Comp.\ Phys.\ Comm.{} {\bf 82},~74~(1994)\relax
\relax
\bibitem{tech:cern-dd-ee-84-1}
R.~Brun et al.,
\newblock {\em {\sc GEANT3}},
\newblock Technical Report CERN-DD/EE/84-1, CERN, 1987\relax
\relax
\bibitem{Amsler:2008zzb}
{Particle Data Group, C.~Amsler \etal},
\newblock Phys. Lett.{} {\bf B 667},~1~(2008)\relax
\relax
\bibitem{Dixon}
P.~Dixon, D.~Kant and G.~Thompson,
\newblock Nucl. Part. Phys.{} {\bf 25},~1453~(1999)\relax
\relax
\bibitem{DeinesJones:1999ap}
P.~Deines-Jones \etal,
\newblock Phys. Rev.{} {\bf C~62},~014903~(2000)\relax
\relax
\bibitem{UA5}
UA5 \coll, G.J.~Alner \etal,
\newblock Z.\ Phys.{} {\bf C 33},~1~(1986)\relax
\relax
\bibitem{Back:2004je}
PHOBOS \coll, B.B.~Back \etal,
\newblock Nucl. Phys.{} {\bf A~757},~28~(2005)\relax
\relax
\bibitem{Bialas}
A.~Bialas and M.~Jezabek,
\newblock Phys.\ Lett.{} {\bf B~590},~233~(2004)\relax
\relax
\bibitem{acta}
T.~Tymieniecka and B.~Brzozowska,
\newblock Acta Physica Polonica{} {\bf 40},~2175~(2009)\relax
\relax
\bibitem{Adeva:2004dh}
SMC Coll., B.~Adeva, et al.,
\newblock Phys. Rev.{} {\bf D~70},~012002~(2004)\relax
\relax
\end{mcbibliography}
\end{document}